\documentclass[fleqn,usenatbib]{mnras}

\usepackage{newtxtext,newtxmath}
\usepackage{mathptmx}

\usepackage[T1]{fontenc}

\usepackage{graphicx}	
\usepackage{amsmath}	
\usepackage{amssymb}	
\usepackage{subcaption}
   \title[FR dichotomy with LOFAR]{Revisiting the Fanaroff-Riley dichotomy and radio-galaxy morphology with the LOFAR Two-Metre Sky Survey (LoTSS)}

\author[B.Mingo et al.]{B. Mingo$^{1}$\thanks{Email: bmingo@extragalactic.info}, J. H. Croston$^{1}$, M. J. Hardcastle$^{2}$, P. N. Best$^{3}$, K. J. Duncan$^{4}$, \newauthor R. Morganti$^{5,6}$, H. J. A. Rottgering$^{4}$, J. Sabater$^{3}$, T. W. Shimwell$^{5,4}$, W. L. Williams$^{4}$, \newauthor M. Brienza$^{7}$, G. Gurkan$^{8}$, V. H. Mahatma$^{2}$, L. K. Morabito$^{9}$, I. Prandoni$^{7}$, M. Bondi$^{7}$, \newauthor J. Ineson$^{10}$, and S. Mooney$^{11}$\\
$^{1}$School of Physical Sciences, The Open University, Walton Hall, Milton Keynes, MK7 6AA, UK\\
$^{2}$Centre for Astrophysics Research, University of Hertfordshire, College Lane, Hatfield AL10 9AB, UK\\
$^{3}$SUPA, Institute for Astronomy, Royal Observatory, Blackford Hill, Edinburgh, EH9 3HJ, UK\\
$^{4}$Leiden Observatory, Leiden University, PO Box 9513, NL-2300 RA Leiden, the Netherlands\\
$^{5}$ASTRON, the Netherlands Institute for Radio Astronomy, Postbus 2, 7990 AA, Dwingeloo, The Netherlands\\
$^{6}$Kapteyn Astronomical Institute, University of Groningen, P.O. Box 800, 9700 AV Groningen, The Netherlands\\
$^{7}$INAF-Istituto di Radioastronomia, Via P. Gobetti 101, 40129 Bologna, Italy\\
$^{8}$CSIRO Astronomy and Space Science (CASS) PO Box 1130, Bentley WA 6102, Perth, Australia\\
$^{9}$Astrophysics, University of Oxford, Denys Wilkinson Building, Keble Road, Oxford, OX1 3RH\\
$^{10}$School of Physics and Astronomy, University of Southampton, Highfield, Southampton SO17 1BJ, UK\\
$^{11}$School of Physics, University College Dublin, Belfield, Dublin 4, Ireland\\
}
\date{Accepted XXX. Received YYY; in original form ZZZ}

\pubyear{2019}

\hypersetup{draft}

\begin{document}
\label{firstpage}
\pagerange{\pageref{firstpage}--\pageref{lastpage}}
\maketitle

\begin{abstract}
The relative positions of the high and low surface brightness regions of radio-loud active galaxies in the 3CR sample were found by Fanaroff and Riley to be correlated with their luminosity. We revisit this canonical relationship with a sample of 5805 extended radio-loud AGN from the LOFAR Two-Metre Sky Survey (LoTSS), compiling the most complete dataset of radio-galaxy morphological information obtained to date. We demonstrate that, for this sample, radio luminosity does {\it not} reliably predict whether a source is edge-brightened (FRII) or centre-brightened (FRI). We highlight a large population of low-luminosity FRIIs, extending three orders of magnitude below the traditional FR break, and demonstrate that their host galaxies are on average systematically fainter than those of high-luminosity FRIIs and of FRIs matched in luminosity. This result supports the jet power/environment paradigm for the FR break: low-power jets may remain undisrupted and form hotspots in lower mass hosts. We also find substantial populations that appear physically distinct from the traditional FR classes, including candidate restarting sources and ``hybrids''. We identify 459 bent-tailed sources, which we find to have a significantly higher SDSS cluster association fraction (at $z<0.4$) than the general radio-galaxy population, similar to the results of previous work. The complexity of the LoTSS faint, extended radio sources demonstrates the need for caution in the automated classification and interpretation of extended sources in modern radio surveys, but also reveals the wealth of morphological information such surveys will provide and its value for advancing our physical understanding of radio-loud AGN.

\end{abstract}  

\begin{keywords}
galaxies: jets -- galaxies: active -- radio continuum: galaxies 
\end{keywords}


\section{Introduction}\label{Intro}

A correlation between the surface-brightness distributions of radio galaxies (hereafter used broadly to encompass radio-loud quasars) and their radio luminosities was established by \citet{FR1974} using the 3CR sample \citep{Mackay1971}. The Fanaroff-Riley (FR) classification has since been widely adopted and applied to many catalogues in the past four decades. Our understanding of how the FR classes relate to source dynamics and active galactic nucleus (AGN) fuelling has evolved considerably over the past few decades. Recent evidence that the FR distinction is important for assessing AGN energy output \citep[e.g.][]{Croston2018} highlights its continuing relevance; however, we still do not have a quantitative understanding of the exact conditions needed to produce a Fanaroff-Riley type I (FRI) or type II (FRII) source. 

Deep, wide-area radio surveys \citep[e.g.][]{EMU2011,MIGHTEE2016,Hurley2017,Villarreal2018,Shimwell2018} are now starting to open up the faint, distant and low surface-brightness radio Universe, and in the process are providing a comprehensive view of the radio-loud AGN population over a wide range in luminosity, with considerably less restrictive selection effects than earlier studies. Automated approaches are required to catalogue, associate and identify host galaxies for the large samples produced by modern radio surveys \citep[e.g.][]{Williams2018b}, and to categorise the resulting samples for scientific analysis \citep[e.g.][]{aniyan17,alhassan18,wu19,lukic19, Ma2019}. However, sensitive low frequency observations are at the same time revealing a more complex extended source population, including candidate hybrid radio galaxies, restarting and remnant radio galaxies \citep[e.g.][]{Kapinska2017,Brienza2016b,Brienza2017,Mahatma2018a,Mahatma2018b}. Simple classification schemes may therefore risk obscuring important physical distinctions. With the availability of large, new samples of extended radio sources, it is timely to revisit the applicability and usefulness of Fanaroff-Riley classifications for 21st-century radio survey populations, and to use these new, large samples to advance our understanding of what determines radio-galaxy physical evolution and its environmental impact.

While there remains considerable debate about the link between accretion mode and jet morphology \citep[e.g.][]{Best2012,Gendre2013,Mingo2014,Ineson2015,Tadhunter2016,Hardcastle2007,Hardcastle2009,Hardcastle2018b}, the FR morphological divide is primarily explained as a difference in jet dynamics: the edge-brightened FRII radio galaxies are thought to have jets that remain relativistic throughout, terminating in a hotspot (internal shock), while the centre-brightened FRIs are believed to disrupt on kpc scales \citep[e.g.][]{Bicknell1995,Laing2002,Tchekhovskoy2016}. It has also long been suggested that this structural difference is caused by the interplay of jet power and (host-scale) environmental density, so that jets of the same power will disrupt (and thus become FRI) more easily in a rich environment compared to a poor one \citep{Bicknell1995,Kaiser2007}. Such an explanation seemed to find support in the discovery by \citet{Ledlow1996} that the FRI/II luminosity break is dependent on host-galaxy magnitude, so that FRIs are found to have higher radio luminosities in brighter host galaxies (where the density of the interstellar medium is assumed to be higher). However, this result was based on highly flux-limited samples, with different redshift distributions and environments for the FRIs and FRIIs, and so serious selection effects have led to some uncertainty as to whether this relation in fact holds across the full population of radio galaxies \citep{Best2009, Lin2010, Wing2011, Singal2014, Capetti2017FRII, Shabala2018}.

An additional complication in using radio observational data to test physical models for jet dynamics and the FR divide is the weak relationship between jet power and radio luminosity. In particular, a systematic difference in the efficiency of producing radio luminosity for a given jet power for FRIs and FRIIs is thought to exist \citep{Croston2018}, caused by the correlation of FR class with lobe particle content. FRI radio galaxies are found to be energetically dominated by heavy particles (protons and ions), while FRII radio galaxies are primarily composed of an electron-positron plasma \citep{Croston2018} -- this situation may be best explained by the role of entrainment of surrounding material into disrupted FRI jets as they decelerate, while undisrupted FRII jets remain more ``pristine''. The combined effect of systematic differences in particle content, environmental effects and radiative losses leads to substantial caveats in the use of radio luminosity as a proxy for jet power \citep[e.g.][]{Croston2018,Hardcastle2018}. This creates challenges for the estimation of radio-source energy output and feedback energetics \citep[e.g.][]{sabater2018,Hardcastle2018c}.

The relevance of morphology to the inference of environmental impact from (jet-driven) AGN populations found in radio surveys is therefore a strong motivation to obtain a better physical understanding of the FR break, and of the full morphological diversity of the radio-loud AGN population. The LOFAR Two-Metre Sky Survey \citep[LoTSS][]{Shimwell2017} provides us with an opportunity to explore these questions in much greater depth than has previously been possible. It is an order of magnitude deeper than previous wide-area radio surveys, with sensitivity to structure on angular scales ranging from 6 arcsec to $\gtrapprox1$ degree, and so comprises the best dataset of radio-galaxy morphological information ever compiled. In this paper we carry out an in-depth morphological examination of the LoTSS AGN population, combining an automated classification algorithm with careful visual analysis. We use our LoTSS morphological catalogue to investigate the relationship between source morphology, radio luminosity, and optical host-galaxy properties. In Section~\ref{Analysis} we provide further details of our new dataset derived from LoTSS Data Release 1 \citep[DR1,][]{Shimwell2018}, followed by an explanation of our methods for morphological classification. In Section~\ref{Results}, we present the overall morphological properties of the sample and their relation to host-galaxy properties, and then in Section~\ref{discuss} discuss our interpretation of results for the FRI and FRII populations, including some interesting subpopulations, before presenting our conclusions in Section~\ref{Conclusions}.

For this paper we have used a concordance cosmology with $H_{0} = 70$ km s$^{−1}$ Mpc$^{−1}$, $\Omega_{m}= 0.3$ and $\Omega_{\Lambda}= 0.7$.


\section{Data and analysis}\label{Analysis}

\subsection{The LoTSS datasets}

We make use of the LOFAR Two-Metre Sky Survey DR1 value-added catalogue \citep[LoTSS-DR1][]{Shimwell2018, Williams2018b} to explore the relationship between radio morphology, luminosity, and host properties for radio-loud AGN. LoTSS-DR1 contains 318,520 sources over 424 deg$^2$ of the northern sky. Of the LoTSS sources, 73 per cent have optical identifications \citep{Williams2018b} and 51 per cent have either spectroscopic or photometric redshifts \citep{Duncan2018b}. Our aim is to investigate the population of radio-loud AGN within this catalogue, so we restrict our analysis to the radio-loud AGN sample of \citet{Hardcastle2018c}, which contains 23,344 sources. The sample in that work was designed to both minimise the contamination by star-forming galaxies and exclude AGN with less reliable redshifts. Both aims are also important considerations for this work, which justify excluding large numbers of LoTSS extended radio galaxies for which it would not be possible to obtain well-determined luminosities or physical sizes. As host-galaxy properties enter into the AGN sample selection, it is possible that the sample is biased against certain sub-populations (e.g. faint radio-loud AGN in highly star-forming hosts). However, for smaller fainter sources we expect that a purely morphological analysis would have difficulty distinguishing the extended emission of FRI radio galaxies from distant star-forming galaxies, and so to avoid contamination from ordinary galaxies we use the AGN sample. Completeness of the AGN sample is discussed in \citet{Hardcastle2018c}, and we comment specifically on the implications for our results in Section~\ref{selection}.

Although our sample is derived from the catalogues of \citet{Shimwell2017} and \citet{Williams2018b}, we make use of images obtained by reprocessing the DR1 area data using a newer version (2.2) of the LoTSS survey pipeline, which makes use of the enhancements that were briefly outlined in \citet{Shimwell2018} (Section 5)
and will be described fully by \citet{tasse19}. This improved imaging has allowed us to include fainter structures, and better characterise the sizes and morphologies of our sources.

Our initial aim is to identify clean samples of FRI and FRII radio galaxies, to enable us to study the relationship between radio luminosity, morphology and host-galaxy properties. In addition to avoiding contamination from nearby star-forming galaxies, it is also necessary to discard objects that are too faint or small to allow morphological classification. After some preliminary visual inspection, we discarded all sources with total flux less than 2 mJy or with catalogued size less than 12 arcsec. LoTSS has a spatial resolution of 6 arcsec, and so a source of 12 arcsec is only two beamwidths across. However, the catalogued sizes (based on \textsc{PyBDSF}, see \citealt{Shimwell2018}) are not always accurate (see the discussion in Section~\ref{SizeFlux}), and so we initially retain sources down to this size for more careful size and flux estimation. The flux cut eliminates sources that would have too few pixels above our noise cut (see Section \ref{Code}) to allow classification, even if their catalogued sizes did pass the 12 arcsec selection. Our initial flux and size filtering leads to a sample of 6850 sources. With further filtering (described in Section~\ref{Code}) we obtain a well-resolved AGN sample of 5805 radio galaxies.

We carried out our morphological classification primarily via a \textsc{Python} code, which automatically classifies sources as FRI, FRII, candidate hybrid (FRI on one side, FRII on the other), or unknown. Extensive visual checking and optimisation led to the conclusion that while our automated method achieves good reliability for some flux and size categories, several types of ``contamination'' of the FRI and FRII classes proved difficult to remove in an automated way. We therefore carried out a further step of visual examination for problematic subsets. We first describe our automated classification in Section~\ref{Code}, followed by a discussion of its reliability in Section~\ref{reliable}, and then, in Section~\ref{vis}, discuss manual adjustments to create a final sample via visual inspection so as to optimise the sample's reliability for science analysis. Finally, selection effects are discussed in Section~\ref{selection}.

Our LoTSS morphological catalogue containing classifications for 5805 extended radio-loud AGN is available from \url{www.lofar-surveys.org/releases.html}.


\subsection{Automated morphological classification}\label{Code}

Our LoMorph \textsc{Python} code\footnote{\url{https://github.com/bmingo/LoMorph/}} takes FITS image cutouts of each source as input, masking all pixels with flux values below a fixed threshold to ensure that only real emission from the source is considered. The choice of RMS noise threshold is not straightforward. Too low a threshold will lead to overestimation of source size, and misclassification particularly of bright, dynamic-range limited sources (where deconvolution artifacts may be present); however, too high a threshold will risk eliminating low surface brightness structures and underestimating source sizes of FRIs with faint edges. This balance must be carefully calibrated according to the characteristics of each dataset - e.g. for data with higher or more uneven noise a higher RMS threshold might work better. After thorough testing we found that the optimal compromise that best exploited the current LoTSS DR2 images was to set the threshold at the higher value of either $4 \times$ the local RMS noise  (determined from the source maps in a box six times the source size, iteratively removing outliers to exclude any sources in the region), or 1/50th of the peak flux.

The basic sizes and shapes of the sources being examined have been catalogued using \textsc{PyBSDF} \citep{PyBDSF2015,Shimwell2018,Williams2018b} and, for most of the sources large enough to be included in our sample, \textsc{PyBDSF} was complemented by Zooniverse visual classification \citep{Williams2018b}. The catalogue size and flux measurements based on single Gaussian components or sums of components provide a good approximation to the source properties, but in a substantial fraction of cases the source's associated components do not encompass the full extent of the source. We therefore adopt a flood-filling procedure to obtain a masked region encompassing the full source extent, prior to carrying out morphological classification. To prevent the flooded region from leaking to adjacent sources we make use of the value-added catalogue information to include all associated components, and mask out any components catalogued to be unassociated with the source being examined. Pixels within any non-associated components at a distance $<90$ arcsec from the optical host are masked out. This distance was chosen to maximise computational speed, without sacrificing precision, as the number of catalogued components $\geq1.5$ arcmin is negligible ($<0.5$ percent), even without considering the probability of them being close to another, non-associated component.

Flood-filling is then carried out on the masked numpy array \citep{numpyArray}, using the \textsc{Python} \texttt{skimage.measure} module of \texttt{Scikit.image} \citep{scikit-image}, specifically the \texttt{label} routine\footnote{\url{http://scikit-image.org/docs/dev/api/skimage.measure.html\#skimage.measure.label}}, which assigns labels to connected islands of pixels on an image. From the image we create a binary mask, with zero values where the pixel fluxes are below 4 RMS (or belong to nearby, unassociated sources), and 1 for pixels above the threshold. As we want (in some cases) to extend the source beyond the catalogued regions, these regions act as a minimum boundary: we pre-fill all the component regions associated to the source of interest with arbitrary flux values above the threshold, to ensure that all the pixels within are included (with values equal to 1) in the mask. We then apply the label submodule, and identify the islands of pixels associated with the source. If there is connected emission above the RMS cut just outside the source components, they will be identified as part of the same island of pixels. These islands are then used to create a new mask for the original image, so that all emission associated with the source is included, and everything else is masked out. We then use the masked region to re-calculate the total flux from all the associated source pixels above the RMS threshold, and the size in arcseconds, from the new maximum extent of the source in pixels. We discuss the implications of these size and flux re-calculations in Section~\ref{SizeFlux}.

\begin{figure*}
	\centering
	\includegraphics[width=0.75\textwidth]{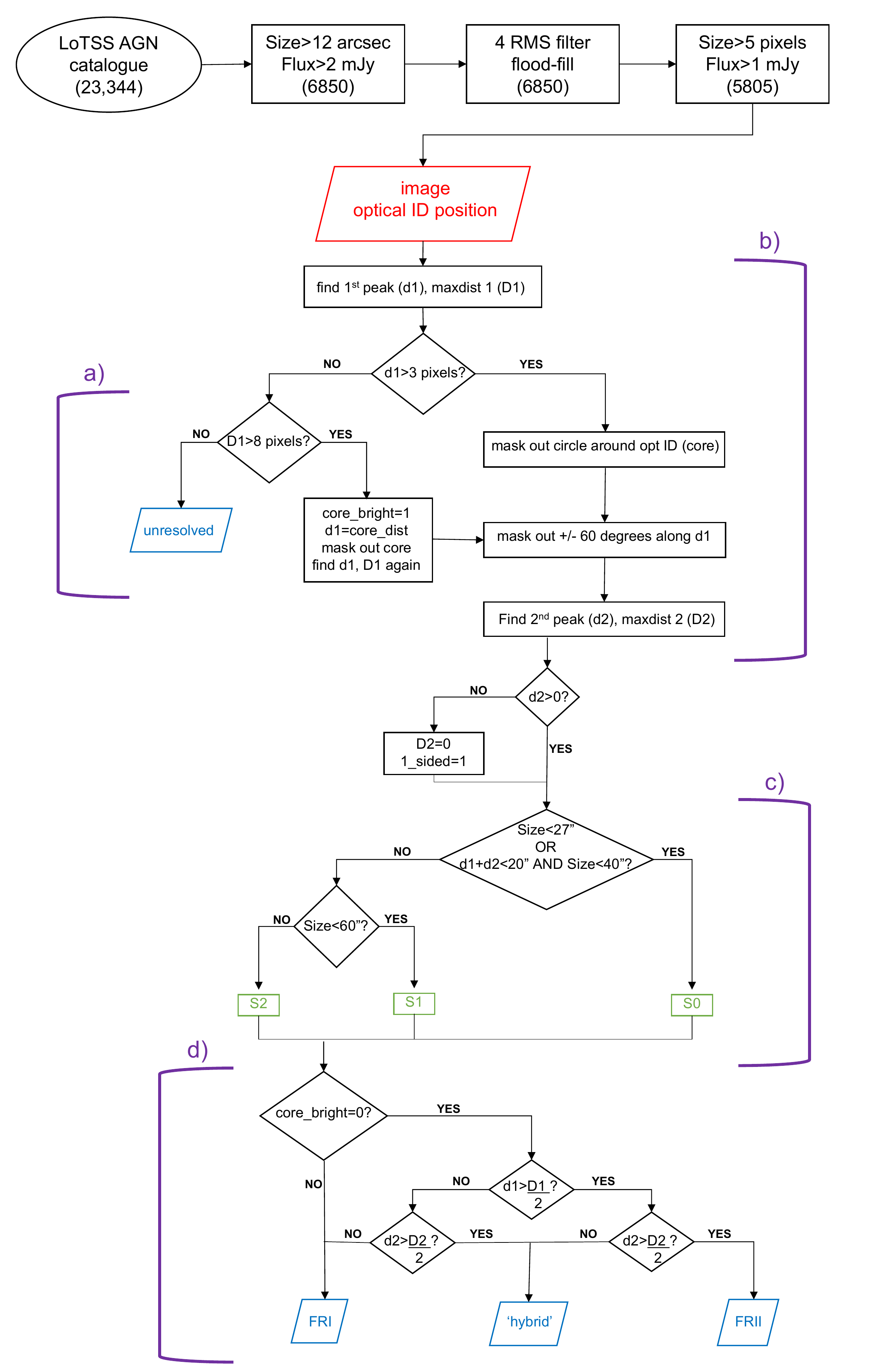}
	\caption{Flowchart for the LoMorph classification algorithm. The sample selection process is summarised at the top of the diagram, with the number of sources in parentheses at the bottom of each box. The code input is described in the red parallelogram, and the classification outputs in the blue parallelograms. The size categories (see Table~\ref{RelLabels}) are highlighted in green. The purple brackets and labels indicate the main four tasks the code carries out: a) filtering out the sources that are too small to be reliably classified; b) finding the peaks of emission and maximum extent of the source; c) sorting the sources in size bins; d) classifying the sources according to their Fanaroff-Riley types. The classification statistics are listed in Table~\ref{Class_stats}. See the main text for a detailed description of the methodology.} \label{Flowchart}
\end{figure*}

Structures with very low surface brightness can fall below the RMS threshold. This is not an issue for our purposes, as, although we want to maximise the number of sources in our samples, the classification of sources with very faint, low dynamic range structures would be less reliable. As such, at this stage a second filter is applied, to eliminate any remaining sources smaller than 5 pixels or with (re-calculated) total fluxes below 1 mJy, as these sources would be too small and faint to classify. This second flux and size filtering leads to a well-resolved AGN sample of 5805 radio galaxies. The full sample selection process is summarised at the top of Fig.~\ref{Flowchart}.

Morphological classification is then carried out on the masked array, making use of the catalogued optical host-galaxy position \citep{Williams2018b} to improve reliability. The incorporation of host-galaxy information limits the applicability of our method to objects that have a host ID; however, unidentified radio galaxies are of very limited use for our science aims as we require luminosity and physical size information. The use of a host-galaxy position enables us to apply the classification to each side of the source separately (for two-sided sources). 

We adopt the traditional definition of FR class \citep{FR1974}: if the brightest region is closer to the core (host) than the midpoint of the source on a given side, then it is an FRI; if the brightest region is more distant than the midpoint then it is an FRII. We use fluxes averaged over 4 pixels (6 arcsec) to calculate the position of the brightest points, to have the best representation of their associated structures, and to minimise the impact of the fact that the pixel size undersamples the beam. If the FR class is determined to be different for each side, the source is classified as a candidate hybrid (we discuss these objects further in Section~\ref{hybrids}).

The full classification algorithm is summarized in Fig.~\ref{Flowchart}, which includes some additional refinements to improve reliability. Steps are included to identify one-sided sources, and size thresholds are used to separate sources whose peaks are too close to enable FRI morphology to be distinguished -- this avoids discarding FRII sources that can reliably be identified at smaller sizes than FRIs, whose peak positions may be consistent with either class. Masking of the core region is used for calculation of the second side of the source, which prevents incorrect identification of the second peak direction. The sources are also categorised into size bins, summarised in Table~\ref{RelLabels}, for use in reliability checking (Section~\ref{reliable}).

For the classification (see Fig.~\ref{Flowchart}), the brightest peaks of emission on each side of the source are identified as d1, d2, and the maximum extent of the source $\pm 60^{\circ}$ along their respective directions as D1, D2. To find d2, D2 a $120^{\circ}$ triangular exclusion mask is drawn along the direction to d1. If the source is one-sided, only d1, D1 are recorded (d2=0, D2=0), and the source is flagged as such; one-sided sources with FRI morphology are classified as FRI (see also the discussion in Section \ref{NAT_WAT}), while those that fulfil the FRII peak distance criteria are classified as hybrid candidates, as they cannot be accurately characterised. If the core is the brightest structure in a source, its distance to the optical position is recorded (core\textunderscore{}dist), it is masked out to identify the remaining structures, and the source is flagged as core-bright. The various distance thresholds in pixels have been optimised for the resolution of the LOFAR beam.

\begin{figure*} 
	\begin{subfigure}[b]{0.5\linewidth}
		\centering
		\includegraphics[width=0.95\linewidth, trim={1.4cm 1cm 2cm 2.4cm},clip]{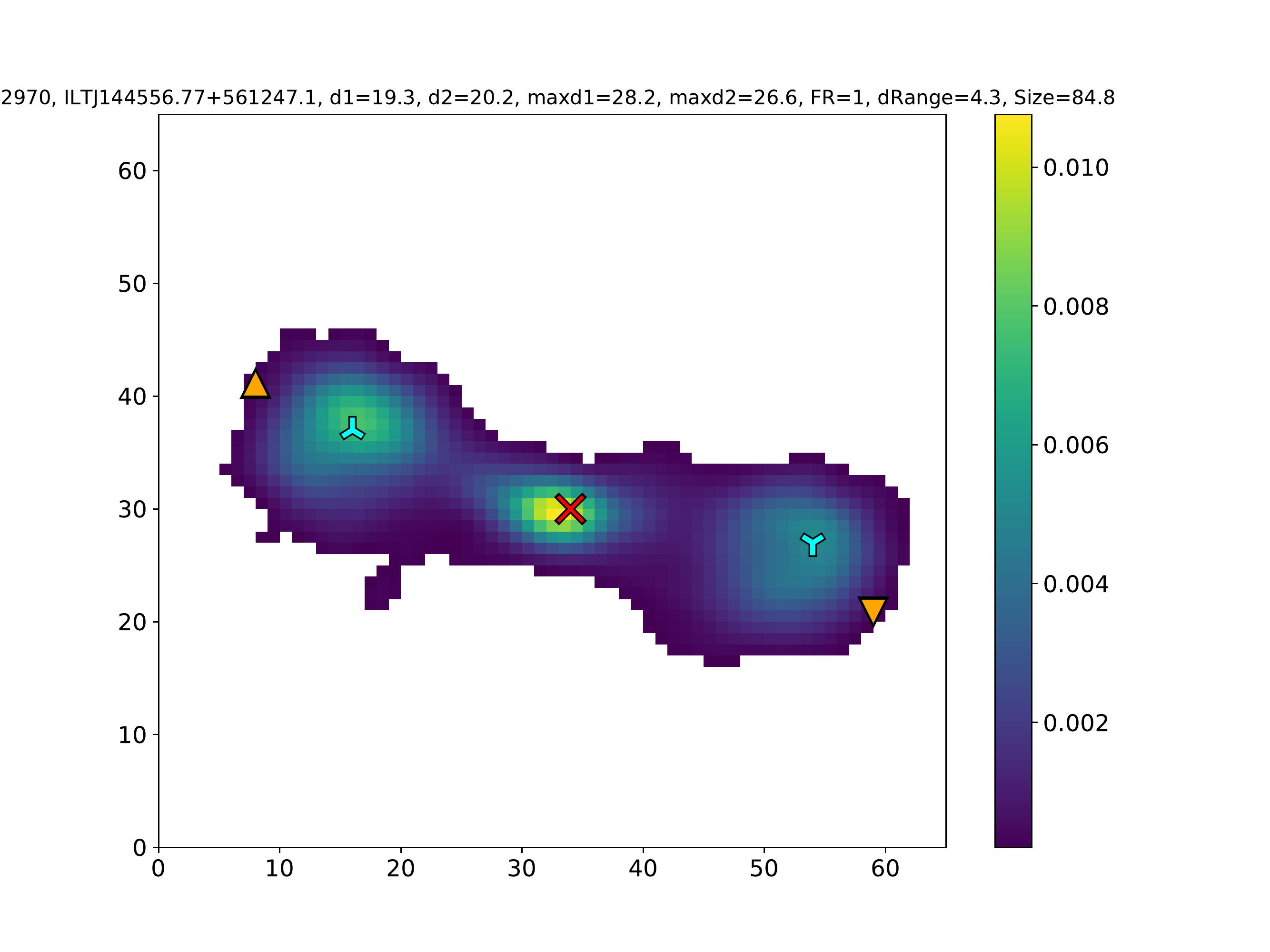} 
		\caption{Lobed FRI} 
		\label{FR1_lobed} 
		\vspace{2ex}
	\end{subfigure}
	\begin{subfigure}[b]{0.5\linewidth}
		\centering
		\includegraphics[width=0.95\linewidth, trim={1.4cm 1cm 2cm 2.4cm},clip]{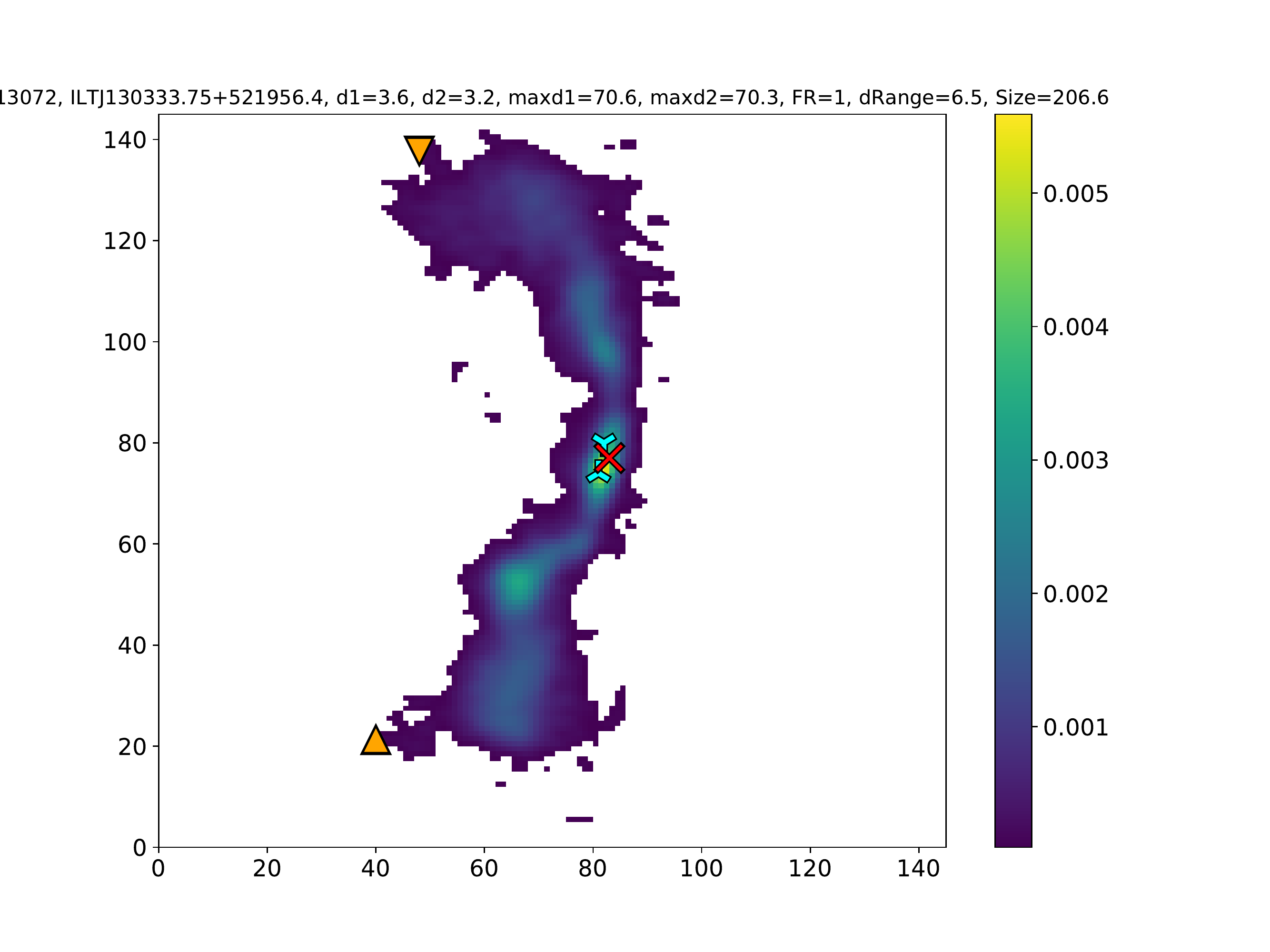} 
		\caption{Tailed FRI} 
		\label{FRI_tailed} 
		\vspace{2ex}
	\end{subfigure}
	\begin{subfigure}[b]{0.5\linewidth}
		\centering
		\includegraphics[width=0.95\linewidth, trim={1.4cm 1cm 2cm 2.4cm},clip]{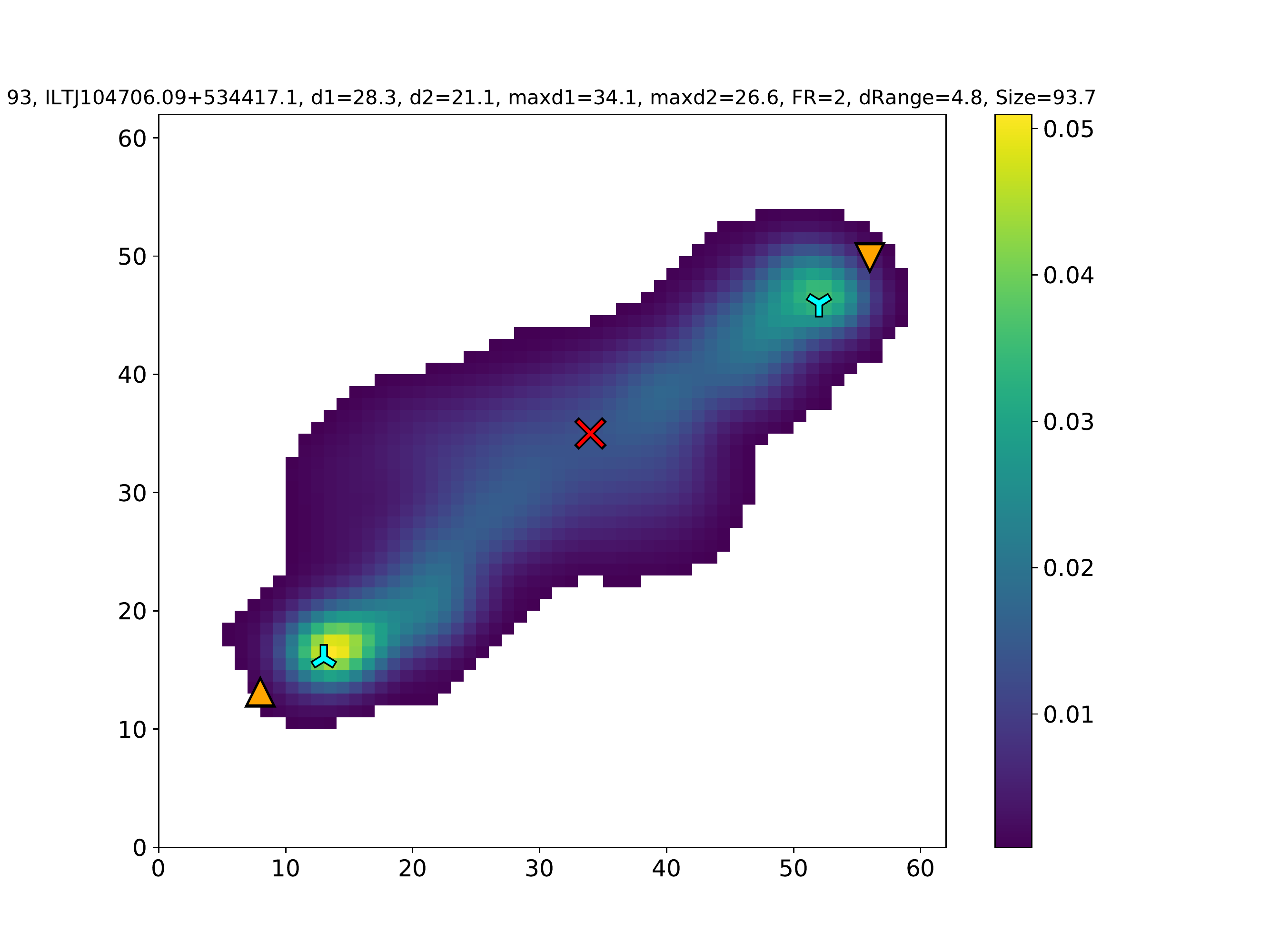} 
		\caption{FRII, all structures connected} 
		\label{FRII_connected} 
	\end{subfigure}
	\begin{subfigure}[b]{0.5\linewidth}
		\centering
		\includegraphics[width=0.95\linewidth, trim={1.4cm 1cm 2cm 2.4cm},clip]{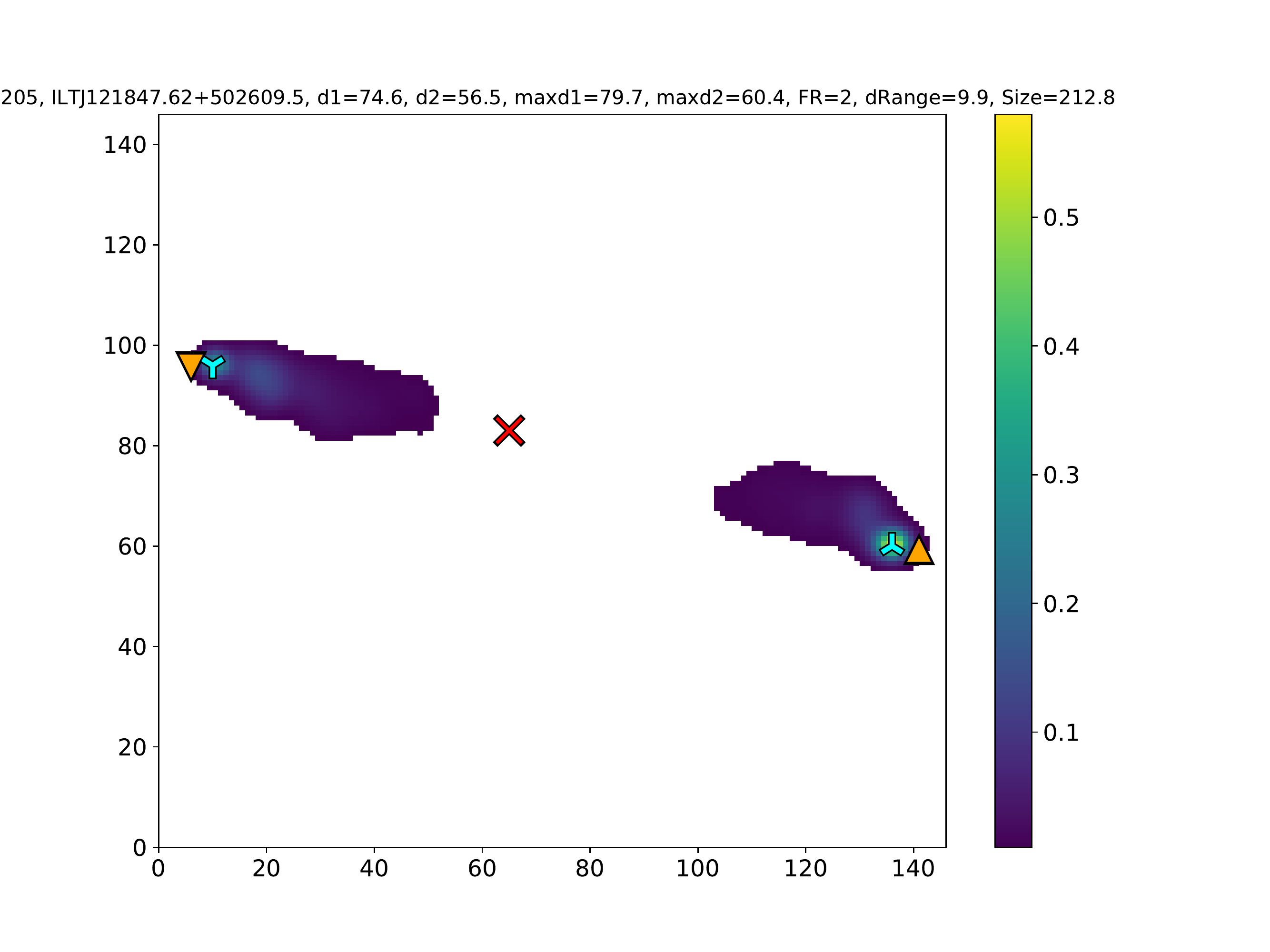} 
		\caption{FRII, structures not connected} 
		\label{FRII_disconnected} 
	\end{subfigure} 
	\caption{Examples of sources classified as FRI and FRII. The plots are produced as output by the classification code, and detail the pixel distances from the optical host (red X) to the first and second-brightest peak of emission excluding the core (d1 and d2 as per Fig.~\ref{Flowchart}; inverted and non-inverted cyan Y, respectively), and the maximum extent of the source in both directions (D1 and D2 as per Fig.~\ref{Flowchart}; up and down pointing orange triangles, respectively for the directions to the brightest and second-brightest peak). The scale is in pixel coordinates, with a scale of 1.5 arcseconds per pixel. The colour bar represents flux units in Jy/beam. }
	\label{Examples} 
\end{figure*}

\begin{table}\small
	\caption{Size and flux bins for the reliability checks, as detailed in the flowchart on Fig.~\ref{Flowchart}. Each combination of labels is applied to both the FRIs and the FRIIs separately (see table~\ref{RelTable}). The definition of the smallest size bin is based on the resolved criteria from \citet{Shimwell2018}.}\label{RelLabels}
	\centering
	\begin{tabular}{cc}\hline
		Label&Size range (arcsec)\\\hline
		S0&size$\le$27 OR (d1+d2$\le20$ AND size$\le40$)\\
		S1&27$<$size$\le$60 AND not in S0\\
        S2&size$>$60\\\hline
		&Flux range (mJy)\\\hline
		F1&$F_{150}\le$10\\
		F2&10$<F_{150}\le$50\\
		F3&$F_{150}>$50\\\hline
	\end{tabular}
\end{table}

Some examples of the classifications and plots produced by LoMorph are shown in Fig.~\ref{Examples}. Fig.~\ref{FRII_disconnected} is also a good example of isolated components and a host galaxy identified and associated thanks to the LOFAR Galaxy Zoo citizen science tool \citep{Williams2018b}. For the following Sections we focus solely on the classification and properties of the FRIs and FRIIs, as we will address the hybrid candidates in a separate work, but we do briefly describe their overall properties in Section~\ref{hybrids}.

It is important to emphasise that our code has been optimised to work on LoTSS images, and incorporates catalogued source and host-galaxy positional information, which unavoidably limits its versatility. While we had initially hoped to develop a more general approach, our preliminary analysis demonstrated that classification reliable enough for our science aims required this information. It will be possible to adapt LoMorph for use with data from other instruments, but it is important to emphasise that the same sources can present very different appearances depending on the frequency, sensitivity and angular scales to which a survey is sensitive. For example, FIRST data generally only show the inner, newer structures of most double-double sources identified in LoTSS \citep{Mahatma2018b,Williams2018b}, and due to the higher frequency and comparatively poorer sensitivity of FIRST to extended structure, sometimes isolated components belonging to the same source may not be correctly identified as such. It is also crucial to note that deep surveys such as LoTSS contain a large number of ordinary galaxies with star-formation associated radio emission, which we have been able to pre-filter by using the sample selection of \citet{Hardcastle2018b}. We have not attempted to separate star-forming galaxies from FRIs and FRIIs using morphology alone, and we believe that achieving high reliability in separating FRIs and galaxy continuum sources is in general unlikely to be possible for deep radio surveys without incorporating multiwavelength data providing host-galaxy information.


\subsection{Classification statistics and reliability}\label{reliable}

\begin{table}\small
	\caption{Classification statistics, before and after the visual adjustment discussed in Section~\ref{vis}. The small categories correspond to the S0 size bin defined in Table~\ref{RelLabels}. Total number of sources: 5805.}\label{Class_stats}
	\centering
	\begin{tabular}{ccc}\hline
		Morphology&Number (LoMorph)&Final sample\\\hline
		FRI&1843&1256\\
		FRII&423&423\\
		Hybrid&427&427\\
		Unresolved&1034&-\\
		Small FRI&1709&-\\
		Small FRII&123&-\\
		Small hybrids&246&-\\\hline
	\end{tabular}
\end{table}

The classification statistics from our automated analysis, categorised as illustrated in Fig.~\ref{Flowchart}, are listed in Table~\ref{Class_stats}. The S0 size bin, containing the smallest cases (corresponding to the three categories at the bottom of Table~\ref{Class_stats}), yielded considerably worse classification statistics, and so we report on this subsample separately from the  main FRI, FRII, and hybrid subsets and do not make use of it in the science analysis of Sections~\ref{Results} and ~\ref{discuss}.

\begin{table}\small
	\caption{Reliability table. The first column shows the subset to which the labels defined on Table~\ref{RelLabels} apply; for example, S2 F2 FR1 refers to sources with sizes greater than 60 arcseconds and fluxes between 10 and 50 mJy (F2), which were automatically classified as Fanaroff-Riley type I (FRI). For clarity, the FRI and FRII subsets are shown separately. Column 2 shows the number of sources in each subset, and columns 3-5 show, respectively, the percentage of sources for which visual inspection has shown the automatic classification to be correct, incorrect, or difficult to determine. The smaller (S0) sources are shown separately at the bottom of the table.}\label{RelTable}
	\centering
	\begin{tabular}{ccccc}\hline
		Subset&Sources&\% Correct&\% Incorrect&\% Uncertain\\\hline
		S1 F1 FR1&107&82&4&14\\
        S2 F1 FR1&50&92&4&4\\
        S1 F2 FR1&459&68&11&21\\
        S2 F2 FR1&488&84&10&6\\
        S1 F3 FR1&210&67&20&13\\
        S2 F3 FR1&430&84&14&2\\\hline
        S1 F1 FR2&41&76&7&17\\
        S2 F1 FR2&17&82&0&18\\
        S1 F2 FR2&39&87&8&5\\
        S2 F2 FR2&56&88&2&10\\
        S1 F3 FR2&71&94&2&4\\
        S2 F3 FR2&199&94&4&2\\\hline\hline
        S0 F1 FR1&484&62&18&30\\
        S0 F2 FR1&735&50&10&40\\
        S0 F3 FR1&490&36&22&42\\\hline
        S0 F1 FR2&82&78&14&8\\
        S0 F2 FR2&23&82&9&9\\
        S0 F3 FR2&18&64&17&17\\\hline
	\end{tabular}	
\end{table}

To verify the reliability of the automatic classifications, we visually inspected 50-100 sources selected at random from each of a series of flux and size bins, as listed in Table~\ref{RelLabels}, determining a by-eye classification for comparison with the automatic classification. Table~\ref{RelTable} shows the results of this comparison. We find that LoMorph is successful at automatically classifying radio galaxies with angular sizes $>27$" -- we obtain an accuracy of 89 per cent for FRIs and 96 per cent for FRIIs, relative to visual inspection, and after eliminating 99 sources with less reliable host IDs. The better classification results for FRIIs than for FRIs are not unexpected, as it is easier to identify an edge-brightened, two-peaked distribution while FRIs are more diverse in surface brightness distributions, including wide-angle tail (WAT) and narrow-angle tail (NAT) sources that can have complex, bent morphologies. The FRI reliability is not high enough to achieve our science aims, and so we discuss manual adjustments to the sample in the next section.

The FRII classifications are, overall, more reliable, but a slight caveat is that the identification of their hosts can be more uncertain, as often there is no radio core to indicate the position of the host relative to the hotspots. They also represent a much smaller subset of the sample, consistent with the fact that FRIs are more common in the local Universe, while the fraction of FRIIs is known to increase at higher $z$ \citep[as  a combination of selection and evolutionary effects, see e.g.][]{Willott2001,Wang2008,Donoso2009,Gendre2010,Kapinska2012,Williams2015, Williams2018a}. 

The key sources of uncertainty for all classifications, which dominate the misclassifications reported in Table~\ref{RelTable}, are:
\begin{itemize}
    \item Issues with noise and noise uniformity, which may artificially extend a source through flood-filling -- $\sim 8$ per cent of FRIs and FRIIs.
    \item Deconvolution limitations, which mostly affect double sources with small angular sizes, making it difficult to interpret whether they are FRIs or FRIIs -- $\sim 4$ per cent of FRIs and FRIIs.
    \item Less reliable host identifications, particularly for more distant sources \citep{Duncan2018b,Williams2018b} -- $\sim5$ per cent of FRIs and FRIIs. 
\end{itemize}

Other, more minor issues that lead to a small number of misclassifications include source asymmetry and projection/orientation effects, complex morphologies (e.g. in dense cluster environments), and intruding sources (either through imperfect component association or inadequate masking).


\subsection{Sample adjustments via visual inspection}\label{vis}

In order to improve the quality of our clean FRI and FRII samples prior to scientific analysis, we made manual adjustments to correct for the most important types of misclassification. Accounting for the ``uncertain'' cases in the FRII sample leaves the overall reliability at 91 per cent, which is still high enough that no cleaning of the sample is needed. The FRIs are more complicated, as there is a much larger percentage of uncertain cases, necessitating further checks. Our visual inspection shows that there are sources that adhere to the FRI classification criteria, but which have a morphology that appears distinct from that of a ``canonical'' FRI with gradually decreasing surface brightness assumed to originate from a decelerating flow. As such, we examined the FRI sample in detail, and excluded the $\sim17$ per cent of the sources that do not exhibit the characteristic FRI lobed or tailed, narrow-angle tail, or wide-angle tail morphologies.

We filtered out five categories of ``contaminating'' source in the automatically classed FRI sample:
\begin{itemize}
\item 19 double-double (restarting FRII) sources. Double-doubles \citep{schoenmakers00,Mahatma2018b} are not thought to have FRI-like decelerating jets, but are automatically classified as FRI by LoMorph, as they have bright inner structures and fainter, old emission further away from the core. These sources are key to understanding radio-galaxy life cycles, and have been discussed in detail by \citet{Mahatma2018b}.
\item 180 sources larger than our S1 threshold of 27 arcsec that consist of a bright core surrounded by a halo-like structure of diffuse emission with no apparent lobe or tail structure (`fuzzy blobs'). Although bright ($90$ per cent have total fluxes above 10 mJy and dynamic ranges $>3.5$), the nature of these sources could not be firmly established, but it is unclear that they possess FRI-like jets.
\item 99 core-bright sources with high dynamic range (75 per cent have dynamic ranges $>4.5$) leading to an automatic FRI classification, but with an anomalous, sharp drop and subsequent rise in brightness beyond the core that makes them appear edge-brightened. These sources also appear distinct from traditional FRIs, and we discuss their nature further in Section~\ref{CoreD}. Some of these sources will be analysed in detail by Jurlin at al. (in prep.)
\item One star-forming galaxy with a bright, compact core likely linked to an AGN (hence its inclusion in the sample), but where the diffuse emission was clearly linked to star formation, based on its correspondence with the optical images.
\item 99 sources where the host ID appeared doubtful. 
\end{itemize}

We exclude these sources for the remainder of our analysis, and list updated classification statistics following this manual adjustment in the third column of Table~\ref{Class_stats}.

In future it may be possible to improve our automated classifications to identify the first four sub-classes of AGN listed above, which meet the traditional FRI definition, but we believe are physically distinct populations that will contaminate any simple population statistical analyses. It may also be possible in future to train machine learning methods to identify them as separate classes. However, for now, we emphasise that automated approaches that assume a simple definition of the FRI source class are likely to suffer significant contamination from sources whose underlying dynamics are distinct from the archetypal decelerating low-power jets, such as those in the 3CRR sample.


\subsection{Improved size and flux estimates}\label{SizeFlux}

\begin{figure*}
	\centering
	\includegraphics[width=0.47\textwidth]{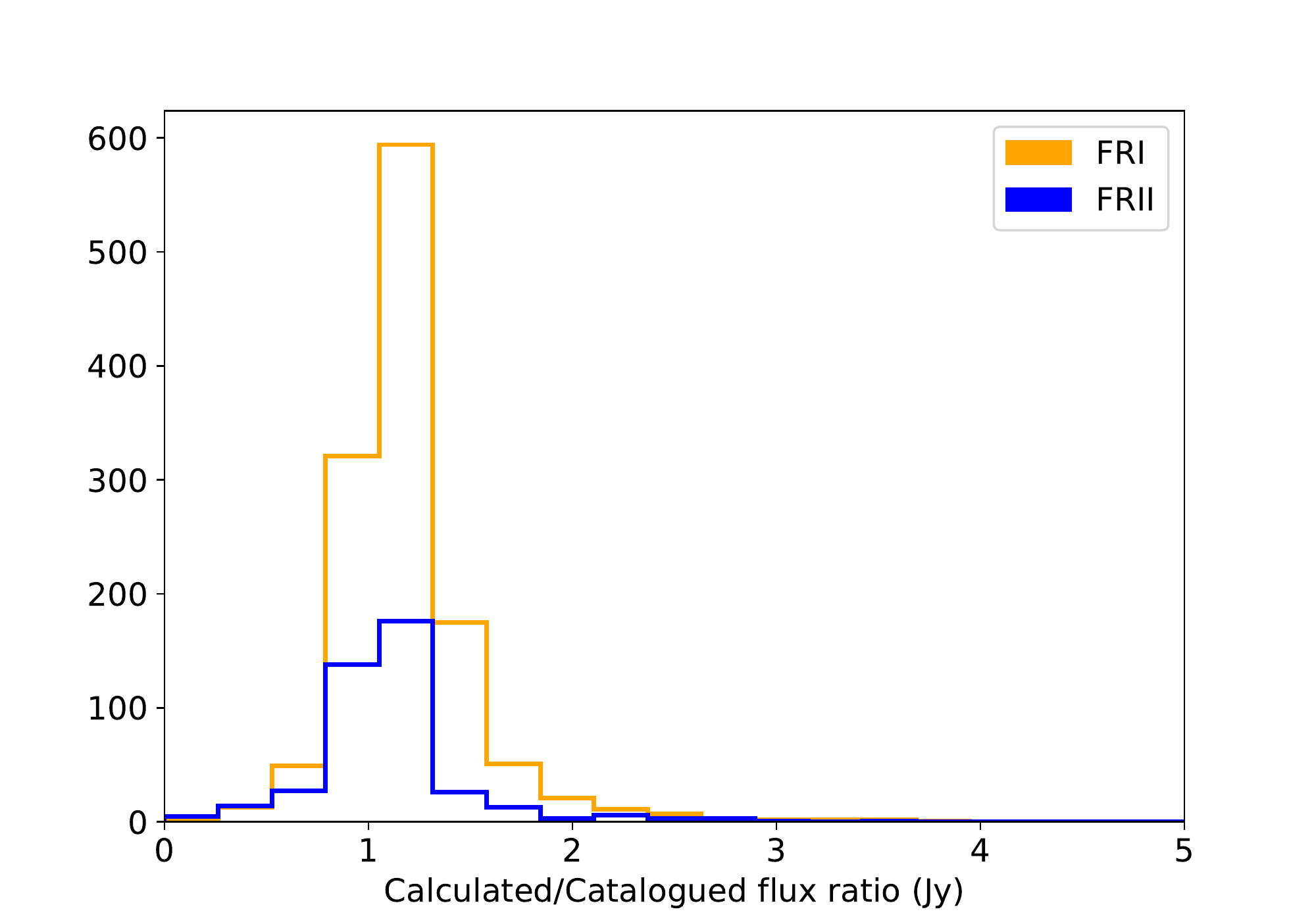}
	\includegraphics[width=0.47\textwidth]{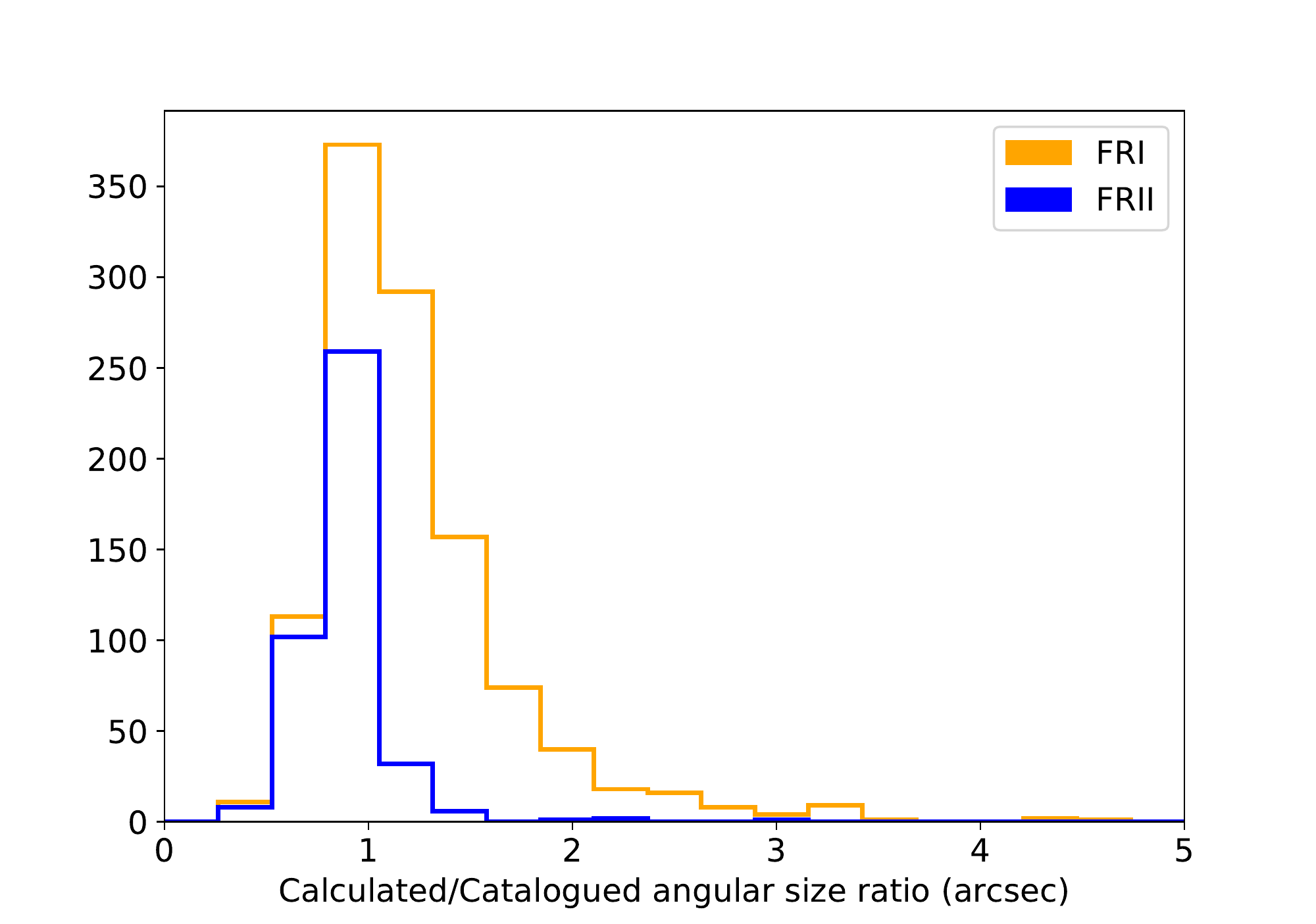}
	\caption{Comparison of the ratio of LoMorph to catalogued source flux (left) and size (right) measurements.}
	\label{Size_comp}
\end{figure*}

As a byproduct of our LoMorph image analysis, we obtain improved total flux and source size estimates that account for emission extending beyond the fitted Gaussian components from \textsc{PyBDSF}, or their aggregation through the LOFAR Galaxy Zoo, in the cases with multiple components \citep[see][]{Williams2018b}. In particular, we have found that the catalogued sizes and fluxes tend to be underestimated for FRI sources where tails gradually decrease in brightness into the noise. The sizes of the FRIIs are slightly overestimated in the catalogue, likely due to small centroid offsets on the \textsc{PyBDSF} regions in asymmetric sources, and to the convex hull method used to group multiple catalogue components, as the FRIIs are often aggregates of multiple components \citep[$51$ per cent of FRIIs, versus $38$ per cent of FRIs, see also][]{Williams2018b}. Fig.~\ref{Size_comp} shows a comparison of LoMorph fluxes and sizes, using our RMS thresholds and flood-filling, with those catalogued by \citet{Shimwell2018}. The median of the size ratios (right panel of Fig.~\ref{Size_comp}) is 1.15 and 0.87 for the FRIs and FRIIs, respectively. We find that 53 per cent of our final FRI sample and 76 per cent of the FRII sample have estimated sizes that agree with those tabulated in the catalogue to within $\pm25$ per cent (increasing to 73 and 94 per cent of the FRIs and FRIIs, respectively, for an agreement to within $\pm50$ per cent). 

In terms of the ratio between the catalogued and calculated fluxes (Fig.~\ref{Size_comp}, left panel), there is agreement within $\pm25$ per cent for 67 per cent of the FRIs and 70 per cent of FRIIs (increasing to 88 and 84 per cent of the FRIs and FRIIs, respectively, for an agreement to within $\pm50$ per cent). The medians of the flux ratio distributions are 1.13 and 1.07 for the FRIs and FRIIs, respectively. On average the LoMorph fluxes are slightly higher than the catalogued values. It is worth noting that our use of the new, more sensitive imaging data may be behind some of the discrepancy, as well as the fact that we focus only on the larger sources for our analysis (the size and flux agreement is much tighter for the small FRIs/FRIIs and unresolved sources described in Section~\ref{reliable}). For very faint sources, small differences in flux and size after the RMS filtering and flood-filling can also have a large impact on the ratios shown in Fig.~\ref{Size_comp}. We have confirmed visually that for sources where the sizes and fluxes diverge from the catalogue value, this is usually because the catalogued components did not fully represent that source structure. A small number of sources ($<2$ per cent) are affected by problems with flood-filling that lead to significant over-estimation of sizes and fluxes, but this does not affect any of the paper results and conclusions. 

In the analysis that follows, we adopt the LoMorph flux and angular size estimates, and use them to obtain luminosities and physical sizes as reported in the next Section. We have checked that using catalogued values does not significantly alter our main results. The luminosity distributions do not change significantly, and our science conclusions are not strongly dependent on the new source sizes, so the larger sizes we measure for a substantial proportion of FRIs do not affect any overall conclusions.


\subsection{Redshift distributions and selection effects}\label{selection}

\begin{figure}
    \centering
    \includegraphics[width=0.47\textwidth]{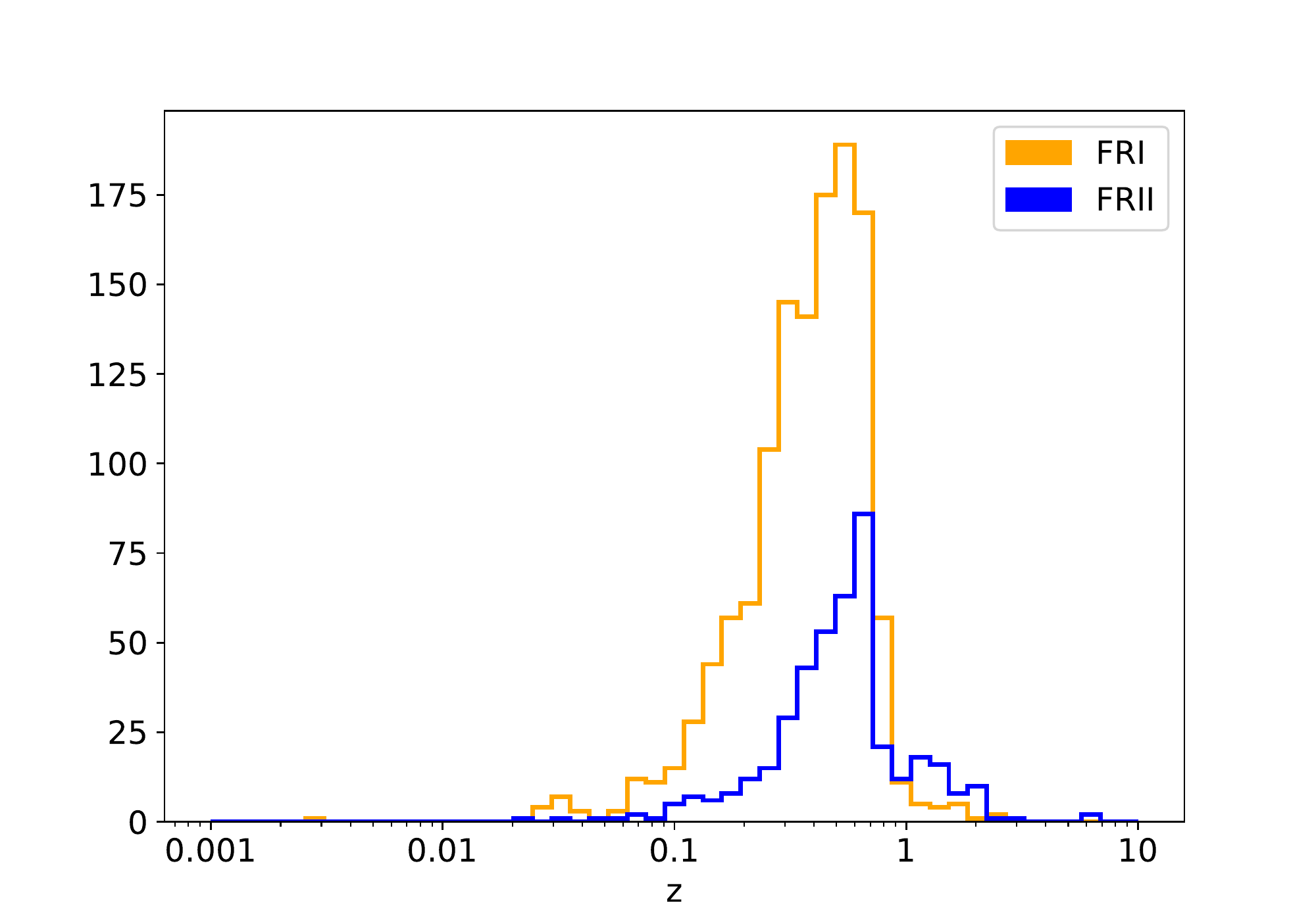}
    \caption{Redshift distribution for the FRI (orange) and FRII (blue).}
    \label{z_histo}
\end{figure}

While our radio-galaxy sample has a lower flux limit and better sensitivity to low surface brightness emission than any previous wide-area survey, it remains essential to consider sample selection effects resulting from both the limitations of the radio data and of the optical and infrared (IR) information used to obtain host-galaxy IDs and redshifts.

The redshift distributions of the parent AGN sample are shown in Fig. 6 of \citet{Hardcastle2018c}: most sources have $z<0.8$, with a tail of objects -- identified with quasars -- extending to $z>2$. Our morphologically classified sample shows similar behaviour, with FRIs and FRIIs in our sample having similar redshift distributions (Fig.~\ref{z_histo}), but with a larger fraction of FRII sources at $z>1$. The redshift distributions are largely a consequence of the available host-galaxy information, with reliable redshifts only available for quasars above $z \sim 0.8$. As discussed in Section~\ref{Results}, we therefore restrict much of our analysis to $z<0.8$.

\begin{figure*}
	\centering
\includegraphics[width=0.8\textwidth]{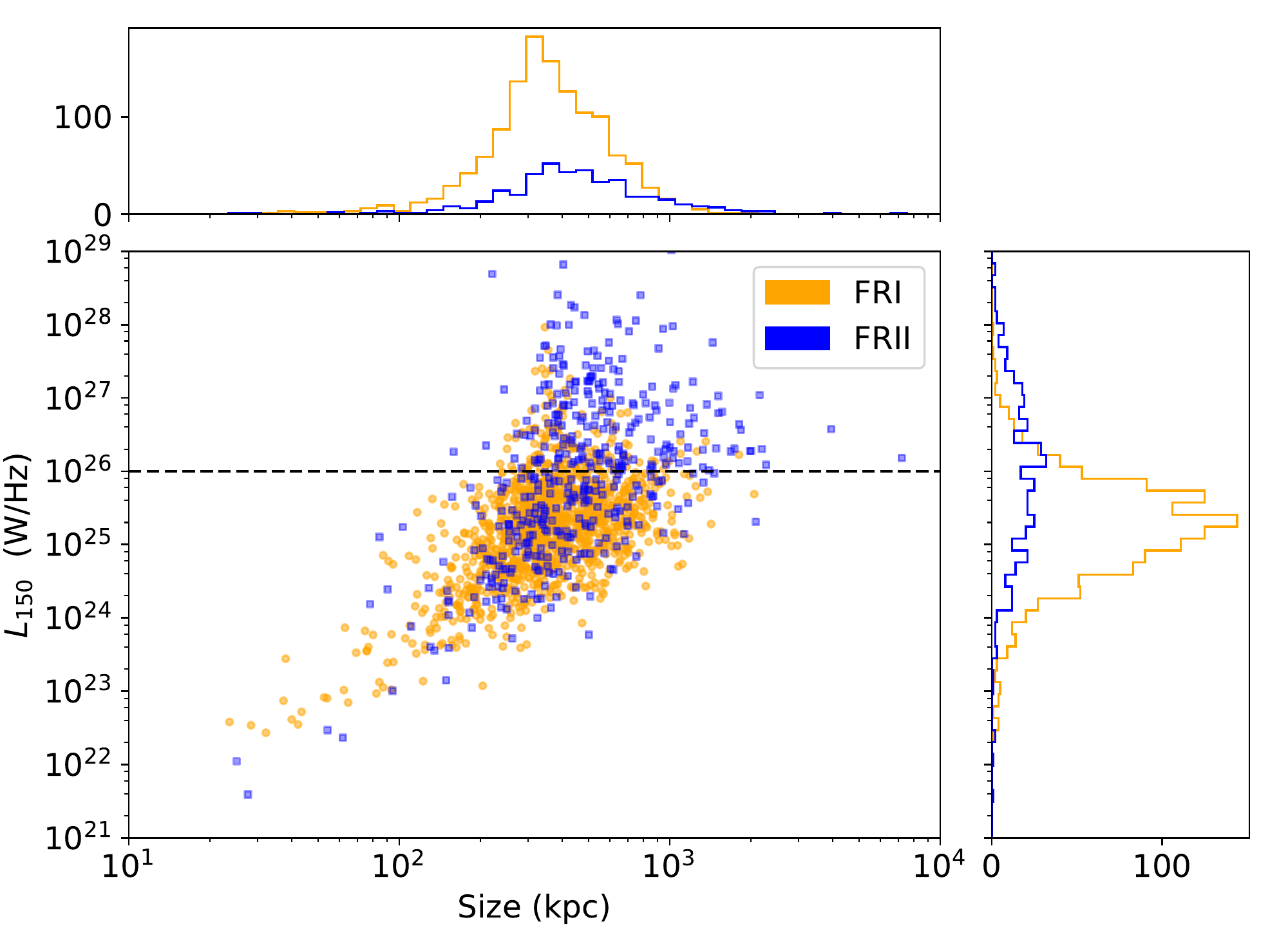}
	\caption{150 MHz luminosity versus physical size for the FRI (orange circles) and FRII (blue squares). The traditional luminosity boundary between both populations is at $\sim10^{26}$ W Hz$^{-1}$ at 150 MHz, indicated by the dashed black line.}\label{L150_size_all}
\end{figure*}

Figure~\ref{L150_size_all} shows the distributions of 150-MHz luminosity ($L_{150}$, K-corrected) versus the physical size (in kpc) for the FRIs and FRIIs in our sample. Histograms for both axes are included to better illustrate the source distributions. We note that the lower right corner of the plot is unoccupied due to surface brightness limits, so that a substantial population of physically large, low luminosity sources could be present, but unobservable \citep[see also the discussion in][]{turner2015,Hardcastle2018c}, while the top left corner is affected by our angular size limit, which gradually increases the physical size lower limit at higher redshifts, where rare luminous objects are more likely. We note that this makes our sample selection very different to, e.g. 3CRR, which contains many compact, physically small luminous radio galaxies that would occupy the top left corner of the plot.

The distributions of radio luminosity and size shown in Fig.~\ref{L150_size_all} are affected both by the survey flux limit, so that low luminosity sources are typically at lower redshift than high luminosity sources, and by surface brightness limitations, so that low luminosity sources are typically smaller, either because large low luminosity sources remain undetected, or because their sizes are underestimated. Additionally, our initial size threshold ($>12$ arcsec, which corresponds to $\sim90$ kpc at $z=0.8$) necessarily eliminates some sources with moderate physical sizes that would be present in the original AGN sample of \citet{Hardcastle2018c}. We explore these effects in more detail throughout Section~\ref{discuss}, and carefully examine the influence of redshift on our conclusions.

It is also important to consider the selection effects imposed by the optical catalogues, and their incompleteness at high $z$. We applied our LoMorph code separately to the sample of LoTSS DR1 sources that otherwise meet our selection criteria, but whose redshifts are poorly constrained, so that they did not meet the criteria for inclusion by \citet{Hardcastle2018c}. After filtering out nearby star-forming galaxies from this sample via radio-to-optical size ratio (Webster et al., in prep.), we found an additional 256 FRIs and 371 FRIIs. These sources are accurately classified by our code, but their poorly-constrained photometric redshifts result in large uncertainties on their sizes and luminosities, making them impossible to include in our science analysis. The majority of these objects have higher redshifts than our main sample, peaking around $z\sim1$ and with a longer tail to higher $z$ in the distribution. The ratio between FRIs and FRIIs is very different for these sources, which is expected because of the evolution of the FRIIs and/or HERG luminosity function to higher redshift, and the fact that unambiguous FRIIs can be identified at smaller angular sizes due to their brightness distribution. As mentioned in Section~\ref{Code}, we do not analyse sources without an identified optical host, which are likely to lie at even higher redshift \citep{Duncan2018b}, but similar selection effects likely apply. We note therefore that our sample is not ``representative'' of the FRI/FRII mix in the full LoTSS catalogue. We emphasise that there is scope for substantially larger FRII samples to be studied once better redshift information becomes available (e.g. via WEAVE-LOFAR, \citealt{smith2016}).

We note that the redshift distributions for the small FRI and FRII candidates listed in Table~\ref{Class_stats} are not significantly different from those of the clean FRI and FRII samples, with just a slightly larger fraction of small FRI candidates found at higher $z$, which may be QSOs with some extended emission (similar to the `fuzzy blobs' we identified as contaminants in Section~\ref{vis}), or small angular sizes due to orientation. Since the distributions for the large and small sources are similar, the dominant selection effect determining the influence of redshift on our catalogue is the depth of the optical catalogues (rather than angular size limitations). 

It is important to emphasise that although our sample has a much lower flux limit than many previous works, it is nevertheless a flux-limited and surface brightness limited sample, and the completeness of host-galaxy identifications as a function of redshift also introduces complex redshift dependences. This problem does not affect the majority of our conclusions, but makes it difficult to investigate trends with host-galaxy brightness, as is discussed further in Section~\ref{ledlow}.

In terms of the size distribution, our selection allows us to partially cover the smaller end of the scale (upcoming work by Webster et al. will explore this area of the LoTSS AGN parameter space further), and we also probe the regime occupied by giant radio galaxies \citep[GRG, typically $>1$ Mpc, see e.g.][and references therein]{Ishwara-Chandra1999,schoenmakers2000b,Machalski2001,Machalski2008,Dabhade2017} The population of GRG discovered by LOFAR is discussed in detail by \citet{Dabhade2019}, \citet{Hardcastle2018c}, and the implications of longer life cycles are discussed by \citet{sabater2018}.


\section{Results}\label{Results}

The main aim of our morphological investigation is to revisit the relationship between FR class (and morphology more generally), radio luminosity and host-galaxy properties. We first report the overall radio properties of our FRI and FRII subsamples (Section~\ref{FR}), before examining their host-galaxy properties in Section~\ref{Hosts}.

\subsection{FRI and FRII radio properties in LoTSS}\label{FR}

Looking at Fig.~\ref{L150_size_all}, it is immediately apparent that a great degree of overlap exists between the FRI and FRII populations. This is contrary to the widely accepted view that luminous sources are FRII and low-luminosity sources are FRI in morphology, as is the case for the 3CRR sample, which contains no FRII sources below $L_{150} \sim 10^{26}$ W Hz$^{-1}$. The overlap in luminosity between FRIs and FRIIs has been seen in previous work using samples with considerably lower flux limits than 3CRR \citep[e.g.][]{Best2009,Miraghaei2017}, but it is particularly striking in the LoTSS dataset.

\begin{figure}
    \begin{subfigure}{1.0\linewidth}
        \centering
        \includegraphics[width=0.95\linewidth]{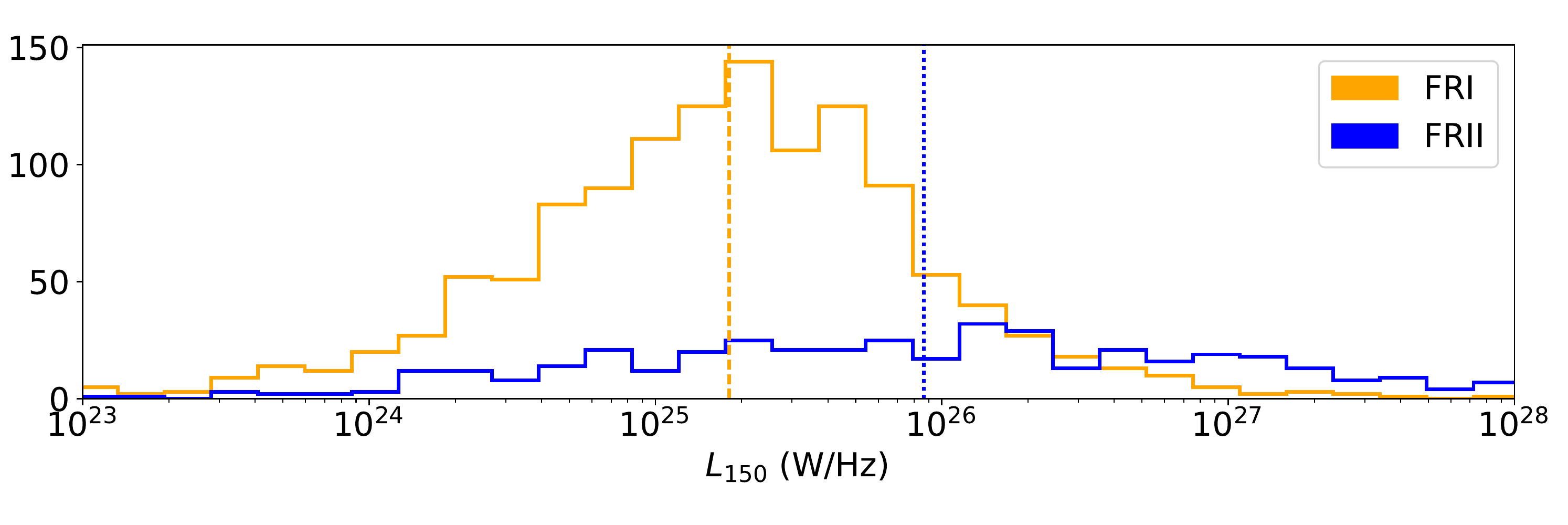}
        \caption{All FRI, FRII}
        \label{L150_histo_all}
    \end{subfigure}\\
    \begin{subfigure}{1.0\linewidth}
        \includegraphics[width=0.95\linewidth]{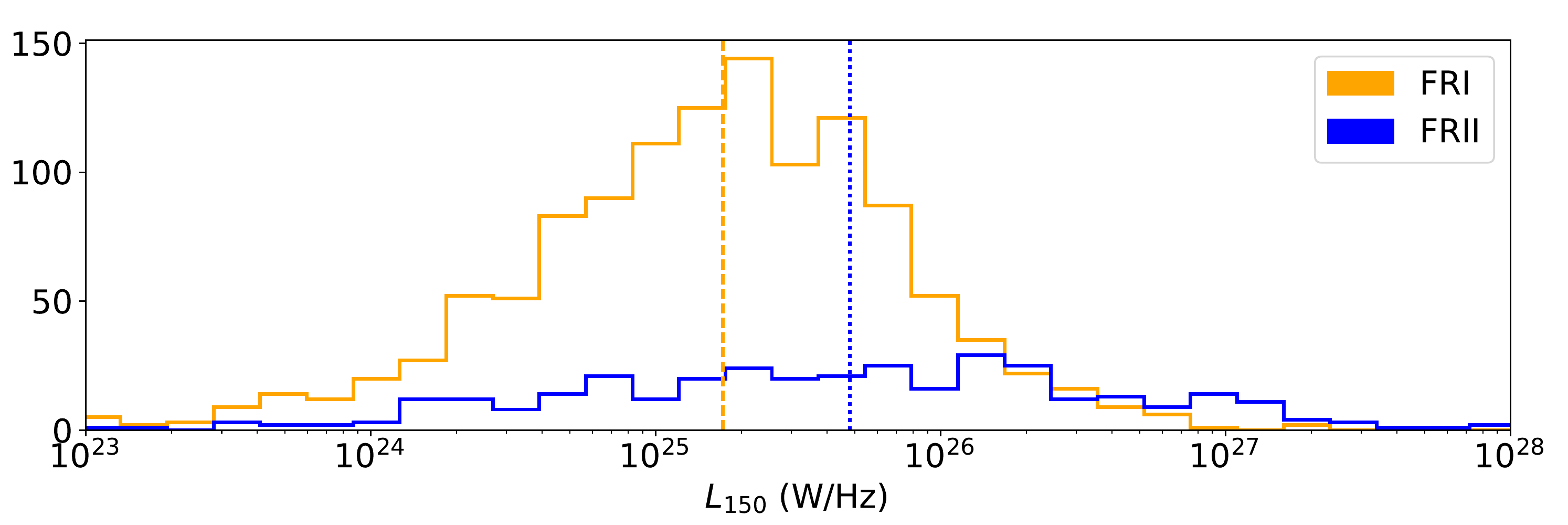}
        \caption{FRI and FRII with $z\leq 0.8$.}
        \label{L150_histo_0p8}
    \end{subfigure}
    \caption{150 MHz luminosity histogram for the FRI (orange) and FRII (blue). The orange dotted and blue dashed lines indicate the median values, respectively, for the FRI and the FRII; (a) includes all sources, while (b) only includes those with $z\leq 0.8$. The luminosity range on both histograms has been slightly restricted with respect to Fig.~\ref{L150_size_all}, to better highlight the differences between the FRI and FRII distributions.}
    \label{L150_histo}
\end{figure}

If we restrict our sample to $z\leq0.8$ (see Section~\ref{selection}), the overlap remains present. Fig.~\ref{L150_histo} shows the histograms and median values for the FRI and FRII, for all sources, and with the sample limited to $z\leq0.8$. The median 150-MHz luminosities for the full $z$ range are $2.0\times10^{25}$ W Hz$^{-1}$, and $8.9\times10^{25}$ W Hz$^{-1}$ for the FRIs and FRIIs, respectively, while at $z\leq0.8$ they are, respectively, $1.9\times10^{25}$ W Hz$^{-1}$, and $4.8\times10^{25}$ W Hz$^{-1}$. Restricting the redshift range to that for which the host coverage is most complete narrows the gap between the two populations: this is because mainly higher luminosity sources are eliminated, which primarily affects the FRII sub-sample, reducing its median luminosity: in terms of source numbers, this restriction eliminates $\sim$3 per cent of the FRIs and $\sim18$ per cent for the FRIIs (see Table~\ref{FR_z_stats}). 

\begin{table}
    \centering
    \begin{tabular}{ccc}\hline
         Subset&Full $z$ range&$z\leq0.8$\\\hline
         FRI&1256&1213\\
         FRII&423&345\\\hline
         NAT&264&251\\
         WAT&195&193\\\hline
         Core-D&99&85\\\hline
    \end{tabular}
    \caption{Top two rows: number of FRI and FRII sources spanning the full redshift range, and for $z\leq0.8$, see also Fig.~\ref{z_histo}. Third and fourth rows: FRI subpopulations (wide- and narrow-angle tails), discussed in Section \ref{subpops}, included in the statistics for the FRIs on the first row. Last row: core-dominated sources, discussed in detail in Section \ref{CoreD}, and not included in the statistics for the FRIs.}
    \label{FR_z_stats}
\end{table}

The canonical FRI/II luminosity break is around L$_{150} \sim 10^{26}$ W Hz$^{-1}$ \citep{FR1974,Ledlow1996}. In our sample a significant minority of FRIs lie above this luminosity (194 sources, or $\sim13$ per cent of the full redshift sample, 106 of which have $\leq0.8$, representing $\sim9$ per cent of the lower $z$ subsample). There are a handful of luminous FRIs in 3CRR, and the existence of bright quasars with FRI morphologies is well known \citep[e.g.][]{Heywood2007,Gurkan2018b}. Roughly $45$ per cent of the luminous FRIs in our sample are indeed quasars with $z>0.8$. The sources with FRII morphologies and very low luminosities present more of a challenge for the traditional paradigm. In jet dynamical models for the FRI/II break, it would be expected that low-power jets must inhabit a very sparse inner environment to avoid disruption turning them into FRI-type jets. The LoTSS FRIIs with luminosities below the traditional break of L$_{150} \sim 10^{26}$ W Hz$^{-1}$, which we refer to as ``FRII-Low'', therefore merit further examination -- we investigate their nature further, and discuss why the relationship between morphology and luminosity is much less clear-cut in LoTSS than in the 3CRR sample, in Section~\ref{friilow}.


\subsection{Host galaxies of the FRI and FRII samples}\label{Hosts}

\begin{figure*}
	\centering
	\includegraphics[width=0.47\textwidth]{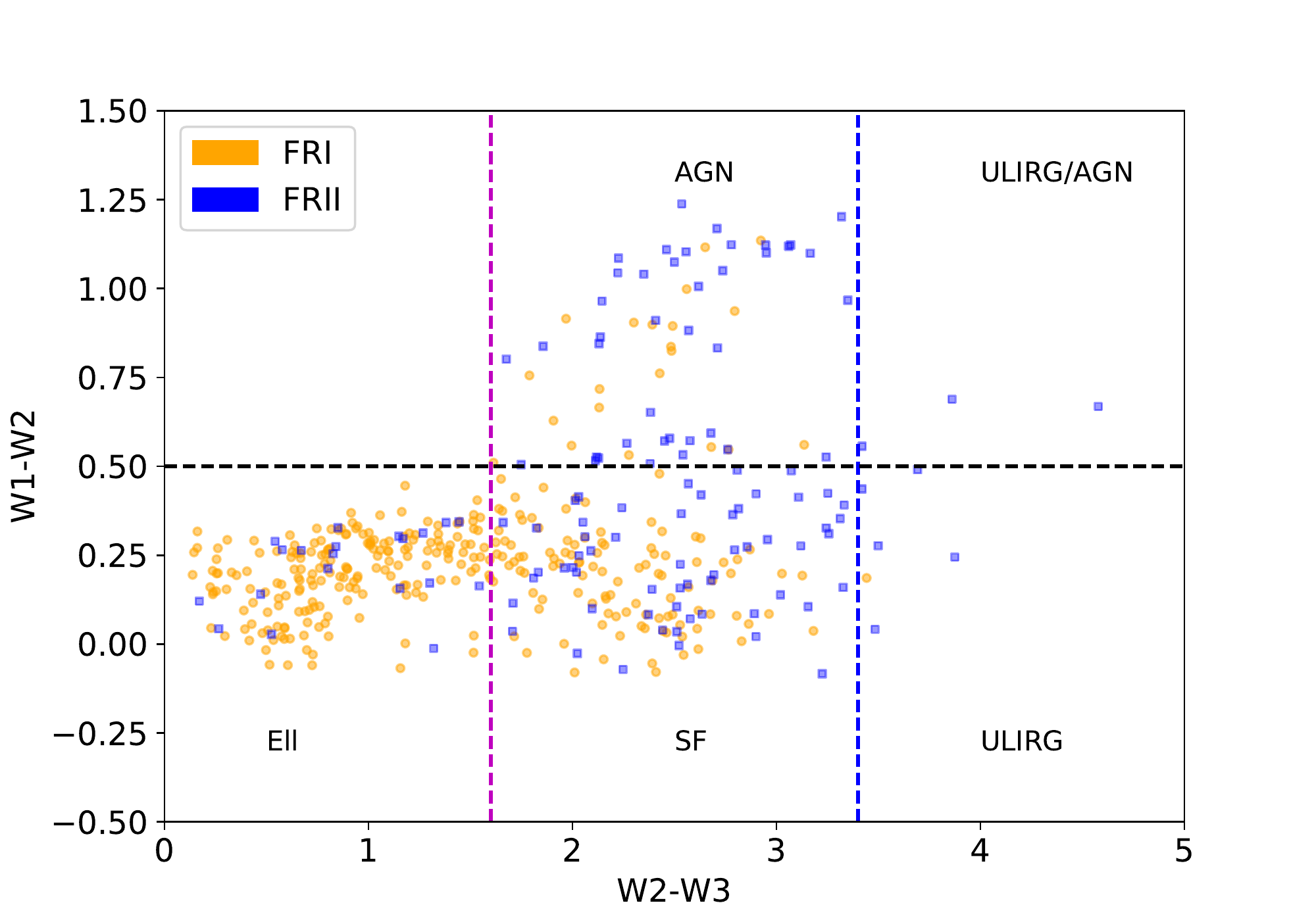}
	\includegraphics[width=0.47\textwidth]{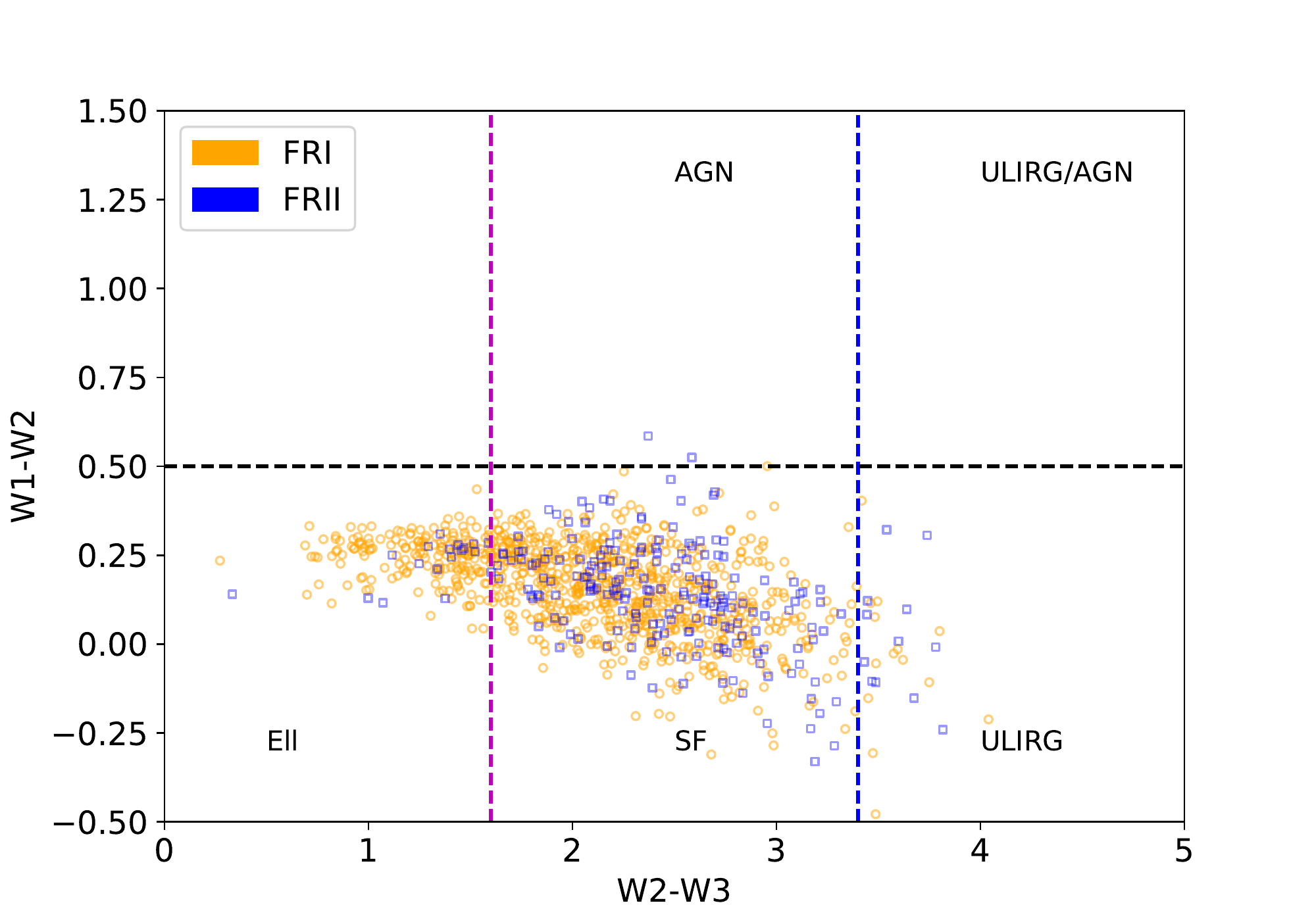}
	\caption{WISE colour-colour plot for all FRI (orange circles) and FRII (blue squares) with $z<0.8$, in Vega magnitudes, with sources detected in all 3 WISE bands shown in the left panel, and sources with a W3 upper limit (so that their position in the horizontal direction may be further left than shown) in the right hand panel. The lines represent rough divisions between host populations, with the $x$ axis being a proxy for star formation prevalence, and the $y$ axis for AGN dominance, as shown in our previous work \citep{Mingo2016}. See the main text for a detailed description.}\label{WISE_c_c}
\end{figure*}

In Fig.~\ref{WISE_c_c} we plot the WISE colours (in Vega magnitudes) for our FRI and FRII sources. The WISE colour-colour plot is a good diagnostic tool to identify some of the properties of the host galaxies of our sample. The synthetic SEDs originally shown by \citet{Wright2010} and \citet{Lake2012} show how the W1, W2, and W3 WISE bands can be used to diagnose the prevalence of star formation and the relative dominance of a radiative AGN. We have used the rough population divisions of \citet{Mingo2016} to identify sources with hosts that are likely to be elliptical galaxies (bottom-left), star-forming galaxies (bottom-centre), starburst/ultra-luminous infrared galaxies (ULIRG, bottom-right and top-right), and AGN-dominated (top-centre and top-right). Given that our sample uses the selection criteria of \citet{Hardcastle2018c} and \citet{Gurkan2018a}, the sparsity of starburst/ULIRG hosts is expected, as we only retain sources for which the radio emission is in significant excess to that expected from star formation. The relative gap between AGN and host-dominated sources (around W1-W2$\sim0.4-0.6$) can be explained through a combination of selection \citep{Hardcastle2018c} and evolutionary effects \citep{Assef2010,Assef2013}.

As discussed by e.g. \citet{Gurkan2014}, \citet{Mingo2016}, it is important to note that selections of AGN based on various cuts on the WISE colour/colour diagram, such as those used by e.g. \citet{Stern2012}, \citet{Mateos2012}, and \citet{Secrest2015}, are very good for selecting clean samples of (optically/mid-IR/X-ray) bright AGN, but they are biased against lower luminosity sources. Even without considering the population of low-excitation radio galaxies (LERGs: radio galaxies with a radiatively inefficient AGN), many Seyferts and high-excitation radio galaxies (HERGs) also lie below the W1-W2=0.5 line. However, the WISE diagram does enable the interplay between AGN, radio and host-galaxy properties to be explored for our sample.

The host distributions for our FRIs and FRIIs are consistent with previous work showing that radiatively inefficient AGN (LERGs) are predominantly hosted by red, elliptical galaxies, while radiatively efficient sources tend to have bluer, more star-forming hosts \citep[e.g.][]{Janssen2012,Gurkan2014,Ineson2015,Ineson2017,Mingo2016,Weigel2017,Williams2018a}. While we do not have excitation class information for our sample, we expect from many previous studies that (excluding the quasars mentioned in Section~\ref{FR}) the FRIs will predominantly be LERGs, while the FRIIs will be a mix of HERGs and LERGs \citep[e.g.][]{Hardcastle2007,Hardcastle2009,Best2012,Mingo2014}.

Fig.~\ref{WISE_c_c} shows a large degree of overlap between FRIs and FRIIs: while it is true that the latter have predominantly bluer hosts, and a significant fraction of them clearly are bright HERGs (W1-W2>0.5), there seems to be a substantial fraction of FRIs with hosts that also seem to be star-forming. Limiting the sample to sources with $z\leq 0.8$ makes very little difference to the plot, other than eliminating some potential QSOs. It is, however, important to note that most of the FRIs in the bottom-centre region of Fig.~\ref{WISE_c_c} have upper limits on W3. The actual W3 values for these sources cannot be arbitrarily low, as they are physically tied to the W1 and W2 measurements through the properties of their spectral energy distributions, but many of these sources may in reality be located further towards the elliptical region. Even accounting for the upper limits, there remains a significant degree of overlap between FRI and FRII host colours. 

Further investigation of host and AGN properties will require additional excitation class information, which does not currently exist for the LoTSS AGN sample, but can be acquired through the future WEAVE-LOFAR optical spectroscopic survey \citep{smith2016}. Our sample of morphologically classified AGN spanning a wide range of radio luminosity will provide an excellent benchmark sample for follow-up studies of the relationship between morphology, AGN accretion mode and host-galaxy properties.


\section{Discussion}\label{discuss}

In the previous Section, we have presented a morphological investigation of extended radio-loud AGN within the LoTSS DR1 catalogue, with an examination of their host properties. Below we consider the interpretation of those results in more detail: specifically we examine the nature of the low-luminosity FRII systems in our sample (Section~\ref{friilow}), we revisit the relation between FR break luminosity and host-galaxy magnitude first reported by \citet{Ledlow1996} (Section~\ref{ledlow}), we discuss the candidate hybrid class from our automated analysis (Section~\ref{hybrids}), and finally we consider the diversity of the FRI population in particular, discussing several specific sub-populations present within the LoTSS sample and the implications of this diversity for future radio surveys work (Section~\ref{subpops}).


\subsection{The nature of the low-luminosity FRIIs in LoTSS}\label{friilow}

Of the FRIIs spanning the full $z$ range, 51 per cent (216 sources) have L$_{150} \leq 10^{26}$ W Hz$^{-1}$, with  a significant fraction (89 sources, $\sim21$ per cent) of FRIIs with L$_{150} \leq 10^{25}$ W Hz$^{-1}$, one order of magnitude below the expected FRI/II boundary \citep{FR1974}. Given that the overwhelming majority of these low-luminosity sources have low redshifts, for the subset of sources at $z\leq0.8$, their relative fraction is even higher, with 214/345 FRIIs (62 per cent) having L$_{150} \leq 10^{26}$ W Hz$^{-1}$, and 89/345 (26 per cent) having L$_{150} \leq 10^{25}$ W Hz$^{-1}$.

In this section we consider the nature of the low luminosity FRIIs, and the apparent discrepancy between our results and the original work of \citet{FR1974}, in detail.

\begin{figure*}
\centering
\includegraphics[width=0.25\textwidth, trim=120 60 90 30, clip=true]{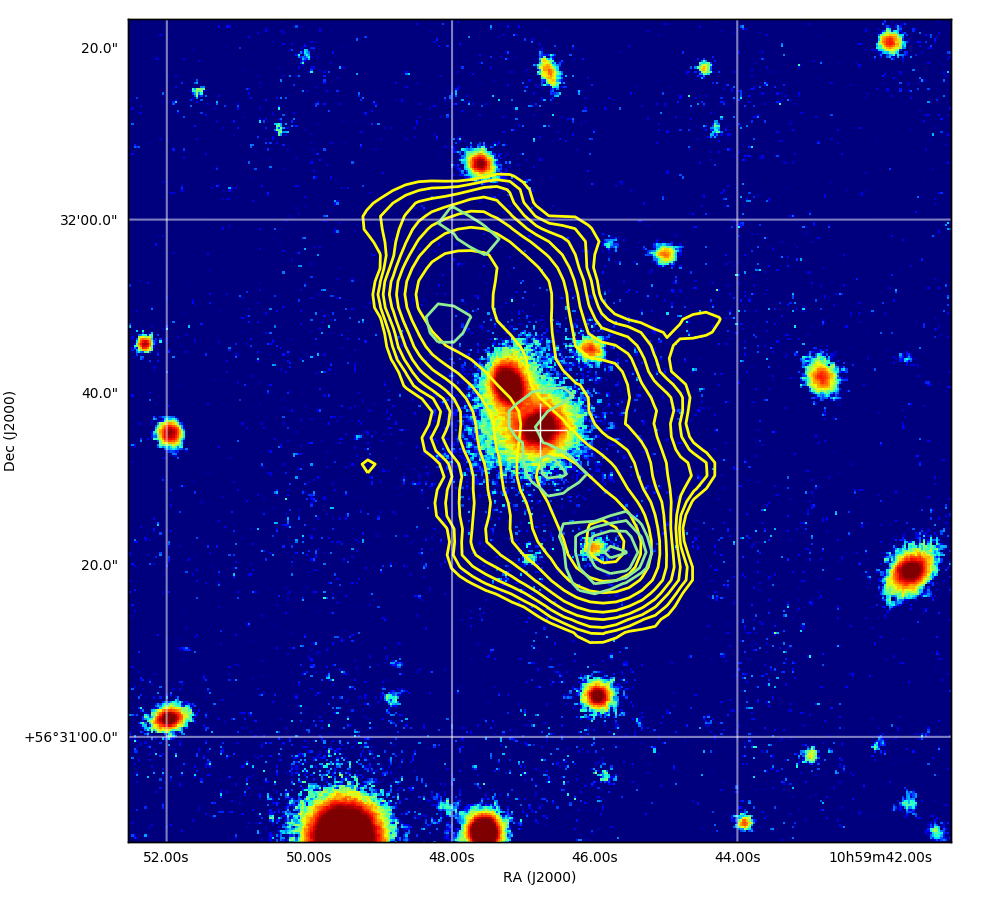} 
\includegraphics[width=0.25\textwidth, trim=120 60 90 30, clip=true]{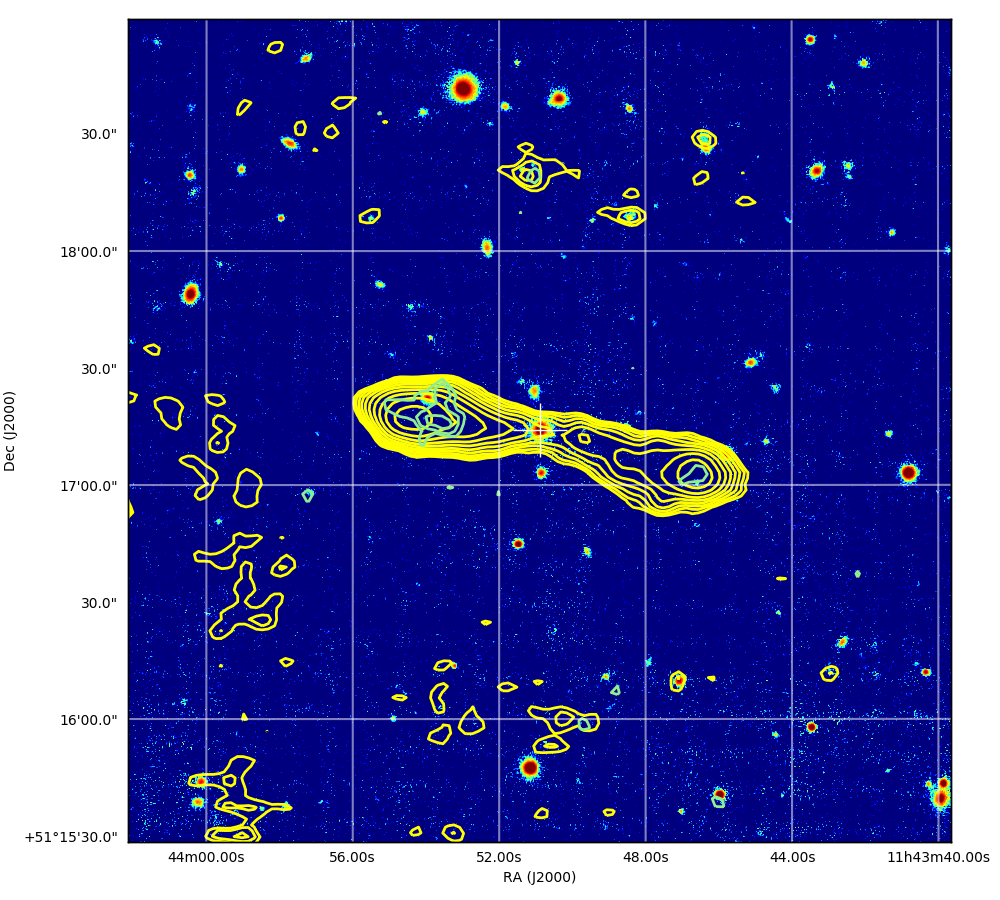} 
\includegraphics[width=0.25\textwidth, trim=120 60 90 30, clip=true]{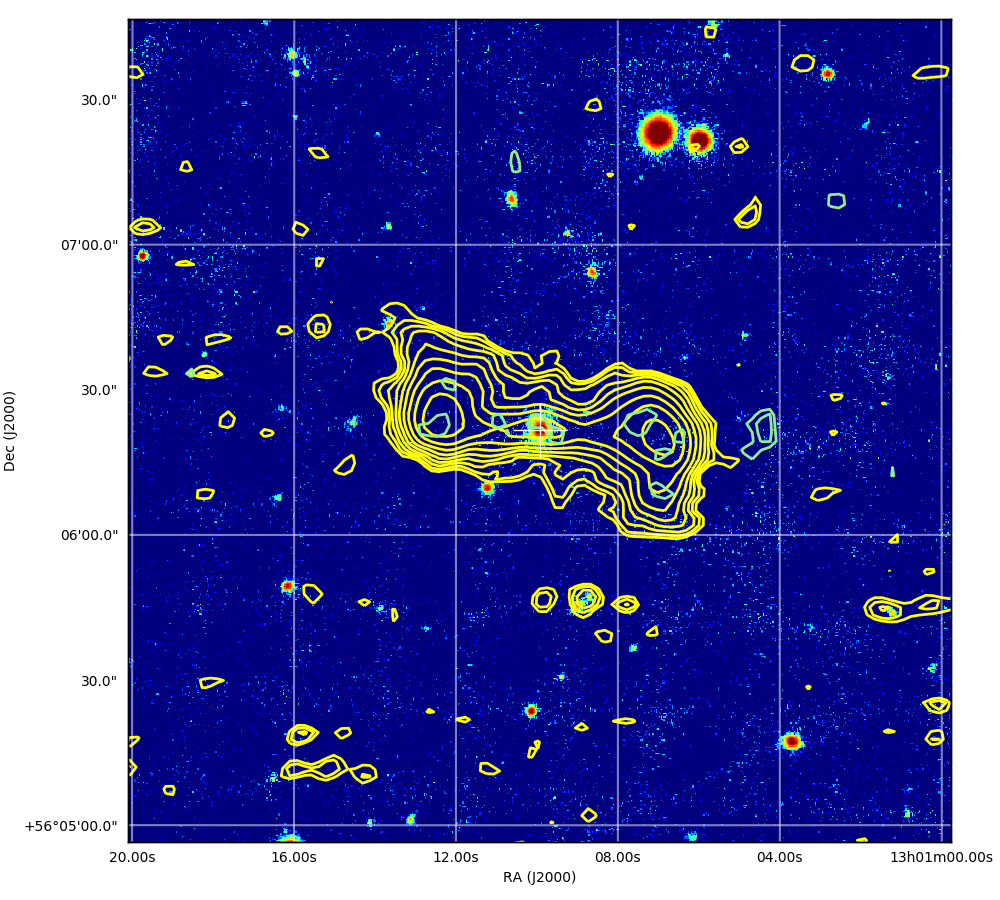} 
\includegraphics[width=0.25\textwidth, trim=120 60 90 30, clip=true]{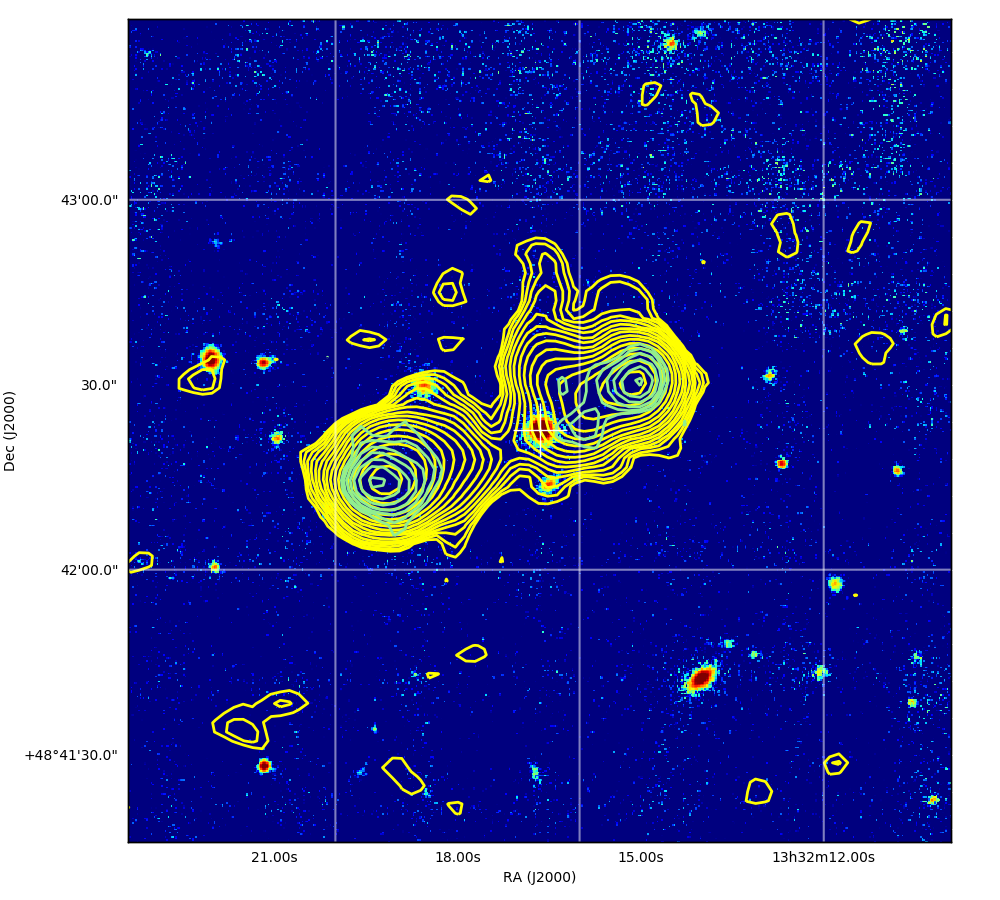} 
\includegraphics[width=0.25\textwidth, trim=120 60 90 30, clip=true]{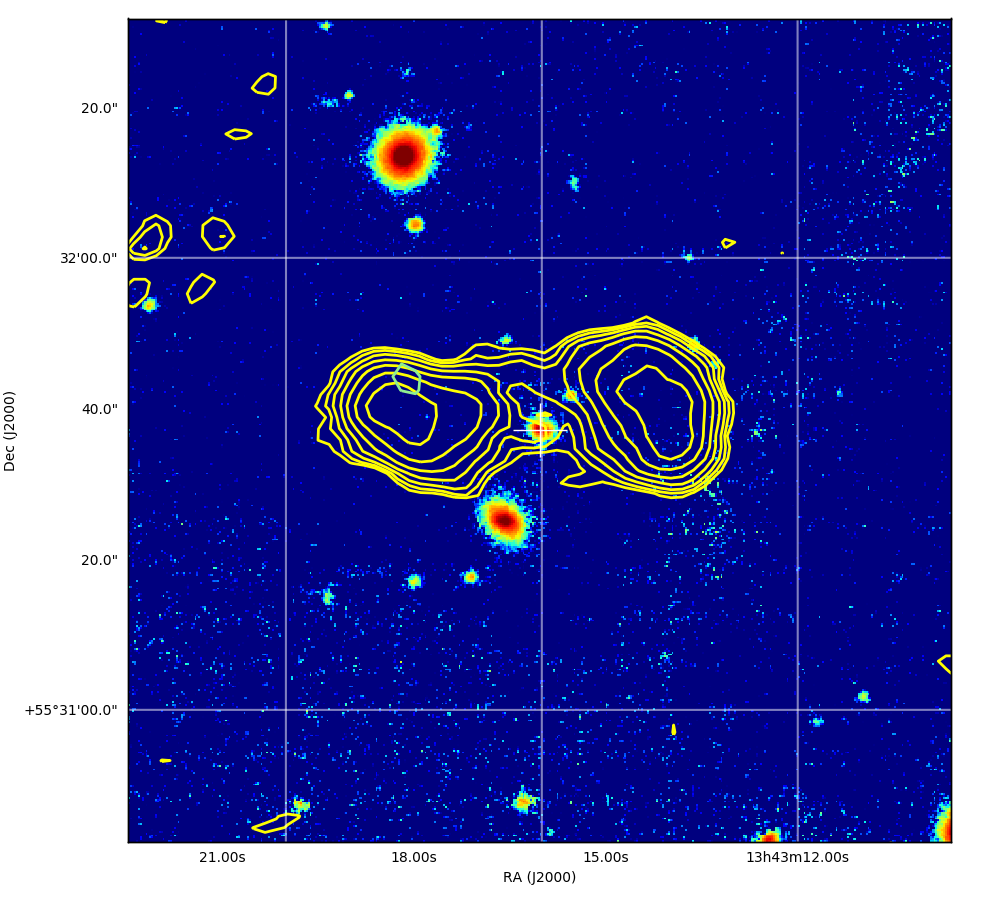} 
\includegraphics[width=0.25\textwidth,trim=120 60 90 30, clip=true]{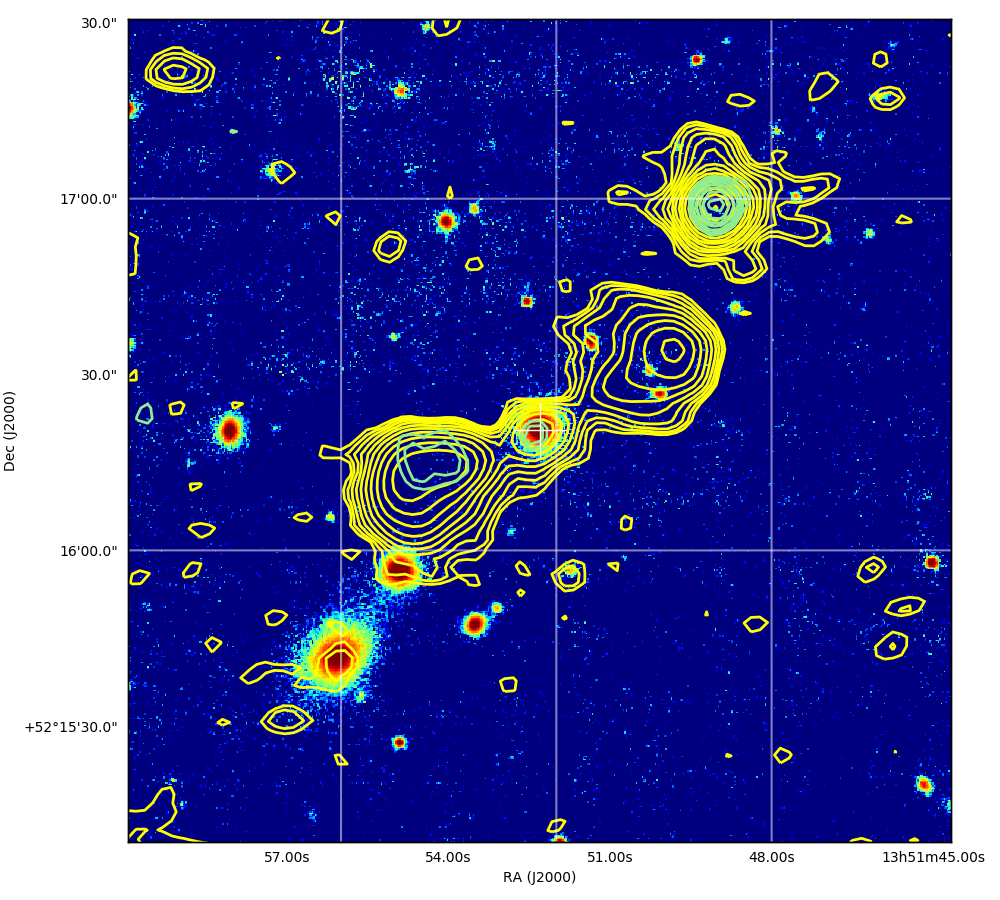} 
\includegraphics[width=0.25\textwidth, trim=120 60 90 30, clip=true]{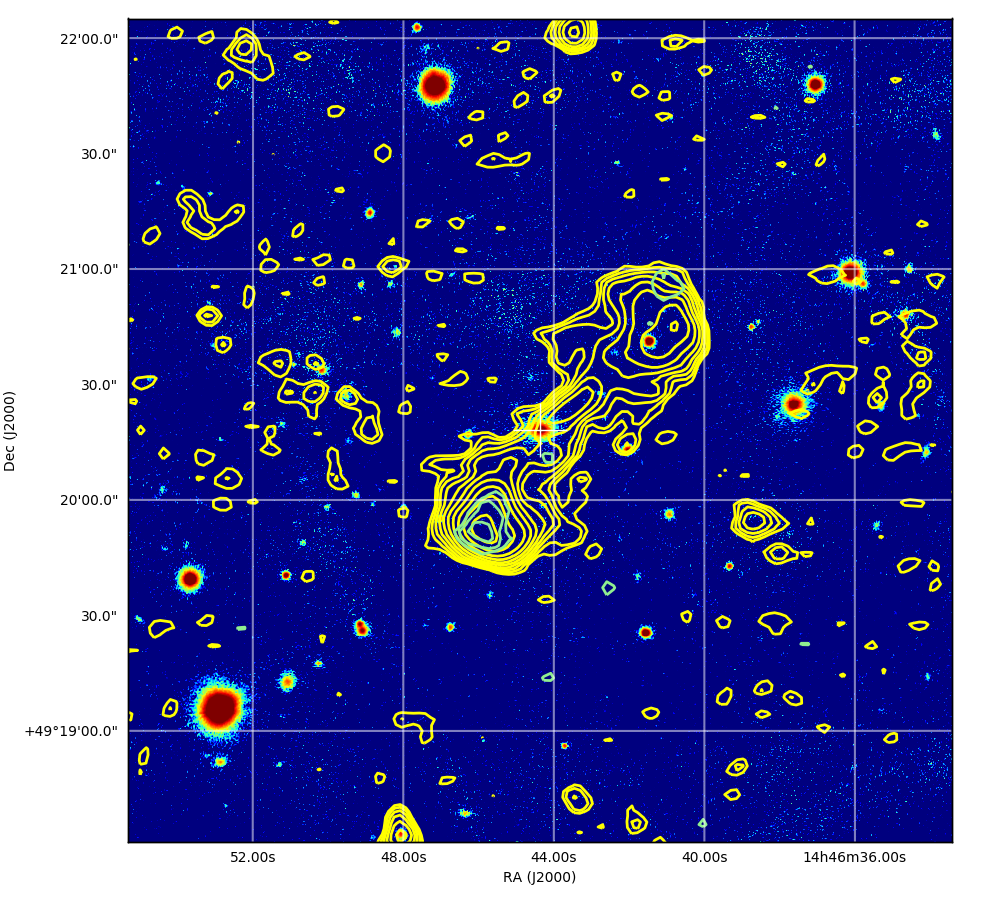}
\includegraphics[width=0.25\textwidth, trim=120 60 90 30, clip=true]{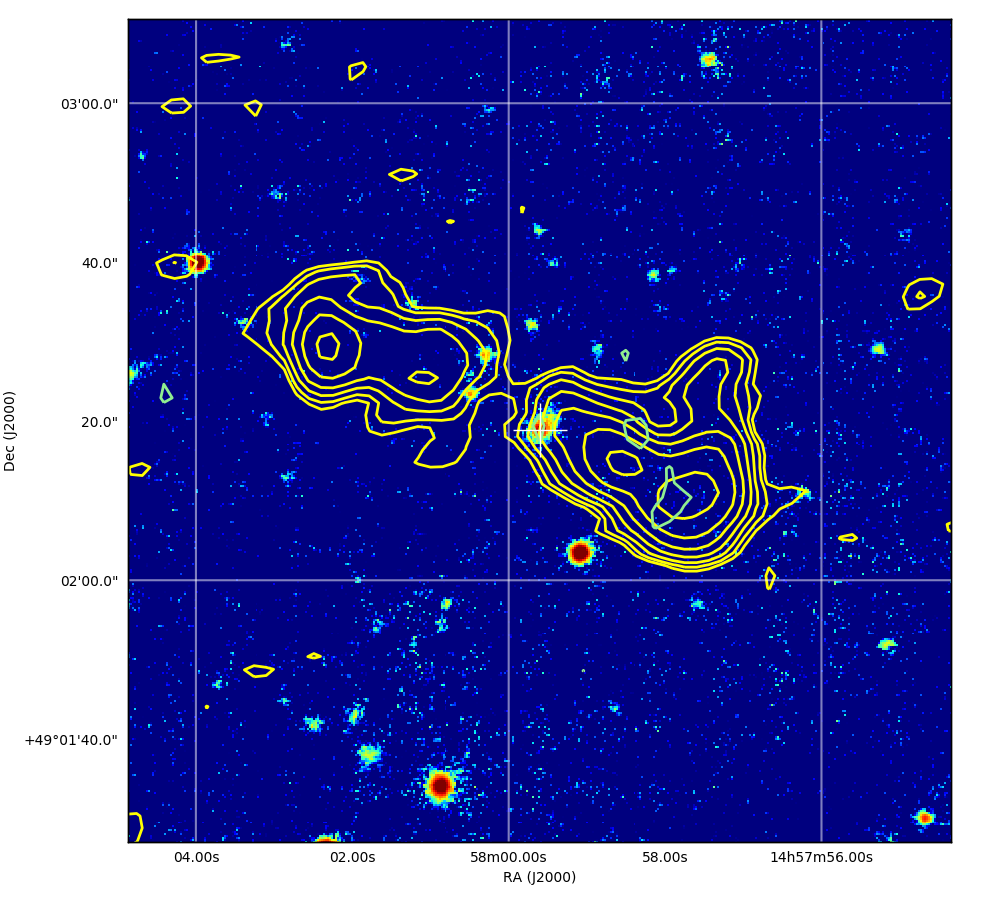}
\includegraphics[width=0.25\textwidth, trim=120 60 90 30, clip=true]{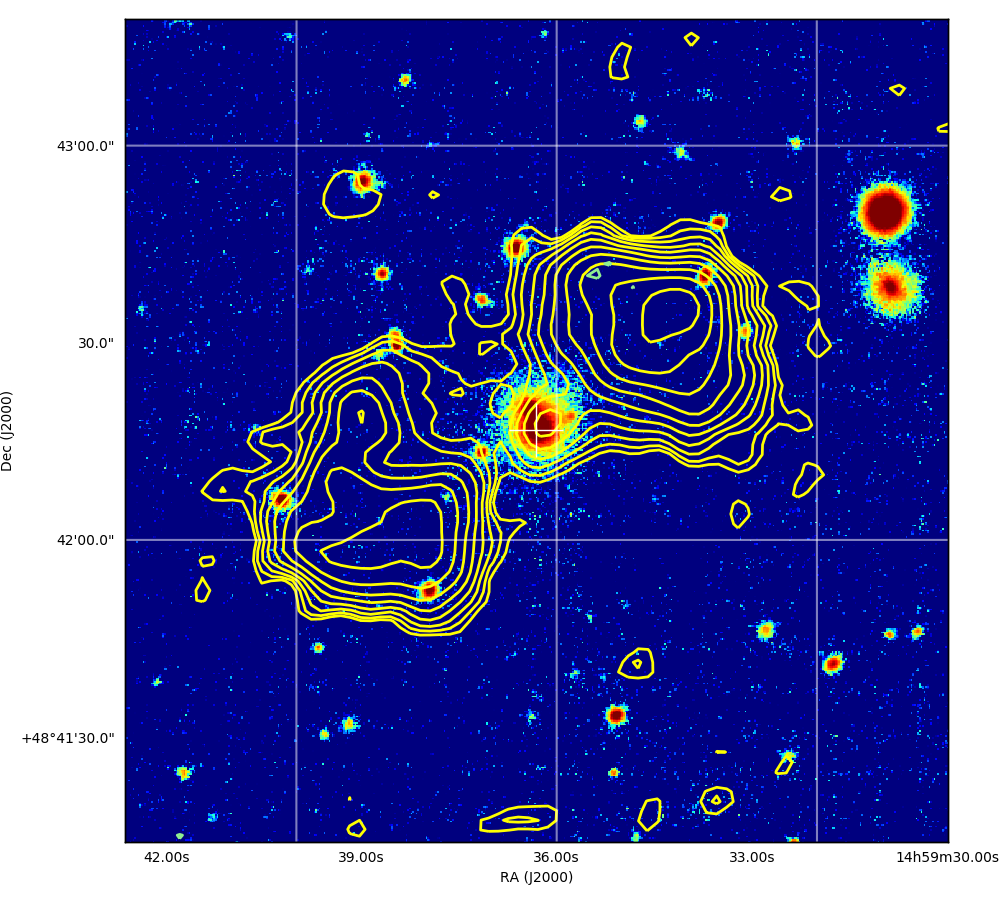}
\caption{\small Examples of the FRII-Low objects, with LoTSS 150-MHz (yellow) and FIRST 1.4 GHz (green) contours overlaid on PanSTARRs i-band images. Vertical grid lines are separated by 1 arcmin.} 
\label{friilow_gallery}
\end{figure*}

Our visual inspection of these sources indicates that, in most cases, their morphology is unambiguously that of an FRII -- we present a gallery of examples in Fig.~\ref{friilow_gallery}. There are, however, particular classes of potential interlopers that could meet our criteria for FRII categorisation. Some low-luminosity FRIIs ($\sim15$ per cent) are bent and could be wide-angle tail sources where we detect emission out to the bright flare-points, but where the tails themselves are too faint for LOFAR to detect.  It is also possible that a subset of these sources have incorrect host identifications or redshift estimates, so that in reality they are at a larger distance than catalogued, and hence are more luminous than reported. However, $\sim 55$ per cent of our low-luminosity FRIIs have a radio enhancement at the centre of the identified host galaxy, suggesting the AGN/jet base is correctly associated with the galaxy. The photometric redshift estimates have an uncertainty of $\sim0.03$ and an outlier fraction of 1.5 per cent, but 51 per cent of the low-luminosity FRIIs (and 54 per cent of those with $L_{150}<10^{25}$ W Hz$^{-1}$) have spectroscopic redshifts, so that, while it is possible that some examples of incorrect host identifications or redshifts are present in our low-luminosity FRII sample, this cannot account for the majority of the low-luminosity FRIIs.

We therefore conclude that a population of low-luminosity radio galaxies with FRII morphology does exist. Two possible theories for the origin of these low-luminosity FRII objects are (1) that they are older sources, which have begun to fade from their peak radio luminosity \citep[e.g.][]{Shabala2008,Hardcastle2018b}, or (2) that they inhabit low-density inner environments so that their jets can remain undisrupted despite having low power. These two explanations may both be relevant to subsets of the FRII-Low population. A third possibility is that there is something more fundamentally different between FRI and FRII jets (and therefore the jet disruption model is wrong). Ongoing, higher resolution JVLA (Karl G. Jansky Very Large Array) follow-up of a sub-sample of low-luminosity FRIIs will enable us to map the hotspot and jet structures in these objects in detail and to establish more firmly whether their jet dynamics appear identical to the higher luminosity FRIIs. However, we can already consider whether the host-galaxy and spectral properties of the FRII-Low sample provide us with clues to why such low-luminosity FRII systems exist. 

Many of the low-luminosity FRIIs have hotspots detected in the FIRST survey, so it is unlikely that all these sources are newly-extinguished, fading FRIIs, although it is possible that a fraction of them may be, and this possibility must be explored further. To compare the properties of the LoTSS low-luminosity FRIIs with canonical high luminosity FRIIs, we selected two samples above and below $L_{150} = 10^{26}$ W Hz$^{-1}$, with similar ranges of angular sizes (40 -- 100 arcsec), physical sizes (200 -- 500 kpc), and $z\leq0.8$. We refer to these subsets, respectively, as FRII-High (49 sources) and FRII-Low (72 sources). 

\begin{figure}
	\centering
\includegraphics[width=0.47\textwidth]{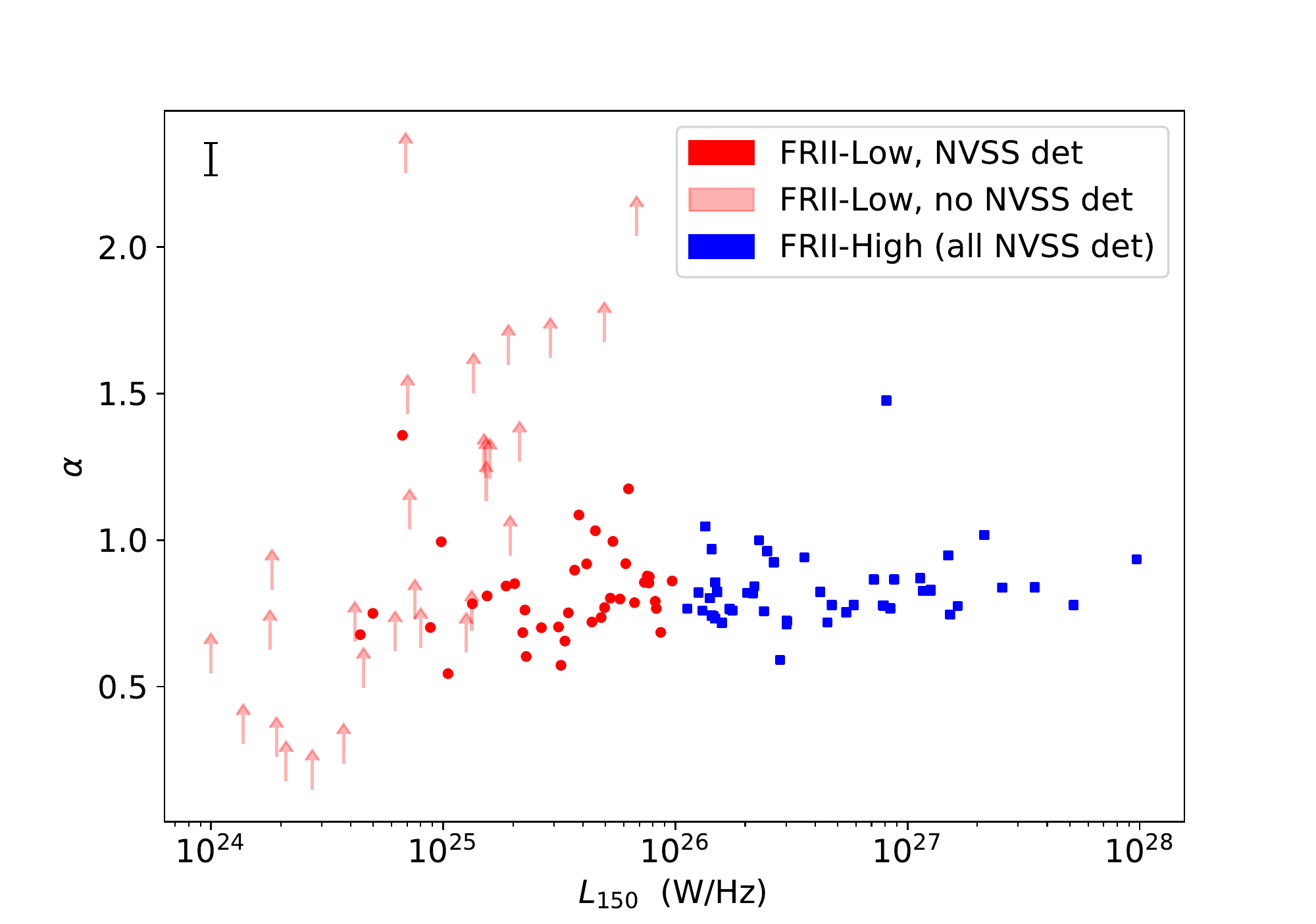}
	\caption{A comparison of LoTSS--NVSS spectral index as a function of 150-MHz luminosity for the FRII-Low and FRII-High subsamples at $z\leq0.8$. Lower limits on the spectral index for the FRII-Low not detected by NVSS are represented with upward-pointing arrows. A representative error bar for $\alpha$ is shown in black on the top left corner of the plot.}\label{NVSS_alpha_L150}
\end{figure}

To test whether FRII-Low are systematically older than FRII-High, we obtained spectral indices where possible using NVSS 1.4-GHz measurements \citep{Condon1992}. 39/72 FRII-Low are detected (at a $3\sigma$ level) by NVSS within 30 arcsec of the LoTSS catalogue position. All 49 FRII-High are detected by NVSS, with separations $<30$ arcsec. For the non-detected sources, we determined a $3\sigma$ upper limit on the 1.4-GHz flux within the area of the detected LoTSS source. Fig.~\ref{NVSS_alpha_L150} shows the distribution of LoTSS-NVSS spectral index ($\alpha$, where radio flux density $S_{\nu} \propto \nu^{-\alpha}$) for the FRII-Low and FRII-High sources. It is apparent that a higher proportion of FRII-Low must have $\alpha>1.0$, indicating that a subset of the FRII-Low are indeed likely to be older sources. However, more than half the FRII-Low have $\alpha$ in the range 0.7 to 1, where nearly all of the FRII-High lie, and so age cannot be the only explanation for the existence of low-luminosity FRIIs. As a further test of this explanation, we considered whether core radio emission in the FIRST survey \citep{becker95} could be used as an additional indicator of currently active jets. However, assuming typical core prominence ratios \citep[e.g.][]{Mullin2008}, the predicted fluxes for the majority of FRII-Low are below the FIRST sensitivity limit, and we cannot perform a useful comparison.

\begin{figure*}
	\centering
\includegraphics[width=0.48\textwidth]{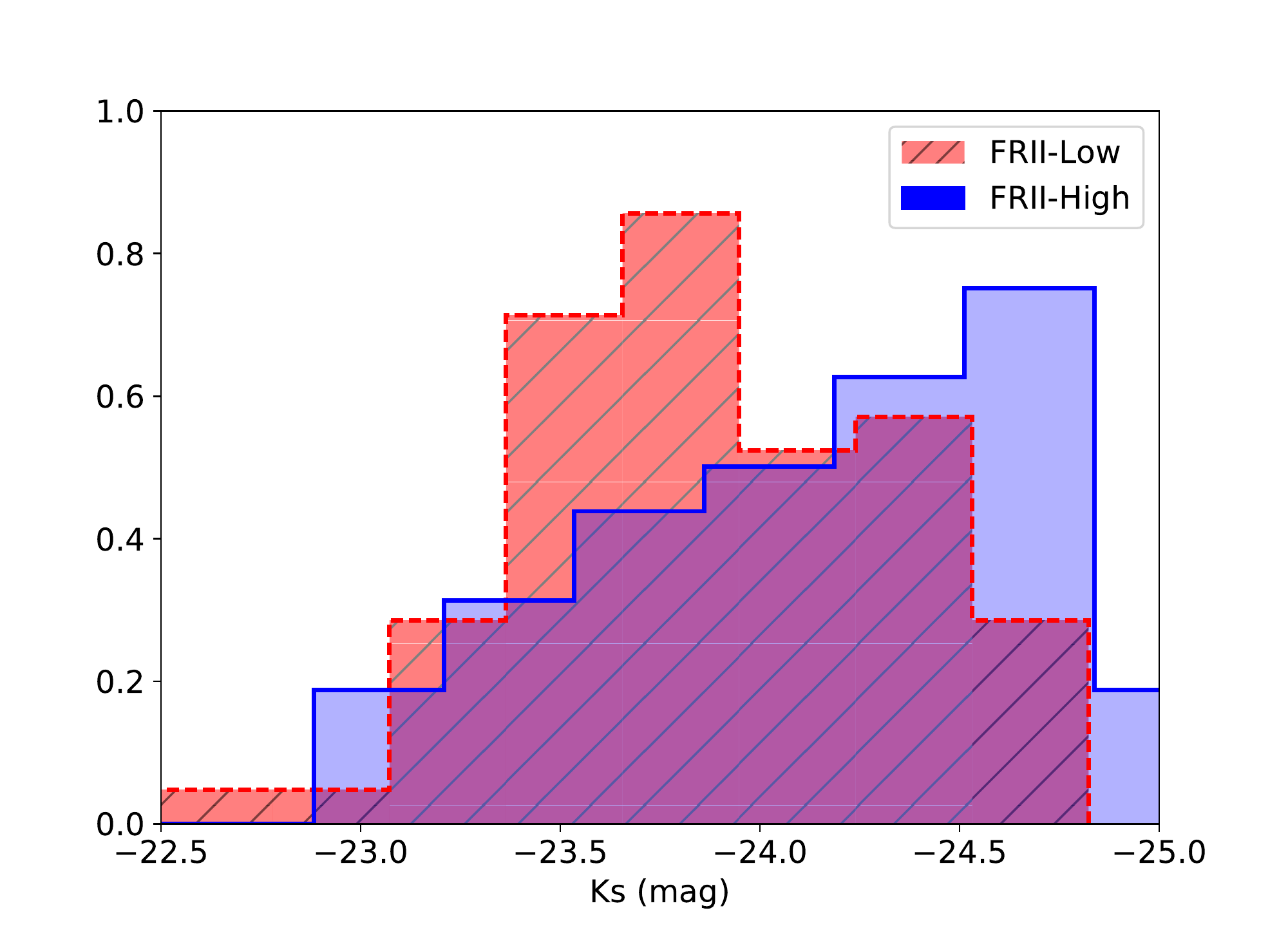}
\includegraphics[width=0.48\textwidth]{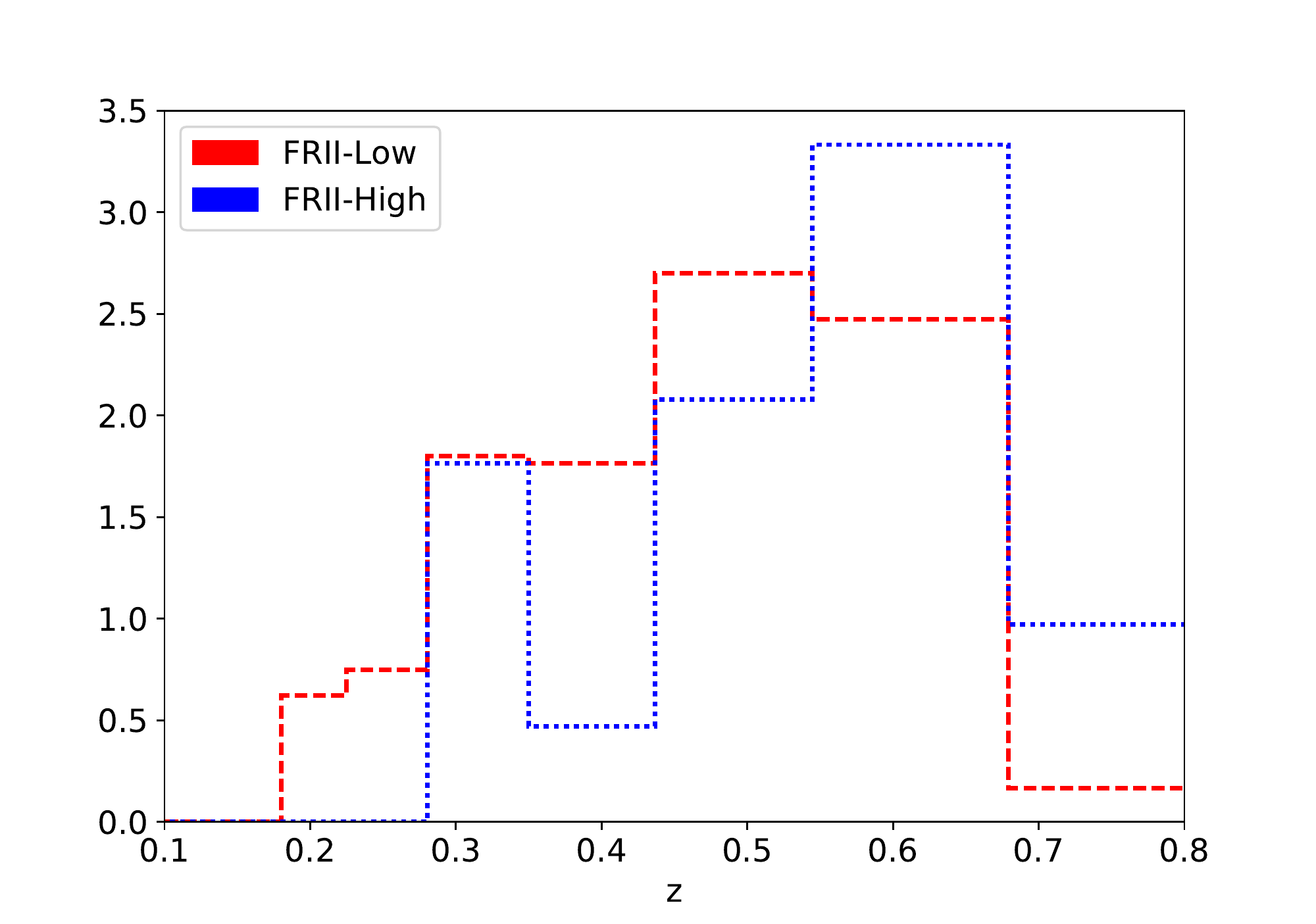}
\includegraphics[width=0.48\textwidth]{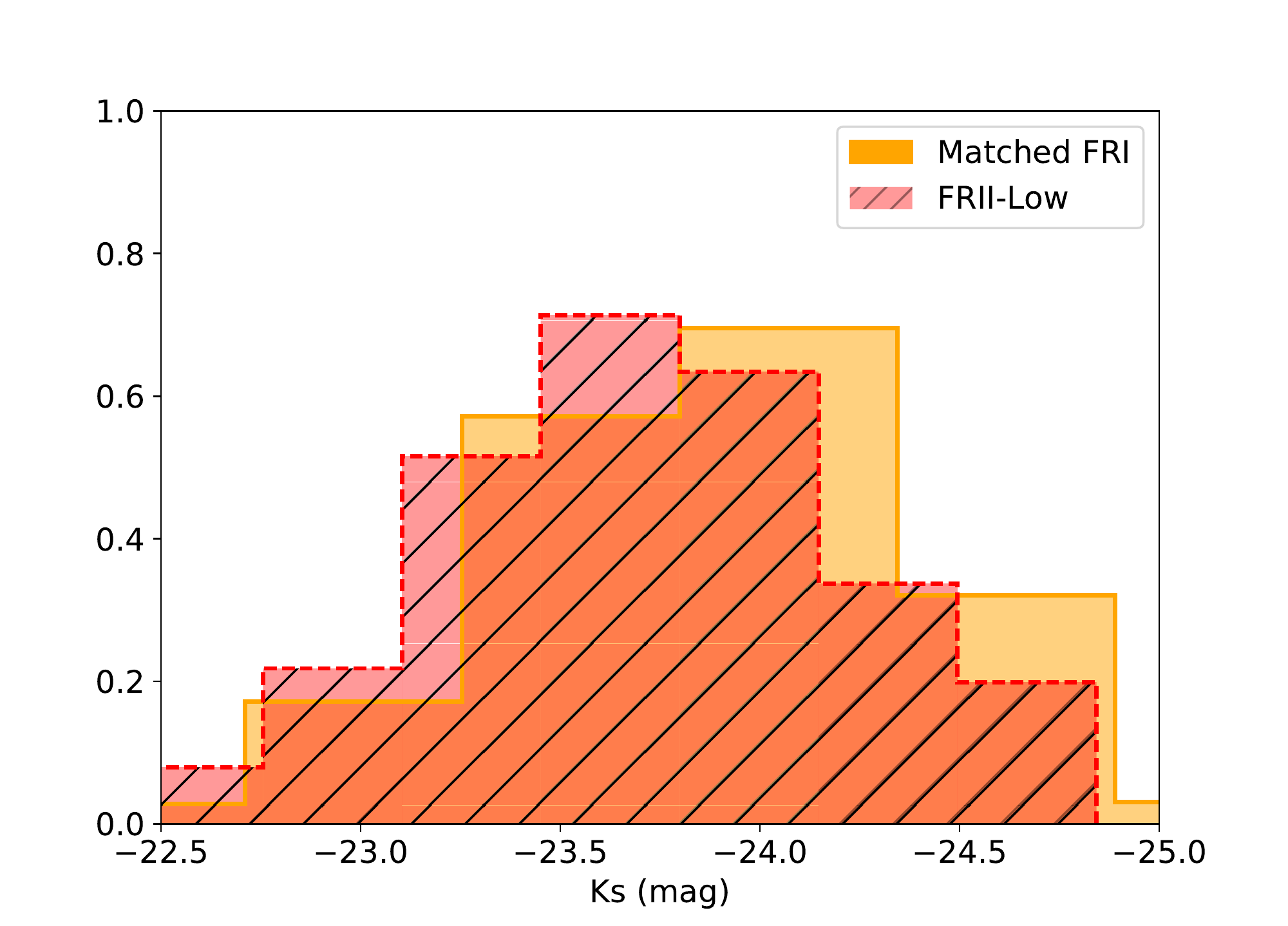}
\includegraphics[width=0.48\textwidth]{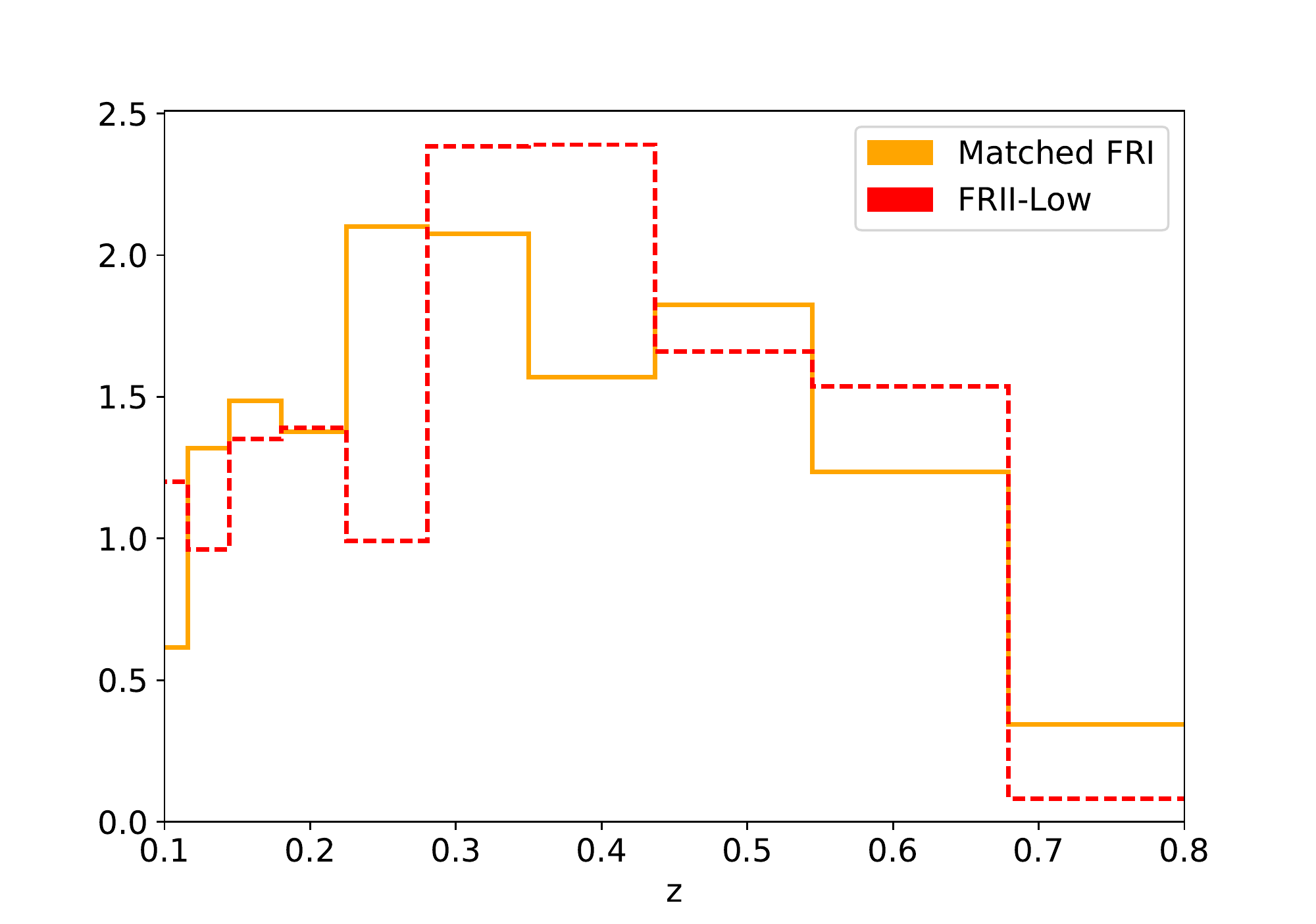}
	\caption{A comparison of host-galaxy $M_{K_{s}}$ and $z$ distibution, at $z\leq0.8$, for the FRII-Low and FRII-High subsamples (top), and for the FRII-Low and FRI subsamples of matched luminosity (bottom).}\label{low_hosts}
\end{figure*}

We next investigated the host-galaxy properties, to test whether the FRII-Low inhabit fainter hosts that are likely to have a lower inner density, reducing the likelihood of jet disruption. In the top panel of Fig.~\ref{low_hosts} we compare the distribution of host-galaxy rest-frame $K_{s}$ band magnitudes \citep[$M_{K_{s}}$,][]{Duncan2018b} of the FRII-Low with the FRII-High subsample (see above), restricting the sample to the range of physical and angular sizes occupied by both populations. It is apparent that the host-galaxy magnitudes are significantly different for the two subsamples: FRII-Low sources inhabit systematically lower luminosity host galaxies. The right-hand panel of the Figure shows the redshift distributions for the two subsamples, which are not significantly different, and so the difference in host-galaxy properties for FRII-Low and FRII-High cannot be explained by selection effects. 

If the jet disruption model for the FR break is correct we would expect that, compared to an FRI source of similar {\it jet power}, an FRII source would reside in a less rich inner environment, and so we would also predict a difference in the host-galaxy properties of FRII-Lows and FRIs of similar luminosity. In the lower panel of Fig.~\ref{low_hosts} we therefore also compared the FRII-Low host-galaxy properties with those of a sample of FRIs selected to have the same range in size and radio luminosity (bearing in mind that luminosity does not equate to jet power). There is a small apparent difference in the distributions, in the expected sense that the matched FRI hosts appear systematically slightly brighter -- we used a Mann-Whitney U test to investigate whether the two samples have the same underlying distribution of $M_{K_{S}}$, and find that the null hypothesis can be ruled out at $>99.9$ per cent confidence. Hence we can conclude that the FRII-Lows have systematically fainter host galaxies than the FRIs in the same radio luminosity range. The right-hand panel demonstrates that the redshift distributions for the two subsamples are indistinguishable, so that the host-galaxy difference cannot be attributed to different redshift ranges for the two sub-samples.  

As a further check we compared the large-scale environments of the matched subsamples, to assess whether this could have an additional influence, but as environmental information is currently only available for systems with $z<0.4$ \citep{Croston2018b}, the fraction of sources in each subsample with a cluster match are consistent to within somewhat large uncertainties. 

We therefore conclude that the low-luminosity FRII population revealed by LoTSS is consistent with the jet disruption model, and that it is likely to be made up of two main categories of object: low-power jets hosted by galaxies of lower mass than the high-luminosity FRIIs and the similar luminosity FRIs, enabling the jets to remain undisrupted; and older FRIIs that are starting to fade from their peak luminosity but retain an edge-brightened morphology. 

A crucial question then is why we see such a substantial overlap in the luminosities for FRI and FRII populations with LOFAR (and previously with FIRST/NVSS samples), whereas \citet{FR1974} saw a much cleaner distinction, with no FRII morphology sources below $L_{150} \sim 10^{26}$ W Hz$^{-1}$. The most obvious difference between the two samples is the strong flux limit of 10.9 Jy at 178 MHz for 3CRR, compared to $\sim 2$ mJy for our sample selected for morphological classification from the more sensitive overall LoTSS catalogue. The high flux limit for 3CRR has a profound effect on the redshift distributions of the FRIs and FRIIs being compared: taking $L_{150} = 10^{26}$ W Hz$^{-1}$ as the FR break value, in the 3CRR sample objects below this luminosity can only be detected to $z>0.06$. Objects significantly below the FR break (e.g. with $L<10^{25}$ W Hz$^{-1}$ cannot be detected in 3CRR beyond $z=0.02$. Only 8 3CRR FRIIs have $z<0.06$, and {\it none} are below $z<0.02$. For the FRIs in 3CRR, 21 have $z<0.06$ and 7 $z<0.02$. If we consider the ratio of FRII-lows to FRIs in our sample, and assume that this ratio will be the same for 3CRR in the redshift range where FRIs and FRII-low can be detected, then we would predict that $3\pm2$ FRII-low might be expected in 3CRR, which is not very different from the observed value of zero. It is also worth noting that only 3/216 of the FRII-low ($L_{150} < 10^{26}$ W Hz$^{-1}$) in our sample have $z<0.06$. We therefore conclude that the absence of FRII-lows in the 3CRR sample can be entirely explained by their rarity in the local Universe together with the high flux limit of 3CRR. In future work it will be interesting to explore how host-galaxy evolution may be relevant for the relative prevalence of FRII-low and FRI radio galaxies.

Finally, we note that we find seven sources with $10^{25}<L_{150}<10^{26}$ W Hz$^{-1}$ and sizes larger than 1 Mpc (all smaller than 2 Mpc), and thus giant radio galaxy (GRG) candidates. Although their redshifts are photometric they are relatively well-constrained, and thus their sizes and luminosities should be reasonably accurate as well \citep[see the discussion by][]{Hardcastle2018c}. They represent a very small fraction of FRII-Lows ($\sim 3$ per cent), which is consistent with the fact that GRGs are believed to grow fast, arising from massive hosts into relatively sparse environments \citep[see][and references therein]{Dabhade2019,Hardcastle2018c,sabater2018}, in contradiction with the lower-mass hosts of the overall FRII-Low population. Optical spectroscopy of their hosts and higher frequency radio data to constrain their ages could shed some light into whether these seven sources are true GRGs and why they are underluminous.

\subsection{Testing the jet disruption model: host-galaxy dependence of the FR break}\label{ledlow}

The apparent existence of an optical-magnitude dependence of the FR break luminosity, reported by \citet{Ledlow1996}, provided a strong piece of supporting evidence for a jet deceleration and disruption origin of the FRI/II dichotomy \citep{Bicknell1995,Kaiser2007}. If jet disruption is caused by the interaction of jet power with environmental density, then a jet of similar power close to the FR break is more likely to get disrupted and become an FRI in a denser environment. Therefore, if optical magnitude is a reasonable proxy for local density on the scale of jet disruption (a few kpc), the FR break luminosity should have an observed dependence. However, the initial result of \citet{Ledlow1996} has since been called into question \citep[e.g.][]{Best2009, Lin2010, Wing2011, Singal2014, Capetti2017FRII, Shabala2018} due to the potential influence of selection effects: for both the literature and Abell cluster samples examined in \citet{Ledlow1996}, the FRIs and FRIIs have significantly different redshift distributions and come from highly flux-limited samples. The large vertical scatter in the original plot of \citet{Ledlow1996} is also important, as highlighted, e.g., by \citet{Saripalli2012}, since it highlights the fact that for a given type of host galaxy, it is possible to produce both FRI and FRII systems, presumably as a result of significantly different jet powers (or other environmental factors less well correlated with optical magnitude). 

There are a number of reasons why the substantial FRI/II luminosity overlap discussed in the previous section may be compatible with the jet disruption model for the FR break. At least an order of magnitude scatter in radio luminosity is likely to exist for a given jet power \citep[e.g.][]{cavagnolo2010,Croston2018}, which already explains some FRI/II overlap in luminosity if an underlying FR break in jet power exists. Another effect acting in the direction of producing overlap is that in denser environments the synchrotron plasma will be better confined, and thus lose less energy by adiabatic expansion, causing it to appear brighter for a given jet power than one in a less rich environment \citep[e.g.][]{Barthel1996}. Therefore, if the FRIs are in denser environments than lower luminosity FRIIs, causing their jets to disrupt, they will also appear more luminous, further enhancing the overlap observed in Fig.~\ref{low_hosts}. 

\begin{figure*}
	\centering
\includegraphics[width=0.55\textwidth]{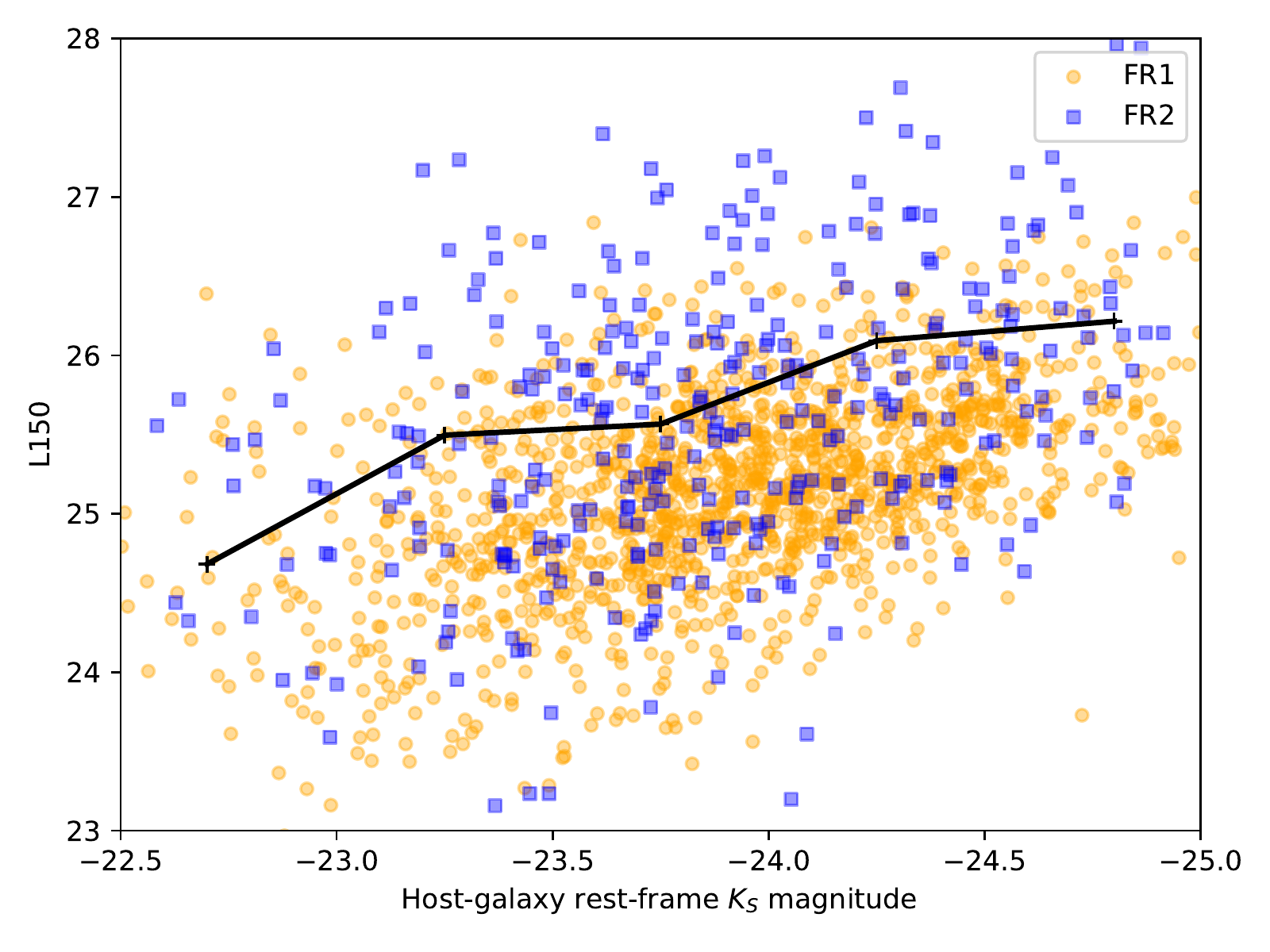}
\includegraphics[width=0.99\textwidth]{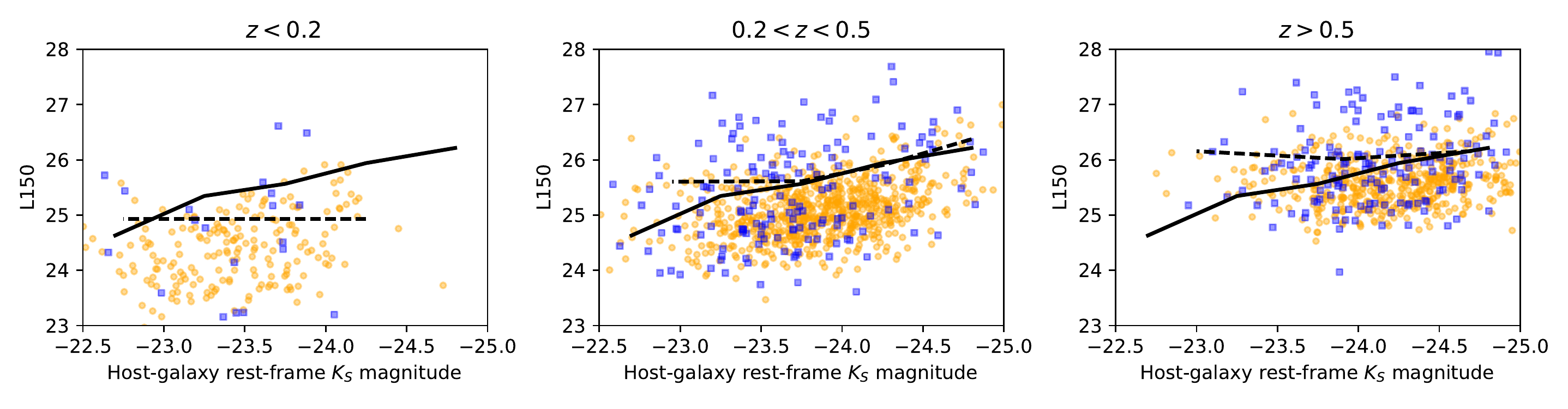}
	\caption{Top: the relationship between morphology, radio luminosity and host-galaxy magnitude (a ``Ledlow \& Owen'' plot). The black line indicates the luminosity above which the normalised probability of finding an FRII exceeds that of finding an FRI. Bottom row: the same sample split into three redshift bins, with dashed lines indicating the break luminosity determined for each redshift slice, and the solid lines showing the full-sample relation as shown in the upper panel.}\label{ledlow_plot}
\end{figure*}

We therefore wanted to revisit the result of \citet{Ledlow1996} and to investigate the dependence of the FR break on host-galaxy properties within our LoTSS sample. Fig.~\ref{ledlow_plot} shows the relationship between morphology, radio luminosity and host-galaxy magnitude for the $z<0.8$ FRI and FRII samples. We use the host-galaxy rest-frame $K_{s}$ magnitudes \citep[$M_{K_{s}}$,][]{Duncan2018b}, as a proxy of overall stellar mass \citep[see e.g.][]{Bell2003b,Caputi2005,Arnouts2007,Konishi2011}. It is important to note, however, that the relationship between $M_{K_{S}}$ and the inner gas pressure distribution -- the quantity of direct influence on jet evolution -- is not well determined. 

The substantial overlap between FRI and FRII populations remains present when radio luminosity is plotted against $M_{K_{s}}$, but both the FRI and FRII samples show a trend of increasing radio luminosity with host-galaxy magnitude. Using six bins in rest-frame $K_{s}$ magnitude, we calculated the luminosity above which the normalised probability of finding an FRII exceeds that of finding an FRI, with errors estimated using Monte Carlo simulations of the two populations with the observed means and dispersion. A strong trend is observed, with the FR break luminosity increasing by over an order of magnitude from the faintest to the brightest host galaxies. To first appearance, therefore, we do see a ``Ledlow \& Owen'' trend in the LoTSS dataset.

However, our sample spans a large range in redshift, out to $z=0.8$, and necessarily suffers from biases due to radio luminosity and host-galaxy flux limits, and radio surface brightness limits. In the lower panels of Fig.~\ref{ledlow_plot} we subdivide the sample into three redshift bins and calculate the break luminosity in bins of host-galaxy magnitude in the same way as for the full sample. Moving from left to right (with increasing redshift) it is clear that the average FR break has a strong dependence on redshift -- although the intermediate redshift slice shows some evidence for a trend partially following that for the full sample, it is evident that the higher break luminosity for bright host-galaxy magnitudes (towards the right-hand side of the top panel of Fig.~\ref{ledlow_plot}) is being driven mainly by high redshift objects, and the lower break luminosity at fainter host-galaxy magnitudes is driven primarily by low redshift objects. There may nevertheless be an underlying dependence of the FR break on host-galaxy magnitude, but with our sample statistics and the strong redshift dependences present in the sample, we must conclude that the observed trend may be induced entirely by selection effects, likely a combination of volume effects, radio surface brightness and host-galaxy magnitude limits. Similar selection effects are likely to have affected previous claims for a host-galaxy dependence.

As LoTSS expands to larger sky areas it will be possible to construct large samples in narrow redshift slices at intermediate redshifts so as to span a wide luminosity range, and so to remove the complications of redshift dependence for this type of comparison. However, given the large FRI/II overlap and the multiple physical explanations for the absence of sharp transitions in the population, it may be that more focused in-depth comparisons of the hosts and environments of the low-luminosity FRIIs with similarly luminous FRIs are the most fruitful route to better physical insights into the origin of the FR break.

\subsection{The candidate hybrid class}\label{hybrids}

\begin{figure*}
    \begin{subfigure}{0.31\linewidth}
        \centering
        \includegraphics[width=0.9\linewidth, trim={1.4cm 1cm 2cm 2.4cm},clip]{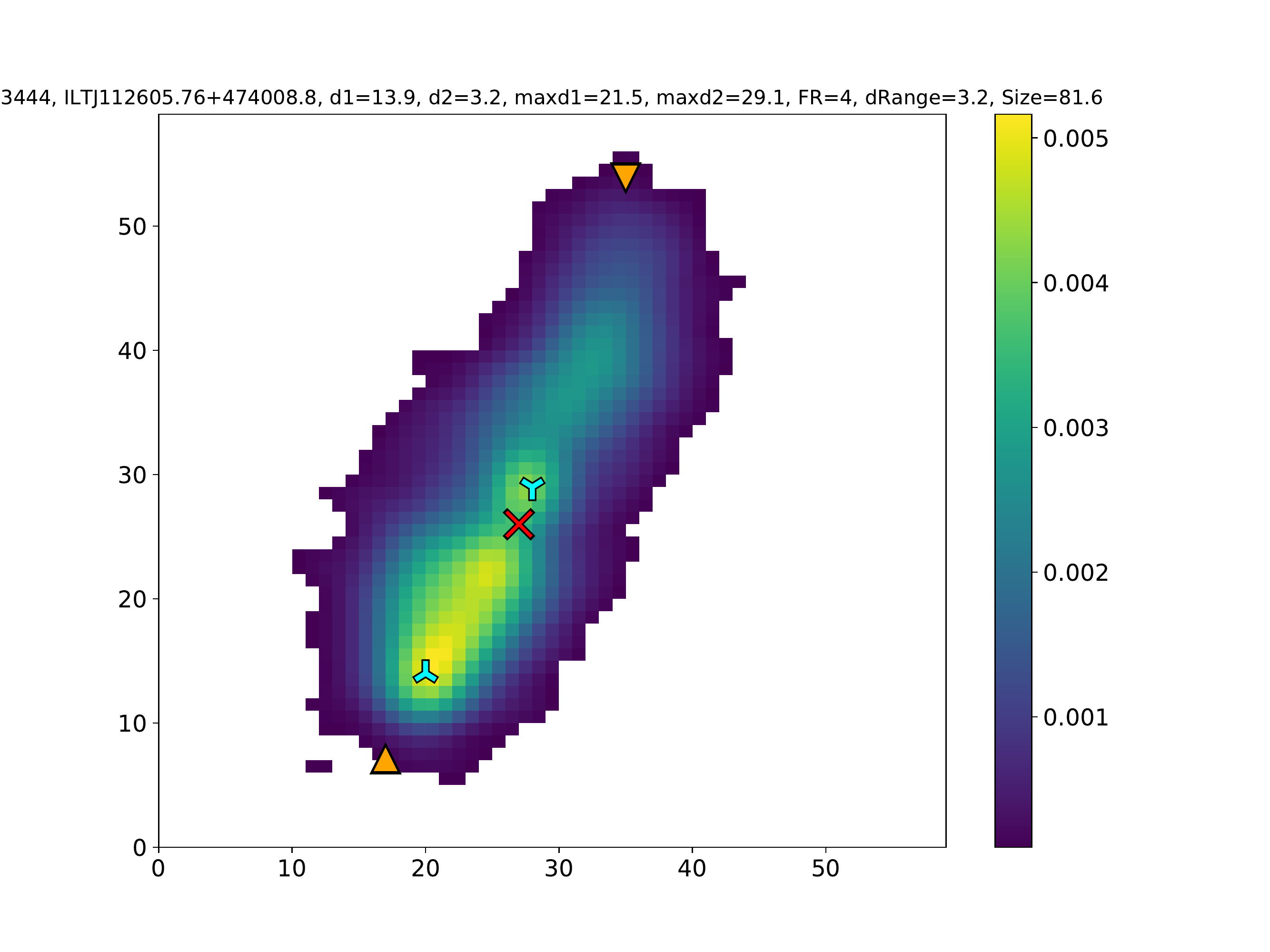}
        \caption{}
        \label{Hybrid_1}
    \end{subfigure}
    \begin{subfigure}{0.31\linewidth}
        \centering
        \includegraphics[width=0.9\linewidth, trim={1.4cm 1cm 2cm 2.4cm},clip]{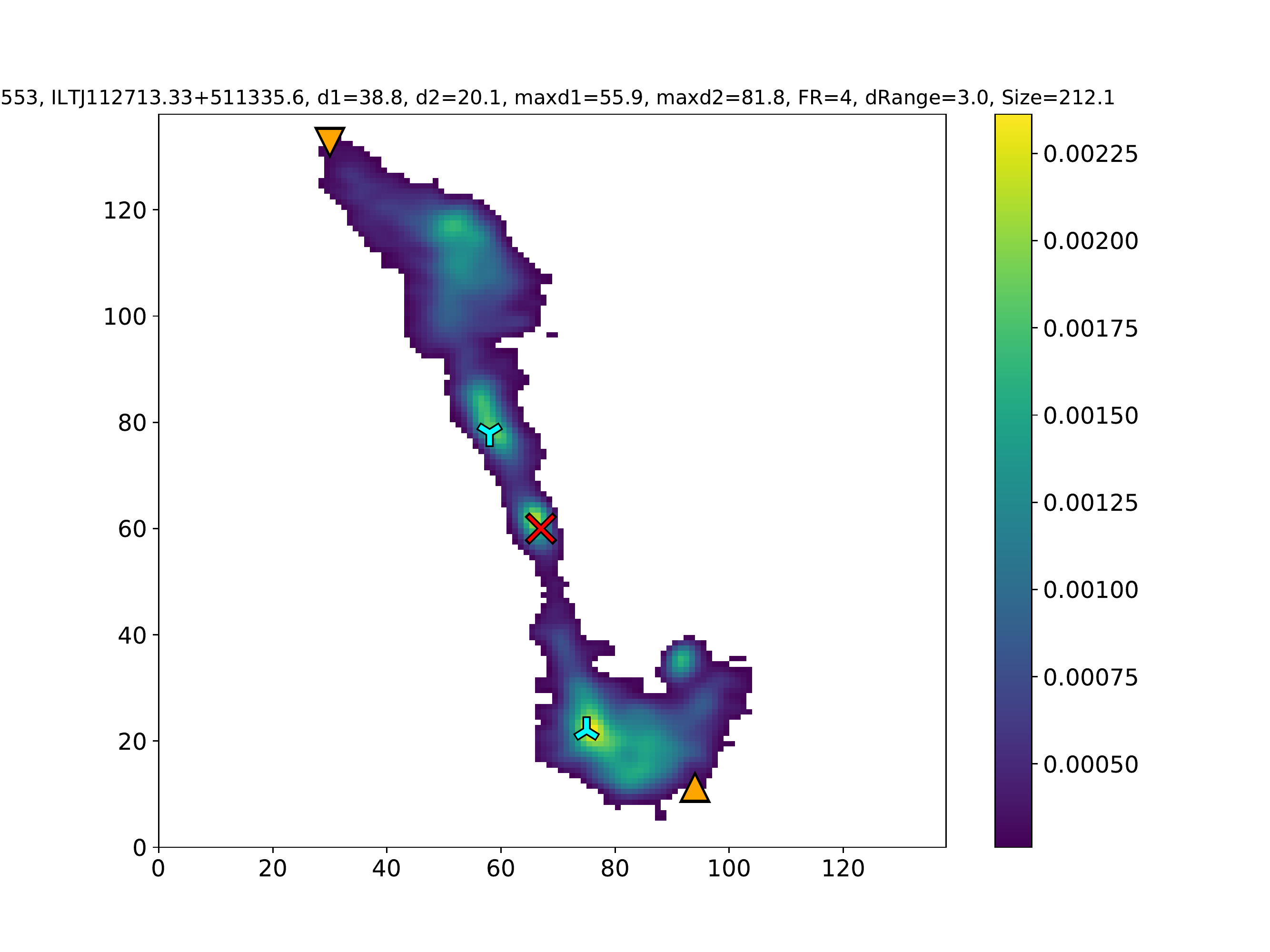}
        \caption{}
        \label{Hybrid_2}
    \end{subfigure}
    \begin{subfigure}{0.31\linewidth}
        \centering
        \includegraphics[width=0.9\linewidth, trim={1.4cm 1cm 2cm 2.4cm},clip]{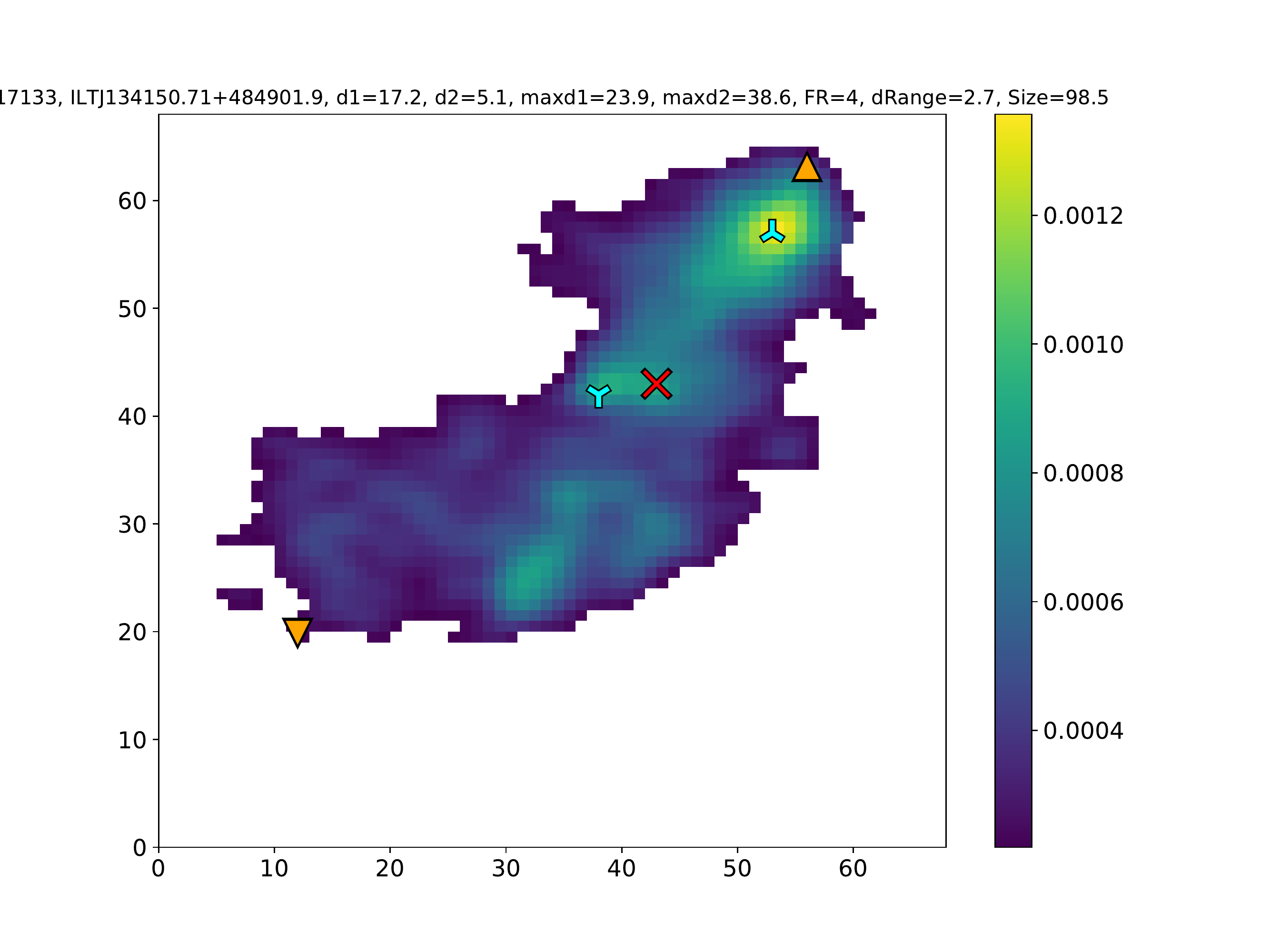}
        \caption{}
        \label{Hybrid_3}
    \end{subfigure}
    \begin{subfigure}{0.31\linewidth}
        \centering
        \includegraphics[width=0.9\linewidth, trim={1.4cm 1cm 2cm 2.4cm},clip]{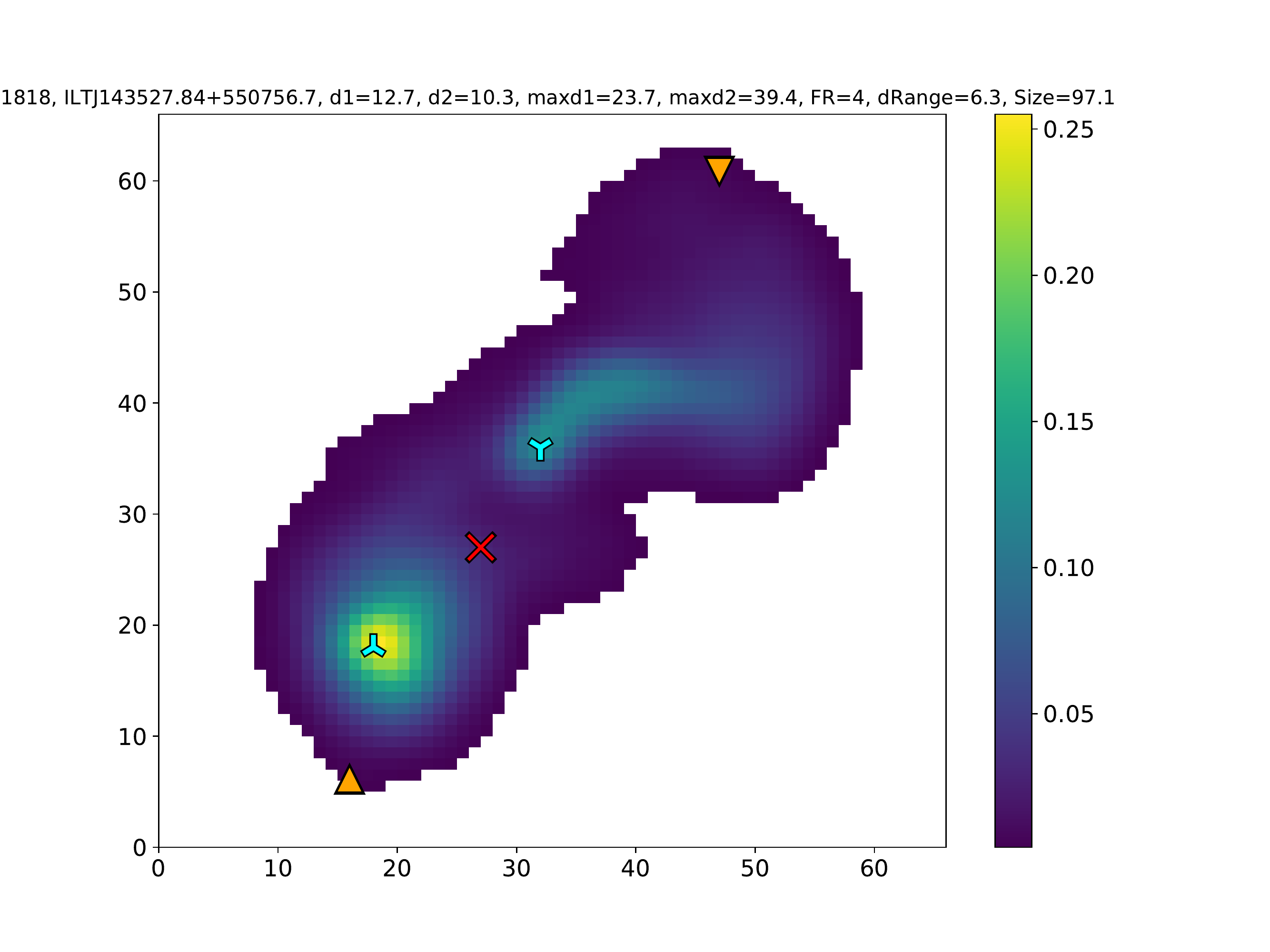}
        \caption{}
        \label{Hybrid_4}
    \end{subfigure}
    \begin{subfigure}{0.31\linewidth}
        \centering
        \includegraphics[width=0.9\linewidth, trim={1.4cm 1cm 2cm 2.4cm},clip]{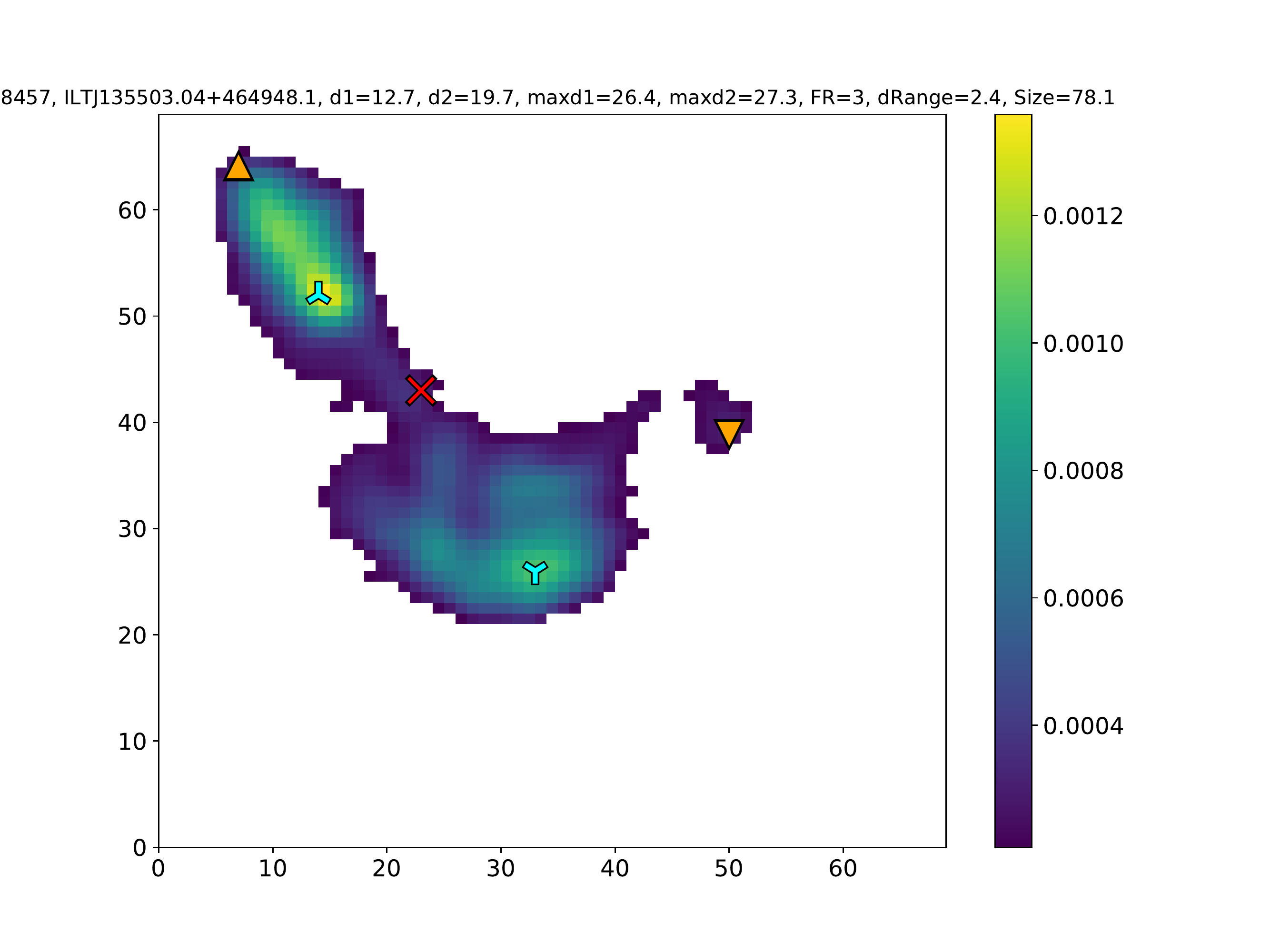}
        \caption{}
        \label{Hybrid_5}
    \end{subfigure}
    \begin{subfigure}{0.31\linewidth}
        \centering
        \includegraphics[width=0.9\linewidth, trim={1.4cm 1cm 2cm 2.4cm},clip]{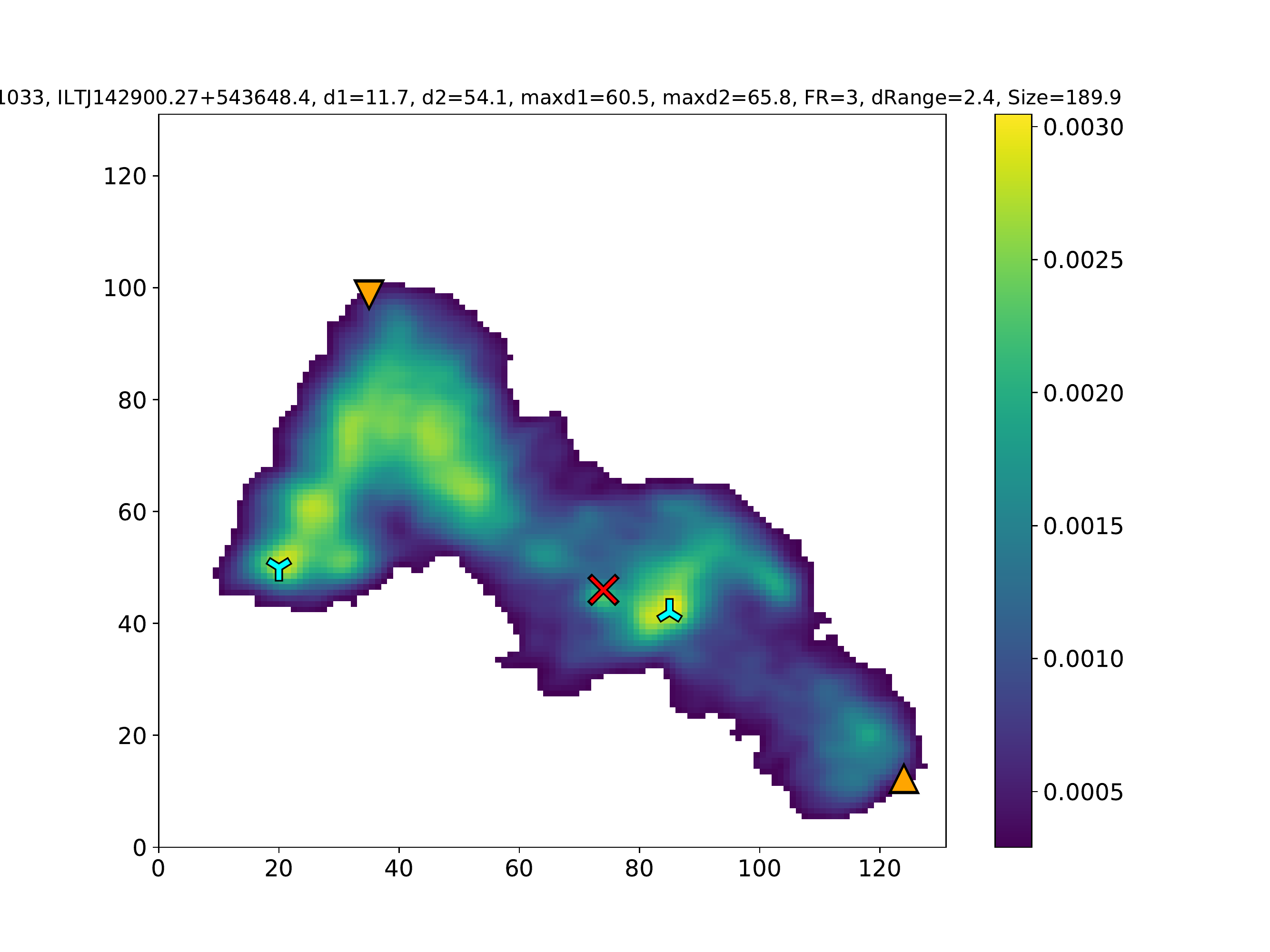}
        \caption{}
        \label{Hybrid_6}
    \end{subfigure}
    \caption{Examples of sources that LoMorph automatically classifies as Hybrids. Colours and symbols as in Fig.~\ref{Examples}.}
    \label{Hybrids_fig}
\end{figure*}

LoMorph classified a substantial subset of the main sample, 422 sources (405 at $z\leq0.8$), as having a candidate hybrid morphology (see Table~\ref{Class_stats}), i.e. the classification on one side is FRI, and on the other side is FRII. A further 209 sources in the S0 size category fell into this class. Visual inspection shows that roughly 75 per cent of the 422 sources in this category can be clearly classified by eye as FRI or FRII, with the automated classification resulting from one side being artificially altered or extended due to one or more of the factors discussed in Section~\ref{vis} (intruding sources, noise, bad host identification, projection effects, deconvolution limitations). Improved imaging and cataloguing for LOFAR surveys data, and/or refinements to the masking and classification algorithms could enable correct classification of many of these sources. We do not believe that the relatively large proportion of sources in this misclassified category ($\sim 5$ per cent of the total sample) introduces any significant biases into our science analysis

Based on visual inspection, we estimate that up to $\sim 25$ per cent of the sources in this class could be true hybrid radio galaxies, or HyMORs \citep[see e.g.][Harwood et al., in prep.]{Gopal-Krishna2000,Gawronski2006,Kapinska2017}. The nature of these systems remains under debate, but it is likely that they remain a heterogeneous class, with the role of projection effects difficult to rule out in many cases. Some examples of LOFAR candidate HyMORs can be seen in Fig.~\ref{Hybrids_fig}. In some cases it seems likely that projection effects may be causing the asymmetry, e.g. Fig.~\ref{Hybrid_1}, which looks like a wide-angle tail in projection, Fig.~\ref{Hybrid_2}, which looks similar to e.g. 3C~465, and perhaps Fig.~\ref{Hybrid_4}. In some cases other factors (jet propagation through an uneven environment, restarting activity, cluster emission) may also be at play, such as in the examples shown in Figs.~\ref{Hybrid_3},~\ref{Hybrid_5}, and~\ref{Hybrid_6}.

While the fraction of hybrid candidates identified by our code is not large, it represents a substantial increase in potential candidates, compared to existing samples, thanks to the ability of LOFAR to resolve fainter, older extended structures compared to previous surveys. We therefore intend to carry out dedicated follow-up to determine the nature of this population.


\subsection{FRI subpopulations}\label{subpops}

Within our FRI class, there are number of sub-categories of physically distinct objects. With many new wide-field surveys coming online, and considerable effort being expended on automated morphological classification, it is important to examine the heterogeneity of our FRI sources and consider the implications for AGN survey science. Below we discuss two FRI subclasses present in significant numbers within our sample, the core-dominated FRIs, and the bent-tailed sources (narrow-angle and wide-angle tails). Table~\ref{FR_z_stats} shows the statistics for each subset.


\subsubsection{Core-dominated `FRI'}\label{CoreD}

\begin{figure*}
    \begin{subfigure}{0.31\linewidth}
        \centering
        \includegraphics[width=0.9\linewidth, trim={1.4cm 1cm 2cm 2.4cm},clip]{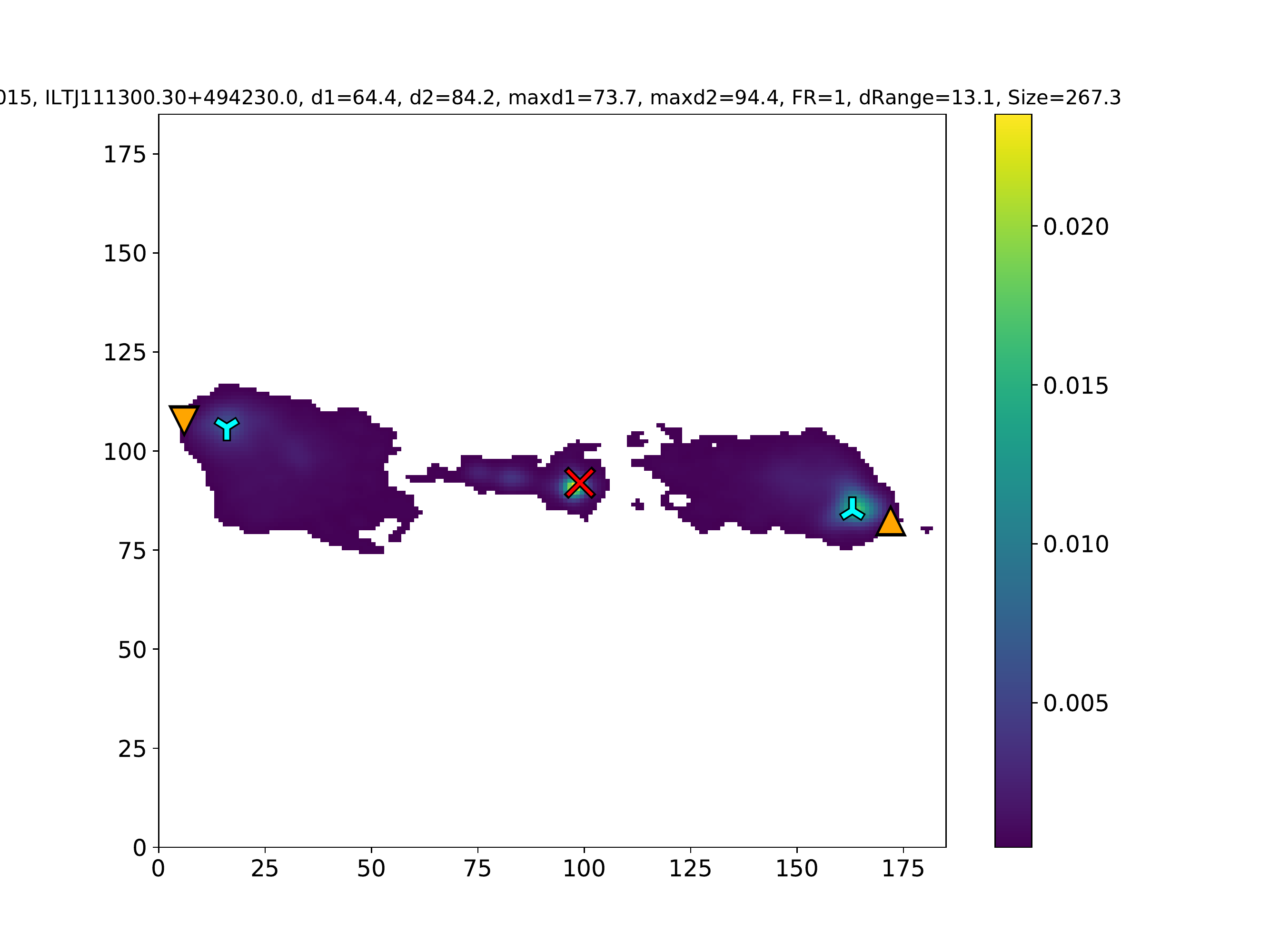}
        \caption{}
        \label{CoreD_1}
    \end{subfigure}
    \begin{subfigure}{0.31\linewidth}
    \centering 
        \includegraphics[width=0.9\linewidth, trim={1.4cm 1cm 2cm 2.4cm},clip]{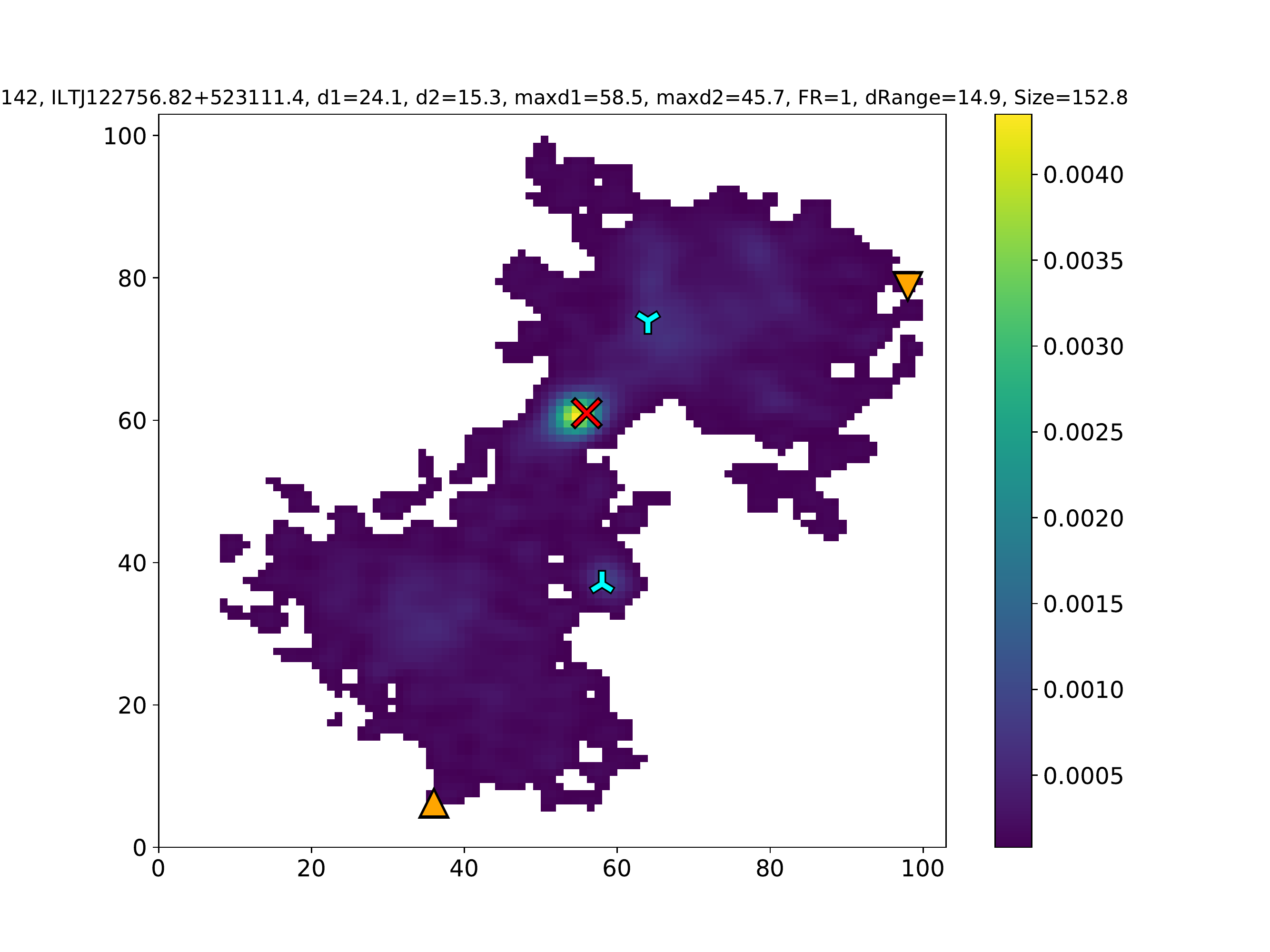}
        \caption{}
        \label{CoreD_2}
    \end{subfigure}
    \begin{subfigure}{0.31\linewidth}
        \centering
        \includegraphics[width=0.9\linewidth, trim={1.4cm 1cm 2cm 2.4cm},clip]{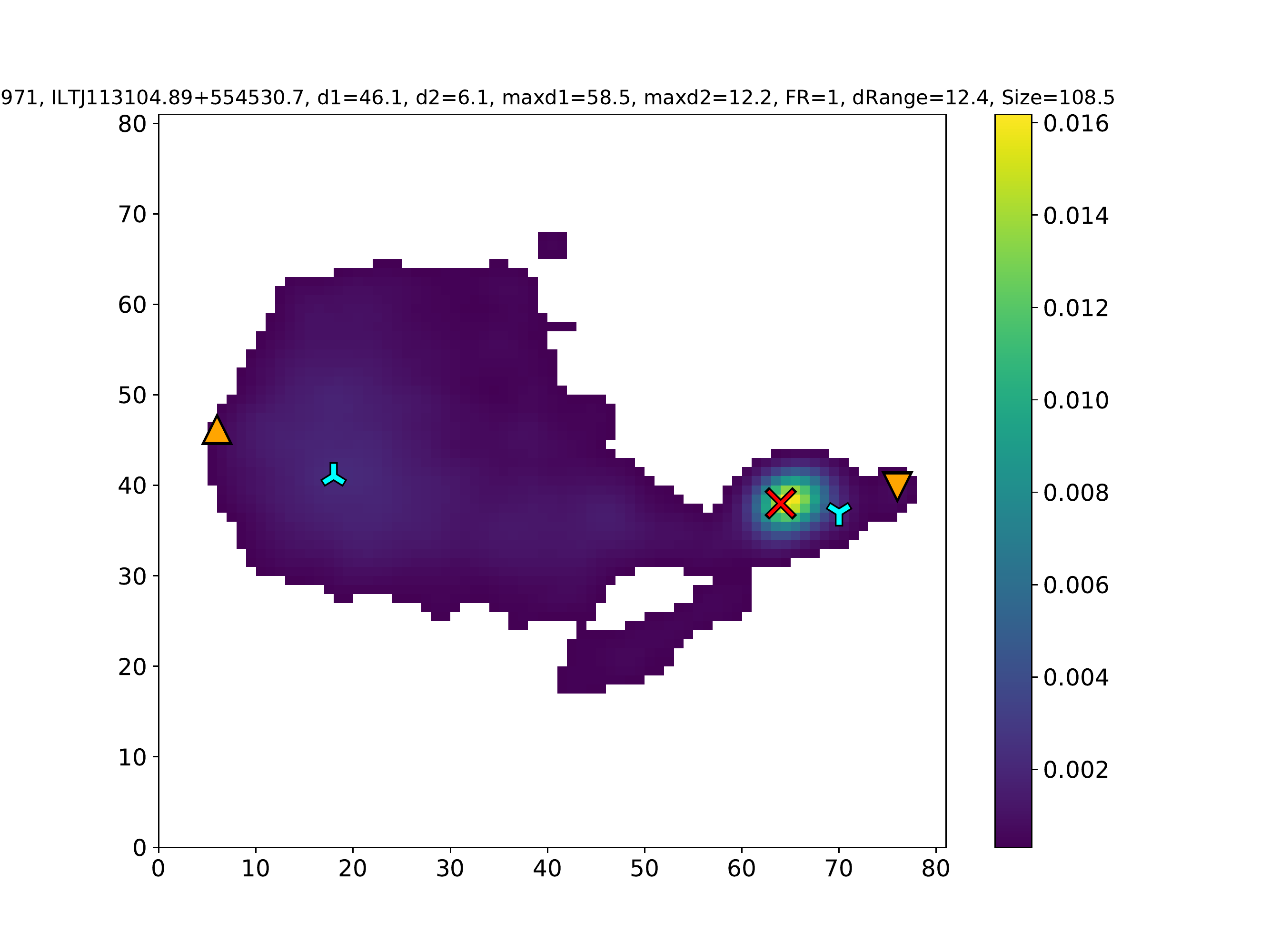}
        \caption{}
        \label{CoreD_3}
    \end{subfigure}
    \begin{subfigure}{0.31\linewidth}
        \centering
        \includegraphics[width=0.9\linewidth, trim={1.4cm 1cm 2cm 2.4cm},clip]{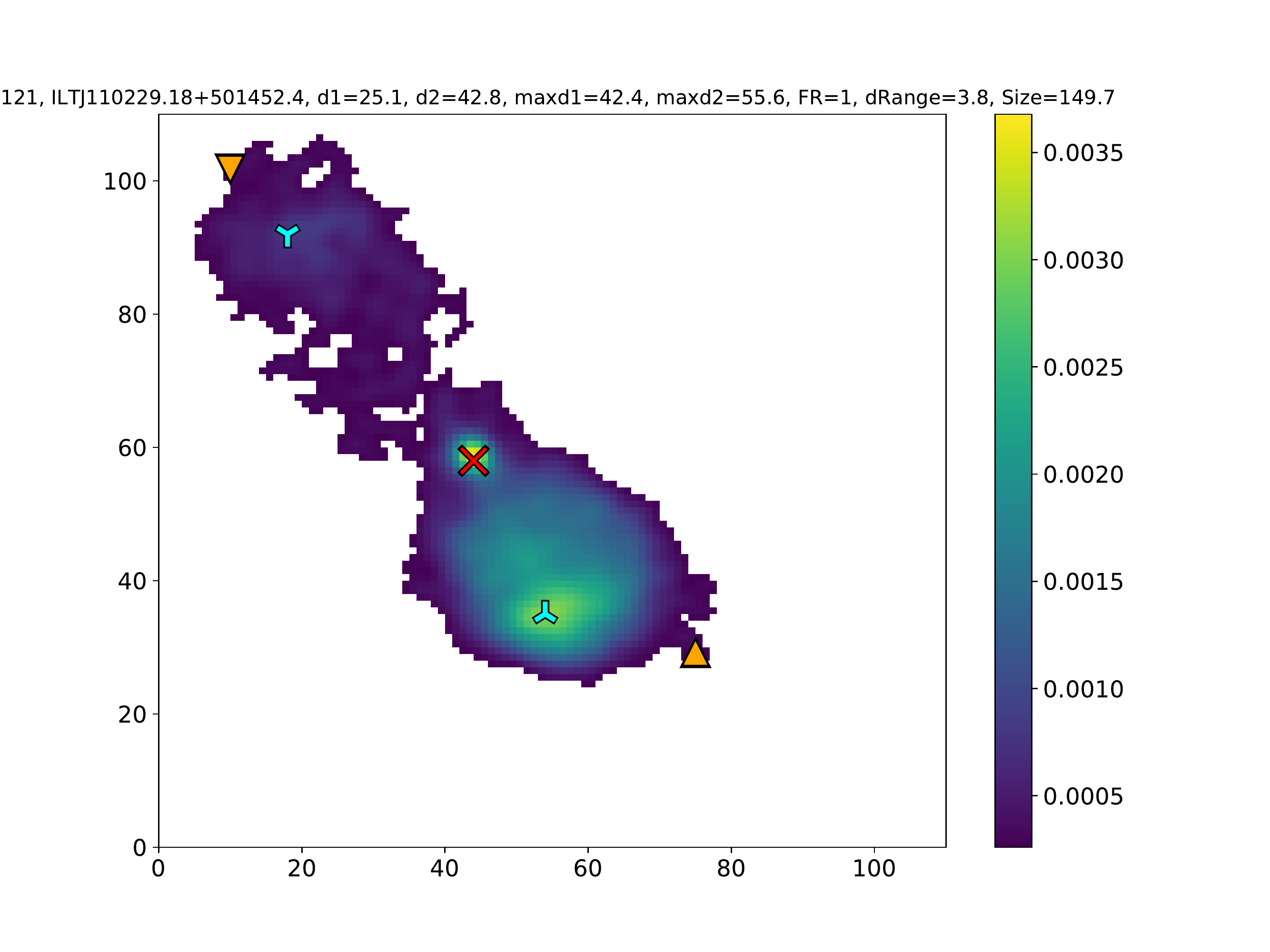}
        \caption{}
        \label{CoreD_4}
    \end{subfigure}
    \begin{subfigure}{0.31\linewidth}
    \centering
        \includegraphics[width=0.9\linewidth, trim={1.4cm 1cm 2cm 2.4cm},clip]{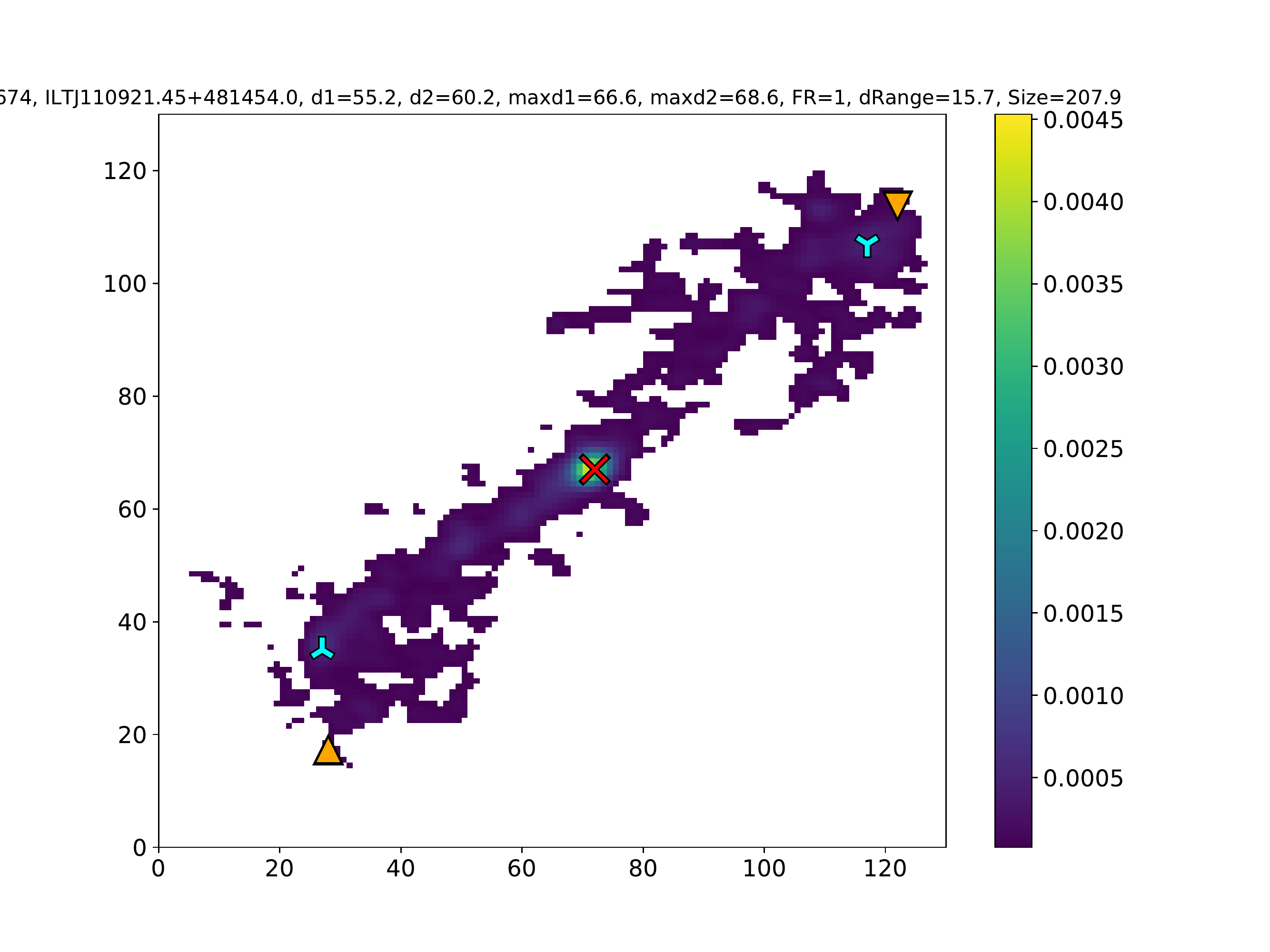}
        \caption{}
        \label{CoreD_5}
    \end{subfigure}
    \begin{subfigure}{0.31\linewidth}
        \centering
        \includegraphics[width=0.9\linewidth, trim={1.4cm 1cm 2cm 2.4cm},clip]{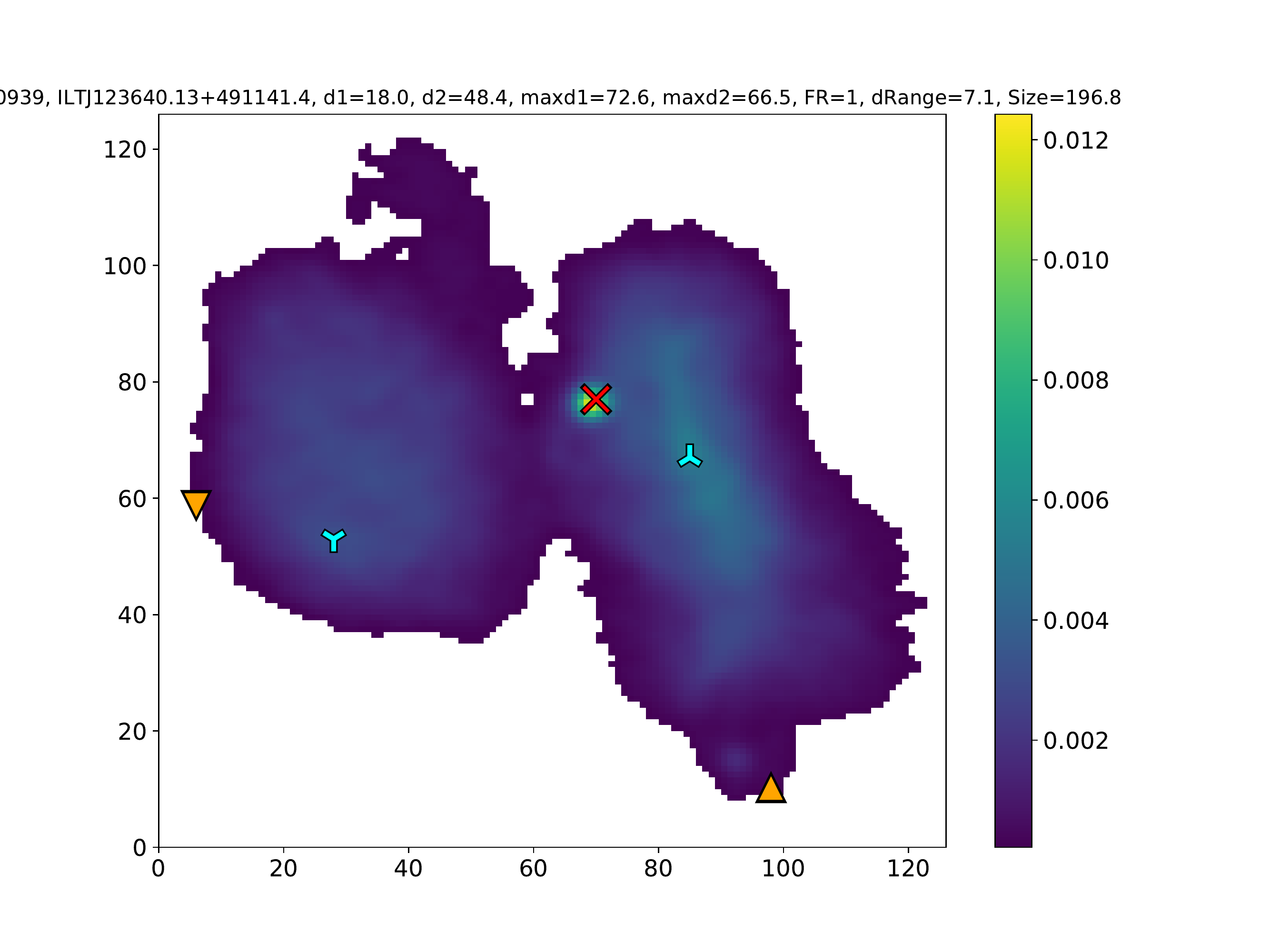}
        \caption{}
        \label{CoreD_6}
    \end{subfigure}
    \caption{Examples of core-dominated sources that LoMorph automatically classifies as FRI. Colours and symbols as in Fig.~\ref{Examples}.}
    \label{CoreD_fig}
\end{figure*}

As described in Section~\ref{vis}, during the visual inspection of the sources automatically classified as FRI we found a population of 99 sources with high dynamic range (75 per cent have dynamic ranges $>4.5$) leading to an automatic FRI classification, but with an anomalous, sharp drop and subsequent rise in brightness beyond the core that does not resemble the behaviour of traditional FRIs. Some examples are presented in Fig.~\ref{CoreD_fig}. For the purposes of the analysis presented in Sections~\ref{Results},~\ref{friilow} and ~\ref{ledlow} we treated these sources as potential contaminants, as discussed in Section~\ref{vis}, and removed them from our sample.

This subset of core-dominated (core-D) AGN is clearly heterogeneous: our visual inspection indicates that it includes sources of intrinsically varied morphology, but is also likely to include sources whose morphology is uncertain due to the resolution and dynamic range limitations of the data. We can see from the statistics on Table ~\ref{FR_z_stats} that $\sim14$ per cent of core-D sources have $z>0.8$, a higher fraction than for the overall FRI population ($\sim3$ per cent), but lower than that of the FRIIs ($\sim18$ per cent), which also indicates that their properties could span both FR populations.

Indeed, one common type of core-D source has edge-brightened lobe structure, but an even brighter core (e.g. Fig.~\ref{CoreD_1},~\ref{CoreD_4}). These undoubtedly meet the traditional definition of being centre-brightened, but their lobes have an FRII-like appearance. In some cases these could be active FRIIs in which orientation leads to an exceptionally bright core \citep[see e.g.][]{Marin2016}, while in other cases they could be sources with ``warm'' rather than ``hot'' spots at the ends of the lobe so that hotspots may be fading, but a jet remains active with a lower power (due to a lower accretion rate from the AGN) or is restarting in the centre. Cases similar to the examples of Fig.~\ref{CoreD_2} and~\ref{CoreD_5} may be candidates for restarting activity in the centre of older lobes. Other cases are more ambiguous (e.g. Fig.~\ref{CoreD_3} and~\ref{CoreD_6}), but overall, the presence and heterogeneity of this class of core-dominated systems, which fall in the FRI class due to being centre-brightened, indicate the need to take great care in assuming that automated classifications will generate physically useful samples of AGN.

\begin{figure}
	\centering
\includegraphics[width=0.47\textwidth]{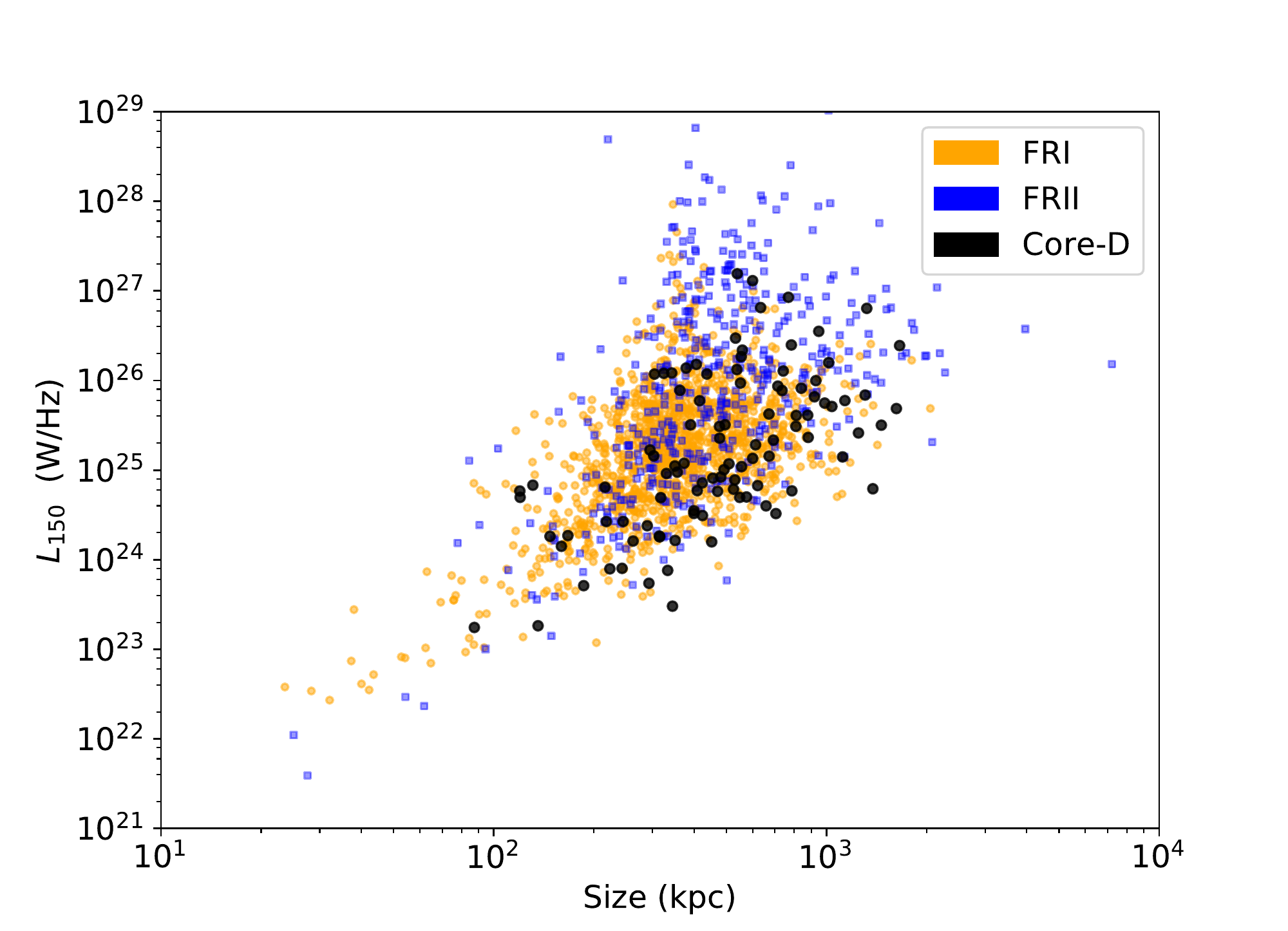}
	\caption{150 MHz luminosity versus physical size showing the core-dominated `FRI' (large black circles), overlaid on the FRI (orange circles) and FRII (blue squares).}\label{L150_size_coreD}
\end{figure}

The heterogeneity of the core-D subpopulation is further shown in Fig.~\ref{L150_size_coreD}, which indicates where these sources are located on the $L_{150}$-size plot. They span a very wide range of luminosities, extending well into the regions dominated by FRIIs at the high end and FRIs at the low end. Interestingly, the host galaxies of core-D sources are more similar to those of the FRII in Fig.~\ref{WISE_c_c}, being bluer on average than the FRIs, and including several clear HERGs. The core-D sources undoubtedly merit further investigation, and detailed studies to identify restarting populations in particular within the LoTSS datasets are ongoing (e.g. Jurlin et al. in prep). The upcoming extension of LoTSS to include the LOFAR international baselines, achieving even higher resolution, is also likely to help identify and characterise this population (e.g. Morabito et al., in prep). 

\subsubsection{Narrow and Wide Angle Tail sources}\label{NAT_WAT}

Bent-tailed sources are of particular interest because they are known to inhabit galaxy clusters and can act as tracers of high-redshift clusters and of cluster mergers \citep[e.g.][and references therein]{Blanton2001,Smolcic2007,Giaciuntucci2009,Mao2010,Garon2019}. The tails of these sources are believed to arise from a combination of the hosts moving through the intracluster medium (ICM), dragging the emission behind them, and strong cluster winds pushing the tails away from their hosts. Our LoTSS morphologically classified sample includes significant numbers of both narrow-angle tail (NAT) and wide-angle tail (WAT) sources (see Table~\ref{FR_z_stats}).

\begin{figure*}
\begin{subfigure}{0.47\linewidth}
        \centering
        \includegraphics[width=0.9\linewidth, trim={1.4cm 1cm 2cm 2.4cm},clip]{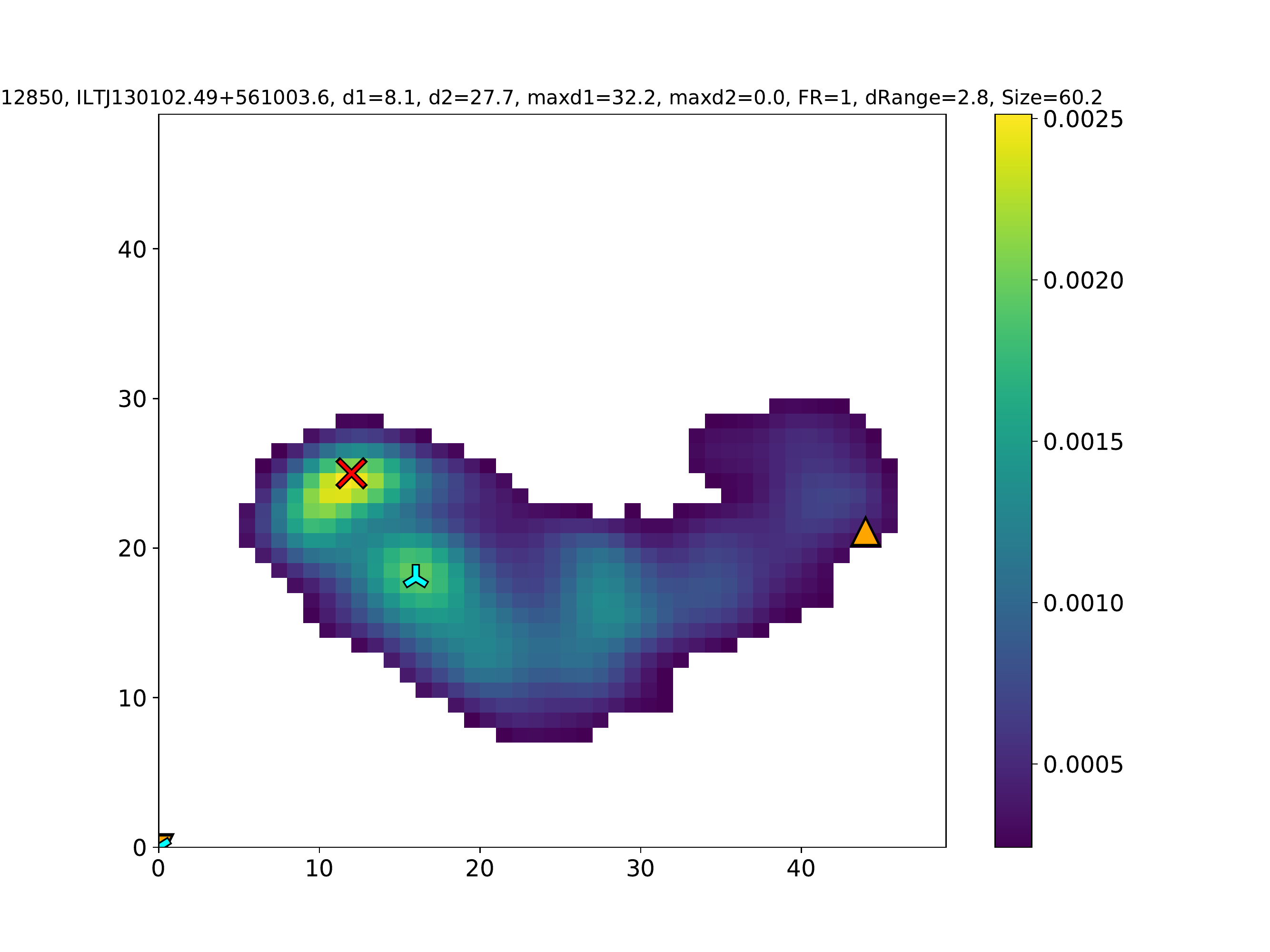}
        \caption{}
        \label{NAT_1}
    \end{subfigure}
    \begin{subfigure}{0.47\linewidth}
        \includegraphics[width=0.9\linewidth, trim={1.4cm 1cm 2cm 2.4cm},clip]{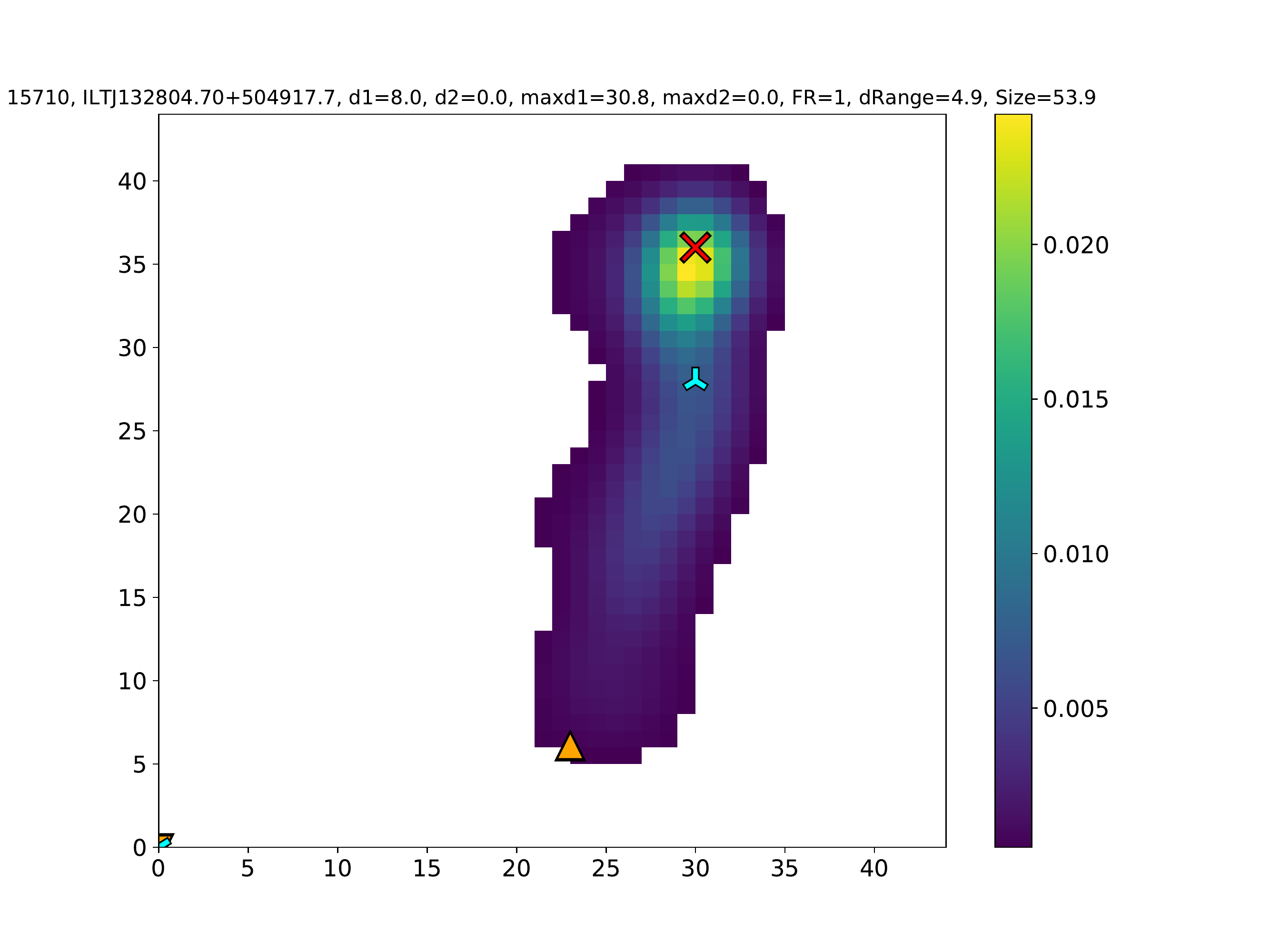}
        \caption{}
        \label{NAT_2}
    \end{subfigure}
    \begin{subfigure}{0.47\linewidth}
        \centering
        \includegraphics[width=0.9\linewidth, trim={1.4cm 1cm 2cm 2.4cm},clip]{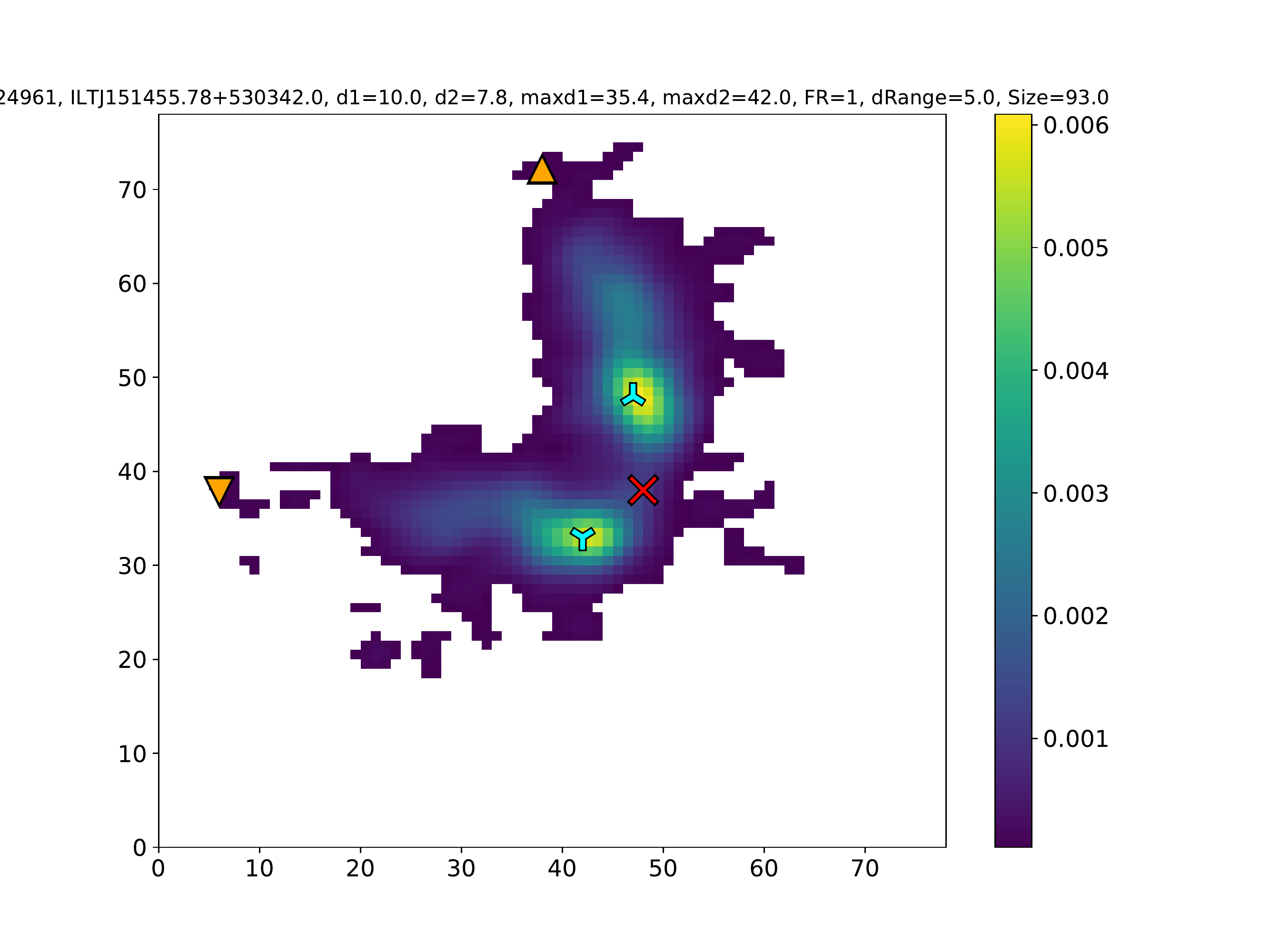}
        \caption{}
        \label{WAT_1}
    \end{subfigure}
    \begin{subfigure}{0.47\linewidth}
        \includegraphics[width=0.9\linewidth, trim={1.4cm 1cm 2cm 2.4cm},clip]{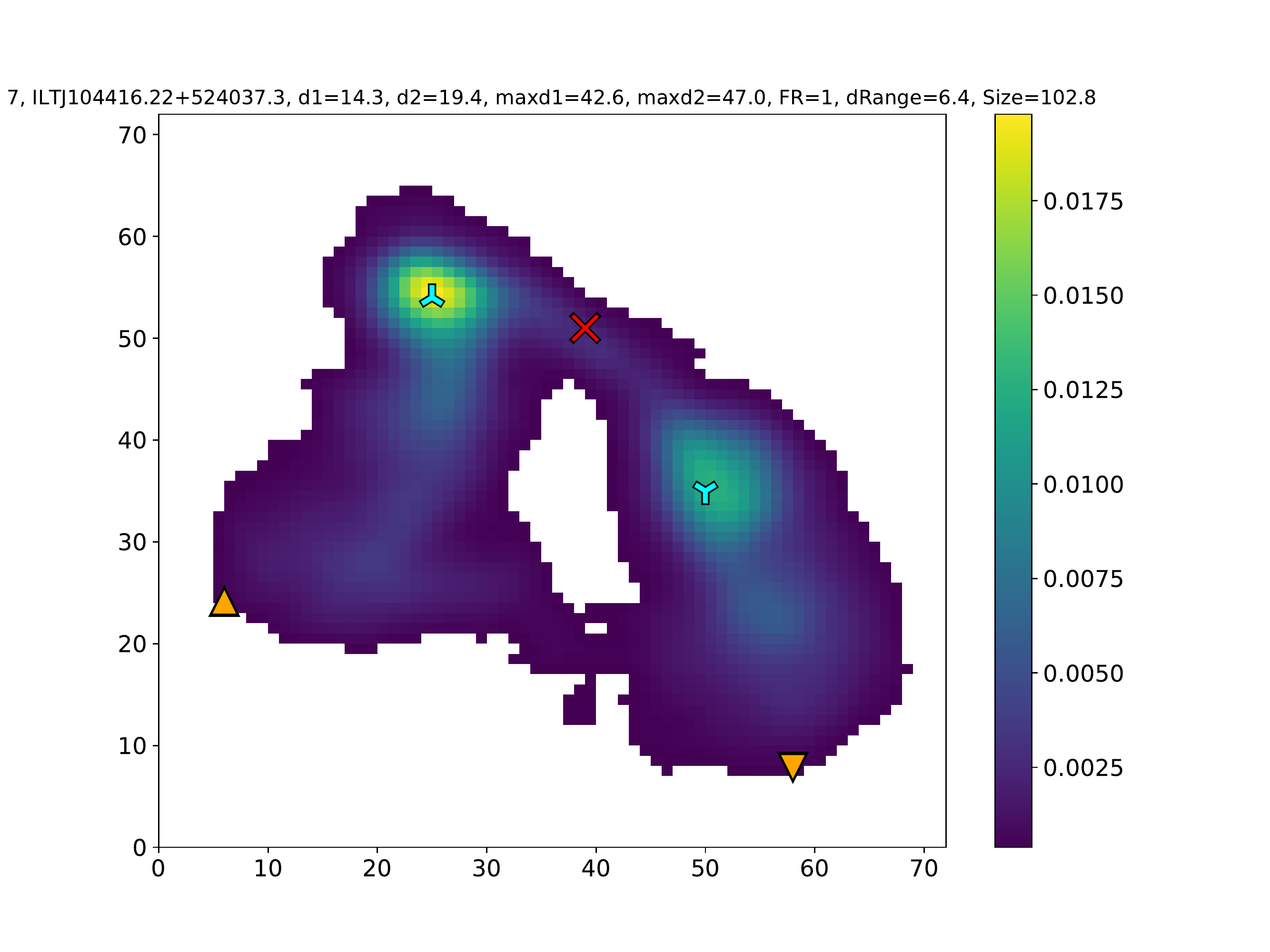}
        \caption{}
        \label{WAT_2}
    \end{subfigure}
    \caption{Examples of narrow-angle tail (NAT, top) and wide-angle tail (WAT, bottom) FRIs from our sample. Colours and symbols as in Fig.~\ref{Examples}.}
    \label{WAT_fig}
\end{figure*}

Fig.~\ref{WAT_fig} illustrates two highly bent examples of NAT found in our sample. Although the definition of what constitutes a NAT varies slightly within the radio astronomy community \citep[see e.g.][]{ODea1985NAT,Terni2017}, we have visually classified as NATs any head-tail sources and any objects with bright cores and an angle $<90^{\circ}$ between both tails. Where it was possible to identify them, we excluded sources where the one-sided emission was likely caused by orientation (core-jet, having one side of the jet highly beamed towards us and the other away from us), rather than a double, unresolved tail. In total we identify 264 NATs, $\sim$95 per cent of which have $z\leq0.8$.

Fig.~\ref{WAT_fig} also illustrates two examples of WATs found in our sample. We identified as WATs bent sources with initial opening angles larger than those of the NAT, and bright compact features less than halfway between the host position and the visible end of the source, giving rise to the tails \citep[Fig.~\ref{WAT_2} is one of the clearest examples; see also e.g.][]{Hardcastle2004WAT}. Roughly 50 per cent of the WATs in our sample have identifiable cores coincident with the optical position, and in some cases ($\sim 30$ per cent) the cores are brighter than the hotspot-like features. For some ($\sim 10$ per cent) these features are barely visible, making them slightly ambiguous cases between NAT and WAT. As discussed in Section~\ref{friilow}, it is also possible that some bent FRIIs are really WATs whose tails are too faint for LOFAR to detect. In total we identify 195 WATs in our sample, with only 3 sources ($\sim1$ per cent) at $z>0.8$.

\begin{figure*}
	\centering
\includegraphics[width=0.47\textwidth]{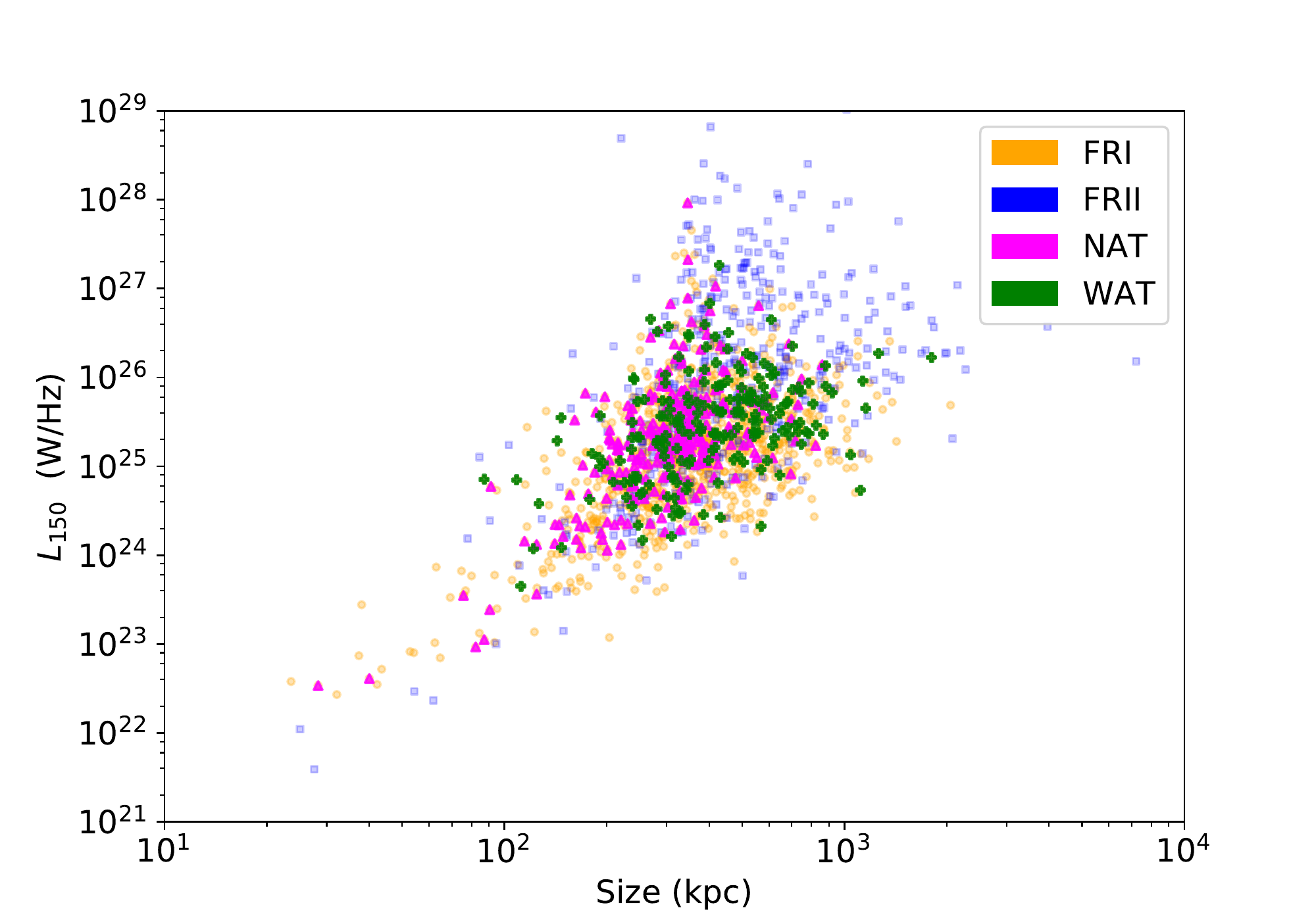}
	\includegraphics[width=0.47\textwidth]{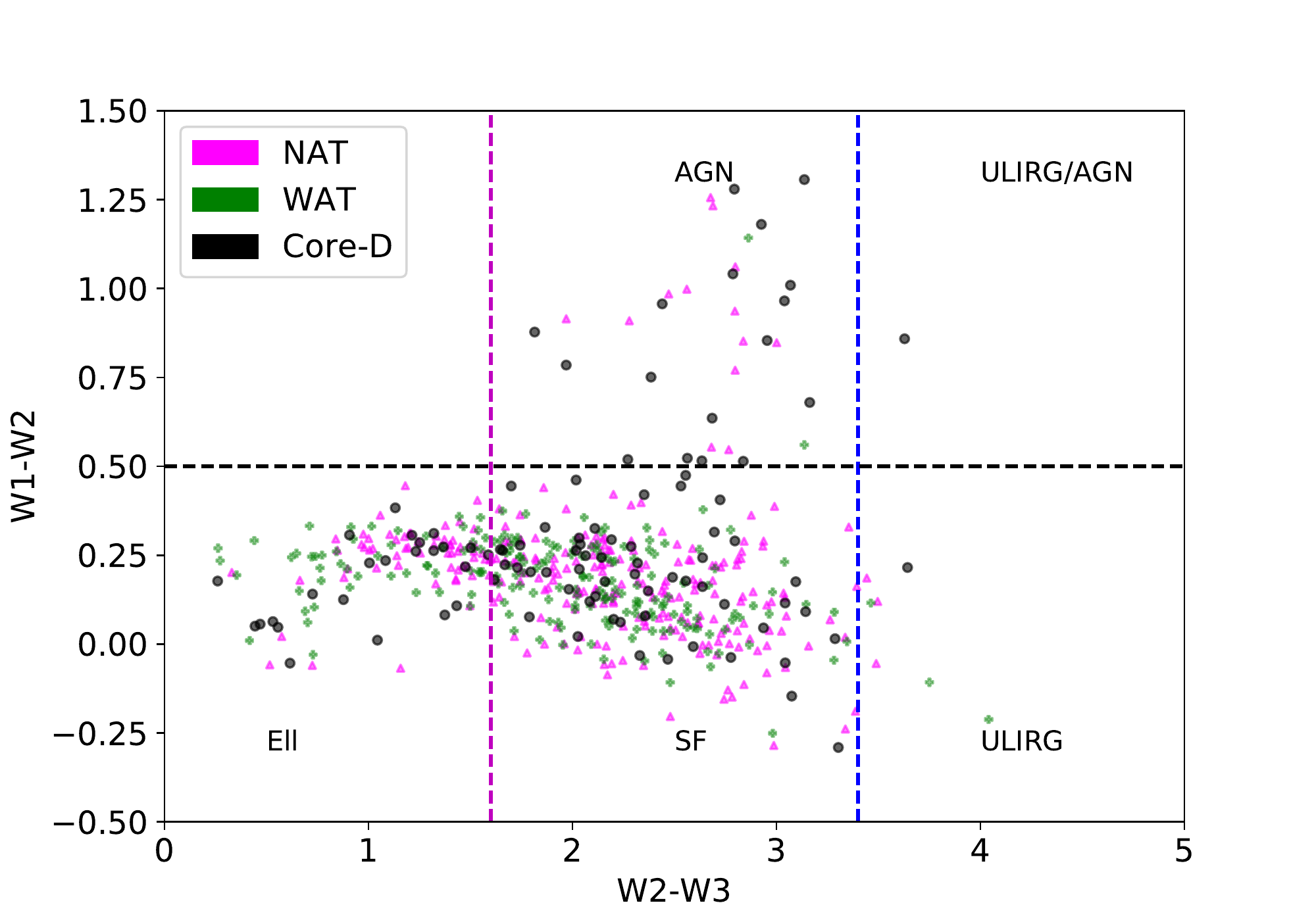}
	\caption{Properties of the NAT and WAT sources in our sample. Left: 150 MHz luminosity versus physical size for the NAT (green `+') and WAT (magenta triangles), overlaid on the FRI (orange circles) and FRII (blue squares). Right: WISE colour-colour plot with the same symbols as the left-hand plot, as well as the core-dominated `FRI' (large black circles). The lines are as in Fig.~\ref{WISE_c_c}.}\label{L150_size_NAT_WAT}
\end{figure*}

The left-hand panel of Fig.~\ref{L150_size_NAT_WAT} shows the $L_{150}$-size distribution for NAT and WAT, with FRI and FRII shown in faded colours for comparison. Although both populations overlap well with the FRI, the size distributions show a significant difference, with the NATs lying closer to the (smaller) left edge of the FRI distribution, and the WATs closer to the (larger) right edge, indicating that the latter reach slightly larger sizes. This is not surprising, as the largest angular size of most NATs will be smaller than that of an equivalent, `unfolded' (tailed) FRI. In terms of luminosity, the NATs and WATs overlap almost completely. 

The right-hand panel of Fig.~\ref{L150_size_NAT_WAT} shows the WISE properties of the NAT and WAT hosts. We see a large degree of overlap, but a few NATs are clearly in the bright HERG region of the plot. While these may be HERG NATs, it may be more likely that these sources are low-inclination, core-jet, beamed QSOs that we have failed to identify as such and exclude from the NAT sample. Apart from those objects, the NATs and WATs have similar host-galaxy colours, though with a slightly redder mean for WATs than NATs, consistent with the usual location of WATs in BCGs and NATs in infalling galaxies more likely to be star-forming.

Finally, we used the LoTSS environmental catalogue of \citet{Croston2018b} to investigate whether, as found by previous studies, the WAT and NAT populations within our sample reside in rich environments. We find cluster match fractions of $48\pm0.7$ per cent, and $49\pm0.7$ per cent for the WAT and NAT subsamples respectively, which are significantly higher than the match fractions of 10 -- 30 per cent found by \citet{Croston2018b} for the LoTSS AGN population as a whole. It is possible that the FRI and FRII environmental difference reported in that work (where FRI radio galaxies show a systematic increase in cluster association fraction with radio luminosity, and a
systematically higher association fraction than FRIIs at high
luminosity) may be entirely explained by the bent-tailed subsample that would have formed part of the FRI sample in that analysis, with non-tailed FRIs having similar environmental properties to the FRIIs.

It is perhaps surprising that the cluster match fraction is not higher than 50 per cent. One possible contributing factor is that a matching radius of 1 Mpc was used, to avoid an unacceptably high level of spurious associations. This may mean some associations have been missed for bent-tailed sources close to cluster outskirts. It is also important to emphasise that the SDSS catalogues are only sensitive down to $M_{500} > 10^{14}$ M$_{\sun}$ and so the majority of the other 50 per cent could be associated with moderately rich galaxy groups. It will be interesting to explore the environments of this sample in more detail in future work.


\section{Conclusions}\label{Conclusions}

We have carried out the largest morphological investigation of radio-loud AGN to date, using the LoTSS Data Release 1 value-added catalogue \citep{Williams2018b}. The survey depth, and unique sensitivity of LOFAR to spatial scales ranging from 6 arcsec to degrees, and the wide range of radio luminosities ($10^{22}$ to $10^{29}$ W Hz$^{-1}$ at 150 MHz), mean we have been able to investigate the diversity of radio-galaxy morphologies and the relationship with luminosity and host-galaxy properties in significantly greater detail than has previously been possible. Although our samples are not complete, they are representative of the overall radio-galaxy population up to $z\sim0.8$.

Our investigation of radio morphology has led to the following conclusions:
\begin{itemize}
    \item Sources of FRI and FRII morphology can both be found over a wide range in radio luminosity. While edge-brightened (FRII) sources on average have higher luminosities than centre-brightened (FRI) sources, as found by \citet{FR1974}, there is a very large overlap in luminosity for the two morphologies.
    \item We identify a subsample of low-luminosity FRIIs (FRII-Low), with luminosities extending three orders of magnitude below the canonical FR break luminosity.
    \item We show that these FRII-Low are likely to comprise a heteregeneous population, some being older FRIIs that are fading in luminosity from their peak, and many being hosted by comparatively low-mass galaxies enabling their jets to remain undisrupted.
    \item We find that the absence of FRII-Lows in previous flux-limited samples (e.g. 3CRR) can be entirely explained by their rarity in the local Universe, and the much higher flux limits of those earlier surveys. In future work it will be interesting to explore how host-galaxy evolution may be relevant for the relative prevalence of FRII-low and FRI radio galaxies.
    \item We find that the centre-brightened AGN population are also heterogeneous, including a population of core-dominated sources likely to include some restarting objects, core-bright/beamed FRIIs, as well as the significant populations of bent-tail sources (NATs and WATs). A small number of double-double sources are also present.
    \item We find that $\sim 50$ per cent of the WAT and NAT sources at $z<0.4$ are associated with SDSS clusters, significantly higher than the cluster match fraction for the general LoTSS radio-galaxy population, consistent with the previously identified preference for bent tailed sources to inhabit rich environments.
\end{itemize}

Our analysis and conclusions have important implications for future radio survey AGN science. We found that our purely automated approach to Fanaroff-Riley classification led to samples that were too physically inhomogeneous to allow for useful science, and found it necessary to supplement our algorithm with visual sorting in some cases. With the enormous datasets coming from the full LOFAR surveys, from other ongoing radio surveys (e.g. VLASS, MeerKAT MIGHTEE), and eventually the Square Kilometer Array (SKA), automated source classification is both essential and inevitable. It is crucial that we fully understand the underlying source populations, and the physical complexity that may be hidden within faint and/or poorly resolved images, before drawing broad scientific conclusions from the application of automated algorithms or machine learning approaches. The LOFAR surveys, and the LoTSS DR1 sample presented here, will provide a powerful basis for further investigation both of AGN populations and the development of robust classification approaches.


\section*{Acknowledgements}

We thank the anonymous reviewer for their helpful and insightful comments.
BM and JHC acknowledge support from the Science and Technology Facilities Council (STFC) under grants ST/R00109X/1 and ST/R000794/1. MJH acknowledges support from the UK Science and Technology Facilities Council [ST/R000905/1]. PNB and JS are grateful for support from the UK STFC via grant ST/R000972/1. KJD, HR, and WLW acknowledge support from the ERC Advanced Investigator programme NewClusters 321271. MB acknowledges support from INAF under the PRIN SKA/CTA ``FORECaST'' project. GG acknowledges the research fellowship from CSIRO. VHM thanks the University of Hertfordshire for a research studentship [ST/N504105/1]. LKM acknowledges the financial support of the Oxford Hintze Centre for Astrophysical Surveys which is funded through generous support from the Hintze Family Charitable Foundation. IP acknowledges support from INAF under the PRIN SKA/CTA “FORECaST” project. SM acknowledges funding through the Irish Research Council Postgraduate Scholarship scheme. The research leading to these results has received funding from the European Union Seventh Framework Programme FP7/2007-2013/ under grant agreement number 607254. This publication reflects only the author's view and the European Union is not responsible for any use that may be made of the information contained therein. This publication arises from research partly funded by the John Fell Oxford University Press (OUP) Research Fund. 

LOFAR, the LOw Frequency ARray designed and constructed by ASTRON, has facilities in several countries, which are owned by various parties (each with their own funding sources), and are collectively operated by the International LOFAR Telescope (ILT) foundation under a joint scientific policy. The ILT resources have benefited from the following recent major funding sources: CNRS-INSU, Observatoire de Paris and Universit{\'e} d'Orl{\'e}ans, France; BMBF, MIWF-NRW, MPG, Germany; Science Foundation Ireland (SFI), Department of Business, Enterprise and Innovation (DBEI), Ireland; NWO, The Netherlands; the Science and Technology Facilities Council, UK; Ministry ofScience and Higher Education, Poland. 
Part of this work was carried out on the Dutch national e-infrastructure with the support of the SURF Cooperative through grant e-infra 160022 \& 160152. The LOFAR software and dedicated reduction packages on \url{https://github.com/apmechev/GRID_LRT} were deployed on the e-infrastructure by the LOFAR einfragroup, consisting of J. B. R. Oonk (ASTRON \& Leiden Observatory), A. P. Mechev (Leiden Observatory) and T. Shimwell (ASTRON) with support from N. Danezi (SURFsara) and C. Schrijvers (SURFsara). 

This research made use of the Dutch national e-infrastructure with support of the SURF Cooperative (e-infra 180169) and the LOFAR e-infra group. The J{\"u}lich LOFAR Long Term Archive and the German LOFAR network are both coordinated and operated by the J{\"u}lich Supercomputing Centre (JSC), and computing resources on the Supercomputer JUWELS at JSC were provided by the Gauss Centre for Supercomputing e.V. (grant CHTB00) through the John von Neumann Institute for Computing (NIC).

This research has made use of the University of Hertfordshire high-performance computing facility (\url{http://uhhpc.herts.ac.uk/}) and the LOFAR-UK computing facility located at the University of Hertfordshire and supported by STFC [ST/P000096/1]. 

This research made use of \textsc{Astropy}, a community-developed core \textsc{Python} package for astronomy \citep{astropy2} hosted at \url{http://www.astropy.org/}, of \textsc{Matplotlib} \citep{Matplotlib}, and of \textsc{topcat} \citep{Taylor2005}. 

This publication makes use of data products from the Widefield Infrared Survey Explorer, which is a joint project of the University of California, Los Angeles, and the Jet Propulsion Laboratory/California Institute of Technology, and NEOWISE, which is a project of the Jet Propulsion Laboratory/California Institute of Technology. WISE and NEOWISE are funded by the National Aeronautics and Space Administration. The Pan-STARRS1 Surveys (PS1) have been made possible through contributions by the Institute for Astronomy, the University of Hawaii, the Pan-STARRS Project Office, the Max-Planck Society and its participating institutes, the Max Planck Institute for Astronomy, Heidelberg and the Max Planck Institute for Extraterrestrial Physics, Garching, The Johns Hopkins University, Durham University, the University of Edinburgh, the Queen's University Belfast, the Harvard-Smithsonian Center for Astrophysics, the Las Cumbres Observatory Global Telescope Network Incorporated, the National Central University of Taiwan, the Space Telescope Science Institute, and the National Aeronautics and Space Administration under Grant No. NNX08AR22G issued through the Planetary Science Division of the NASA Science Mission Directorate, the National Science Foundation Grant No. AST-1238877, the University of Maryland, Eotvos Lorand University (ELTE), and the Los Alamos National Laboratory. Funding for SDSSIII has been provided by the Alfred P. Sloan Foundation, the Participating Institutions, the National Science Foundation, and the U.S. Department of Energy Office of Science. The SDSS-III web site is \url{http://www.sdss3.org/}. SDSSIII is managed by the Astrophysical Research Consortium for the Participating Institutions of the SDSS-III Collaboration including the University of Arizona, the Brazilian Participation Group, Brookhaven National Laboratory, Carnegie Mellon University, University of Florida, the French Participation Group, the German Participation Group, Harvard University, the Instituto de Astrofisica de Canarias, the Michigan State/Notre Dame/JINA Participation Group, Johns Hopkins University, Lawrence Berkeley National Laboratory, Max Planck Institute for Astrophysics, Max Planck Institute for Extraterrestrial Physics, New Mexico State University, New York University, Ohio State University, Pennsylvania State University,
University of Portsmouth, Princeton University, the Spanish Participation Group, University of Tokyo, University of Utah, Vanderbilt University, University of Virginia, University of Washington, and Yale University.


\bibliographystyle{mnras}
\bibliography{bmingo}

\label{lastpage}
\end{document}